\documentstyle[aps,epsfig,floats,amssymb]{revtex}
\draft

\newcommand{\qslash}{\mathbin{q\mkern-10mu\big/}}
\newcommand{\pslash}{\mathbin{p\mkern-10mu\big/}}

\begin{document}
\twocolumn[\hsize\textwidth\columnwidth\hsize\csname
@twocolumnfalse\endcsname

\title{Chiral freedom and electroweak symmetry breaking}

\author{C. Wetterich\footnote{}}

\address{
Institut f{\"u}r Theoretische Physik,
Philosophenweg 16, 69120 Heidelberg, Germany}

\maketitle

\begin{abstract}
Antisymmetric tensor fields with chiral couplings to quarks and leptons may induce  spontaneous electroweak symmetry breaking in a model without a ``fundamental'' Higgs scalar. No microscopic local mass term for the chiral tensors or ``chirons'' is allowed by the symmetries and our model exhibits only dimensionless couplings. However, the chiral couplings are asymptotically free and therefore generate a mass scale where they grow large. We argue that at this scale mass terms for the chiral tensor fields are generated non-perturbatively - the chirons appear as new massive spin one particles. Furthermore a scalar top-antitop condensate forms, giving mass to the weak gauge bosons and fermions. In this scenario the longstanding gauge hierarchy problem finds a solution similar to the mass generation in QCD. We compute the general form of the effective action for the chiral tensors and sketch several possibilities of their detection at LHC or through precision tests of the electroweak standard model.
\end{abstract}

 ]

\section{Introduction}
\label{intro}
Spontaneous symmetry breaking in the standard model of weak and electromagnetic interactions \cite{GSW} can be described in a  consistent way by the Higgs mechanism \cite{H}. Still, doubts have been raised concerning the naturalness of a fundamental Higgs scalar within a unified theory of all interactions. The gauge hierarchy problem centers around the question: why is the Fermi scale characteristic for electroweak symmetry breaking so many orders of magnitude smaller than the scale of unification or the Planck scale $M_p$ characteristic for gravity? While the ``fine tuning problem'' can be avoided by the use\footnote{The renormalized scalar mass term $\mu^2_\varphi$ is a ``relevant para\-meter''. In the renormalization flow $\mu^2_\varphi=0$ is an infrared-unstable partial fixed point characterized by the effective dilatation symmetry realized on the critical hypersurface for an (almost) second order phase transition \cite{CWH}. Small deviations from this fixed point remain small during the renormalization flow, evolving according to an anomalous dimension. Including higher loops changes the quantitative precision of the anomalous dimension, but no tuning order by order of a small renormalized microscopic mass term is necessary. \\$^*$Electronic address: C. Wetterich@thphys.uni-heidelberg.de} of a proper renormalization group improved perturbation theory \cite{CWH}, the basic problem remains to understand why the scale of spontaneous electroweak symmetry breaking is so small. This issue does not change in models where the Higgs scalar arises as a composite top-antitop bound state in models with microscopic four-fermion interactions \cite{BL}. For a wide class of such models it has been shown \cite{GJW} that the microscopic parameters must be extremely close to critical values in order to obtain a small value of the $W$-boson mass $M_W$. The problem of a genuine understanding of the tiny ratio $M_W/M_p$ also remains in many supersymmetric models.

On the other side, a naturally small ratio between the nucleon mass $m_N$ (or $\Lambda_{QCD}$) and the unification mass is realized in QCD through the mechanism of dimensional transmutation. In asymptotically free theories a dimensionless running coupling grows large at a certain momentum scale - in QCD this scale determines the mass of the proton. Technicolor models employ this mechanism for the generation of the Fermi scale via an additional strong gauge interaction for unknown particles. Characteristic problems in these models concern the explanation of the chiral structures in the standard model. Indeed, in the standard model the Higgs scalar is not only responsible for the masses of the weak gauge bosons, but also for the masses of quarks and leptons through their Yukawa couplings. These Yukawa couplings connect left handed and right handed fermions and carry the ``chiral information'' of the model, i.e. they specify the amount of explicit violation of the global chiral flavor symmetries of a gauge theory. The chiral information is closely connected to flavor violation, in particular to the issue of flavor changing neutral currents. In the standard model the necessary suppression of such currents is naturally achieved through the GIM-mechanism \cite{GIM}, but this is far from automatic in models with additional particles.

The recently proposed idea of chiral freedom \cite{CWCH} exploits dimensional transmutation for a natural solution of the gauge hierarchy problem. New strong interactions arise from the asymptotically free chiral couplings of antisymmetric tensor fields to quarks and leptons. These chiral couplings also carry the chiral information. Flavor changing decays proceed only through the CKM-matrix \cite{CKM} for the coupling of the $W$-boson and the GIM-mechanism is realized. Thus flavor- and CP-violation are in close analogy to the standard model, while the generation of the electroweak scale resembles more the technicolor idea.

Our model contains quarks, leptons, gauge bosons and antisymmetric tensor fields, but no fundamental scalar fields. With respect to the Lorentz-symmetry antisymmetric tensor fields belong to the complex irreducible representations $(3,1)$ or $(1,3)$, denoted here by $\beta^+_{\mu\nu}$ and $\beta^-_{\mu\nu}$, respectively. In distinction to a scalar field, a mass term can be forbidden by the symmetries. This is in close analogy to mass terms for chiral fermions which only can arise through electroweak symmetry breaking. Indeed, local mass terms of the type $(\beta^+)^*\beta^+$ are forbidden by the Lorentz symmetry since the corresponding bilinear is not a Lorentz-singlet but rather transforms as (3,3). In our model the chiral tensors $\beta^\pm$ are weak doublets carrying hypercharge, with the same internal quantum numbers as the Higgs scalar. Therefore mass terms $\sim(\beta^+)^2$ are not allowed by the gauge symmetries. Finally, mass terms of the type $(\beta^+)^*\beta^-$ will be excluded by a discrete symmetry $\beta^-\rightarrow-\beta^-$. In consequence, no local mass term is allowed in absence of spontaneous symmetry breaking, just as for the gauge bosons and chiral fermions. All renormalizable couplings of our model are dimensionless and no explicit mass scale appears in the microscopic (classical) action.

Quantum effects result in running couplings. In particular, the chiral couplings are asymptotically free in the ultraviolet and grow large near the Fermi scale. Effective strong interactions between the top quarks are induced by the exchange of chiral tensors. We argue that the corresponding strong attractive force leads to the condensation of top-antitop pairs in the vacuum. Such top condensates \cite{BL} provide masses for the weak gauge bosons and the fermions through the usual Higgs mechanism. The ``fundamental'' Higgs scalar is replaced here by a composite top-antitop bound state which transforms as scalar weak doublet. At a similar scale mass terms for the chiral tensor fields are generated. The chiral tensors appear as new massive spin one fields which may be detected at future colliders as the LHC. 

Antisymmetric tensor fields have been discussed earlier \cite{KEM,WH,CH}. While ``vectorlike'' antisymmetric tensors allow for an additional gauge symmetry and a consistent free theory, the ``chiral tensors'' or ``chirons'' of our model do not admit additional local symmetries. The free theory for chiral tensors is in a sense on the borderline between stability and instability: the energy density for some plane waves vanishes, whereas others have positive energy. (There are no standard ghosts or tachyons.) Due to ``higher order instabilities'', i.e. secular classical solutions, the free theory is actually instable, both on the classical and quantum level \cite{CWQ}. As typical for a borderline situation, one may expect that the interactions decide on which side of the divide between stability and instability the model lies. The interactions of the chiral tensors comprise self interactions as well as the chiral interactions with the fermions and the electroweak gauge interactions with the gauge bosons. 

If the instability is cured by the interactions one concludes that the vacuum cannot be the perturbative vacuum and the propagator for the chiral tensors must be modified by non-perturbative effects. Indeed, stability depends on the properties of the effective action including quantum effects,not on the classical action. We propose in this paper a mechanism that generates the appropriate mass term in the chiral tensor propagator non-perturbatively. The size of the chiron-mass is given by the ``chiral scale'' where the chiral couplings grow large. It is therefore of the same order as the Fermi scale. In presence of this mass term the theory is stable and contains no ghosts or tachyons.

The three independent components of $\beta^+_{\mu\nu}$ are then associated to the three components of a massive spin one particle. This matching of components occurs without the need of additional Goldstone boson degrees of freedom - the massless spin one and spin zero states of the free theory combine to a massive particle in presence of the mass term. Furthermore, the terms in the effective action allowed by the symmetries will lead to a mixing between the various spin one bosons, including the gauge fields. Such mixings result in additional deviations from the standard model already for energies below the mass of the chiral tensors. The new effective couplings are constrained by the LEP-experiments and may show up in precision experiments, as for example the measurement of the anomalous magnetic moment of the muon. We outline various possibilities of ``low energy detections'' of the effects of our model, including anomalies of radiative pion decays. 

This paper is organized as follows: In sect. II we introduce the chiral tensors of our model and describe the most general action containing only dimensionless couplings. (Couplings with positive mass dimension are excluded by the the symmetries.) 
One expects this theory to be renormalizable. In sect. III we discuss the classical propagators and address the issue of instability of the free theory. Sect. IV displays the evolution equations for the running chiral couplings and the generation of the Fermi scale by dimensional transmutation. In sect. V we present solutions to gap equations which suggest that the top quark becomes massive through a non-vanishing top-antitop condensate. Composite Higgs fields corresponding to quark-antiquark bound states are introduced in sect. VI. There we derive the most general local effective interactions with dimension up to four (dimensionless couplings) for the effective low energy comprising fermions, gauge bosons, chirons and composite scalars. 

In sect. VII we turn to the issue of mass terms for the chiral tensor fields. We argue that non-perturbative effects generate a nonlocal correction to the chiron propagator which acts as a mass term. This mass term is local in a different field basis where the particle content of the chiral tensor fields becomes manifest: massive spin-one bosons. Indeed, written in terms of a vector field the mass term and kinetic term take a standard local form, such that the nonlocality only concerns the transition between the vector-field and tensor-field basis through appropriate projectors. In sect. VIII we discuss the mass matrix for the spin-one bosons in more detail, in particular the mixing between chirons and gauge bosons. This mixing contributes to the anomalous magnetic moment for the muon and consistency with observation requires a chiron mass larger than around $300$ GeV. Sect. IX turns to a first sketch of the phenomenological implications of our model. We discuss the effective interactions between the quarks, leptons and gauge bosons which are mediated by the exchange of the massive chiral tensor fields. This provides for an effective theory where the chiral tensor fields are ``integrated out''. The electroweak precision tests at LEP are consistent with our model provided the mass of the chirons is above around $300$ GeV. Sect. X presents our conclusions.

\section{Symmetries for antisymmetric tensor doublets}
The principal idea of this paper relies on the observation that the chiral information can also be carried by an antisymmetric tensor field\footnote{Lorentz indices $m,n$ are converted to world indices $\mu,\nu$ by the vielbein $e^m_\mu$, i.e. $\beta_{\mu\nu}=e^m_\mu e^n_\nu\beta_{mn}$. The use of $\beta^\pm_{mn}$ as basic fields facilitates the computation of the energy momentum tensor. Otherwise the distinction between Lorentz- and world indices is unimportant in absence of gravity. The Lorentz-indices $m,n...$ are raised by $\eta^{mn}=diag(-1,+1,+1,+1)$.} $\beta_{mn}(x)=-\beta_{nm}(x)$. While the renormalizable interactions between the fermions $\psi$ and the gauge bosons preserve a large global flavor symmetry, the observed fermion masses imply that this flavor symmetry must be broken. This explicit breaking must be mediated by couplings connecting the left- and right-handed fermions $\psi_L$ and $\psi_R$ which belong to the representations $(2,1)$ and $(1,2)$ of the Lorentz symmetry. Bilinears involving $\psi_L$ and $\bar{\psi}_R$ transform as $(2,1)\otimes(2,1)=(3,1)+(1,1)$ and similar for $\psi_R\otimes\bar{\psi}_L=(1,3)+(1,1)$. The violation of the flavor symmetry may therefore either involve the Higgs scalar $(1,1)$ or the antisymmetric tensor $\beta_{mn}~(3,1)+(1,3)$. In fact, the six components of $\beta_{mn}=-\beta_{nm}$ can be decomposed into two inequivalent three dimensional irreducible representations of the Lorentz group
\begin{equation}\label{1}
\beta^\pm_{mn}=\frac{1}{2}\beta_{mn}\pm\frac{i}{4}\epsilon_{mn}\ ^{pq}\beta_{pq},
\end{equation}
with $\epsilon_{mnpq}$ the totally antisymmetric tensor $(\epsilon_{0123}=1)$. We observe that the representations $(3,1)$ and $(1,3)$ are complex conjugate to each other. 

The complex fields $\beta^\pm_{mn}$ should belong to doublets of weak isospin and carry hypercharge $Y=1$. The explicit breaking of the flavor symmetry can then be encoded in the ``chiral couplings'' $\bar{F}_{U,D,L}$, i.e.
\begin{eqnarray}\label{2}
-{\cal L}_{ch}&=&\bar{u}_R\bar{F}_U\tilde{\beta}_+q_L-\bar{q}_L
\bar{F}^\dagger_U\stackrel{\eqsim}{\beta}_+u_R\nonumber\\
&&+\bar{d}_R\bar{F}_D\bar{\beta}_-q_L-\bar{q}_L\bar{F}^\dagger_D\beta_-d_R\nonumber\\
&&+\bar{e}_R\bar{F}_L\bar{\beta}_-l_L-\bar{l}_L\bar{F}^\dagger_L\beta_-e_R.
\end{eqnarray}
Here $u_R$ is a vector with three flavor components corresponding to the right handed top, charm and up quarks and similar for $d_R$ and $e_R$ for the right handed $(b,s,d)$ and $(\tau,\mu, e)$ or $q_L,l_L$ for the three flavors of left handed quark and lepton doublets. Correspondingly, the chiral couplings $\bar{F}_{U,D,L}$ are $3\times 3$ matrices in generation space. We have used the short hands $(\sigma^{mn}\gamma^5 =\frac{i}{2}\epsilon^{mn}\ _{pq}\sigma^{pq})$ 
\begin{eqnarray}\label{3}
\beta_\pm &=&\frac{1}{2}\beta^\pm_{mn}\sigma^{mn}=\frac{1}{2}\beta_{mn}\sigma^{mn}_\pm 
=\beta_{\pm}\frac{1\pm\gamma^5}{2}\nonumber\\
\bar{\beta}_\pm&=&\frac{1}{2}(\beta^\pm_{mn})^*\sigma^{mn}
=D^{-1}\beta^\dagger_\pm D =\bar{\beta}_\pm \frac{1\mp\gamma^5}{2}\nonumber\\
\tilde{\beta}_+&=&-i\beta^T_+\tau_2~,~ \stackrel{\eqsim}{\beta}_+=i\tau_2\bar{\beta}_+
\end{eqnarray}
with $\sigma_{\pm}^{mn}=\frac{1}{2}(1\pm\gamma^5)\sigma^{mn},~\sigma^{mn}=\frac{i}{2}
[\gamma^m,\gamma^n]$ and $\bar{\psi}=\psi^\dagger D,~D=\gamma^0,~\gamma^5=-i\gamma^0\gamma^1\gamma^2\gamma^3,~\psi_L=\frac{1}{2}(1+\gamma^5)\psi$. The transposition $\beta^T$ and $\tau_2$ act in weak isospin space, i.e. on the two components of the weak doublet $\beta^+_{mn}$.

The Lagrangian (\ref{2}) accounts for the most general renormalizable interactions between $\beta$ and the fermions which are consistent with the gauge symmetries. Beyond conserved baryon number $B$ and lepton number $L$ it is further invariant with respect to a discrete $Z_2$-symmetry that we denote by $G_A$. It acts 
\begin{equation}\label{A2a}
G_A(d_R)=-d_R,~G_A(e_R)=-e_R,~G_A(\beta^-)=-\beta^-
\end{equation}
while all other fields are invariant. We will require that our model preserves the $G_A$-symmetry. The $G_A$-symmetry has an important consequence: no local mass term is allowed for the fields $\beta^\pm$! Indeed, a Lorentz singlet is contained in $\beta^+\beta^+$ or $\beta^+(\beta^-)^*$, the first being forbidden by hypercharge and the second by $G_A$-symmetry. More generally, mass terms are forbidden by any symmetry under which $\beta^+$ and $\beta^-$ transform differently such that $\beta^+(\beta^-)^*$ is not invariant. As an example, we note that the Lagrangian (\ref{2}) is also invariant under a continuous global axial $U(1)_A$ symmetry with charges $A=1$ for the right handed fermions, $A=-1$ for the left handed ones, $A=-2$ for $\beta^-$ and $A=2$ for $\beta^+$. A mass term for the chiral tensor fields is also forbidden by $U(1)_A$ or by any discrete $Z_N$-subgroup of $U(1)_A, N>4$.

The absence of an allowed mass term is in sharp contrast to the Higgs mechanism and constitutes  a crucial part of the solution of the hierarchy problem. On the other hand, there exist kinetic terms $\sim(\beta^+)^*\beta^+$ and $(\beta^-)^*\beta^-$. The most general one allowed by all symmetries reads 
\begin{eqnarray}\label{4}
-{\cal L}^{ch}_{\beta,kin}&=&-\{Z_+ (D_\mu\beta^+_{mp})^\dagger D_\nu\beta^+_{nq}\nonumber\\
&&+Z_-(D_\mu\beta^-_{mp})^\dagger D_\nu\beta^-_{nq}\}
e^{m\mu}e^{n\nu}\eta^{pq}.
\end{eqnarray}
Here $D_\mu$ is the covariant derivative involving the gauge fields of $SU(2)_L\times U(1)_Y$ and the gravitational spin connection and $e^m_\mu$ denotes the vielbein. The action 
\begin{equation}\label{No1}
S_M=\int d^4xe{\cal L}~, ~e=\det(e^m_\mu)=\sqrt{g}
\end{equation}
involves in addition the usual covariant kinetic term for the fermions and gauge fields of the standard model, as well as gravitational parts. 

By a suitable rescaling of the fields $\beta^\pm$ we can always obtain $Z_+=Z_-=1$. For vanishing gauge fields and flat space the kinetic term (\ref{4}) can then be written in the form
\begin{equation}\label{No2}
-{\cal L}^{ch}_{\beta,kin}=\frac{1}{4}\int d^4x
\{(\partial^\rho\beta^{\mu\nu})^*\partial_\rho\beta_{\mu\nu}-4
(\partial_\mu\beta^{\mu\nu})^*\partial_\rho\beta^\rho\ _\nu\}.
\end{equation}
In this form it was encountered in extended conformal supergravity \cite{WH} and discussed in \cite{CH}.  
In Minkowski space ${\cal L}^{ch}_{\beta,kin}$ is real for real $Z_\pm$. More precisely, the Lagrangian is invariant under hermitean conjugation provided $\beta_{\mu\nu}$ is transformed into its complex conjugate. (Hermitean conjugation includes a transposition - i.e. reordering - of the Grassmann variables $\psi$.)

Dimensionless quartic couplings appear in the form 
$(\beta^\dagger\beta)^2, (\beta^\dagger\vec{\tau}\beta)^2$ with appropriate contractions of Lorentz indices and an even number of factors involving $\beta^+$ or $\beta^-$. More in detail, the quartic interactions are described by 10 real dimensionless parameters
\begin{eqnarray}\label{QI}
-{\cal L}_{\beta,4}&=&\frac{\tau_+}{16}\big[(\beta^+_{\mu\rho})^\dagger\beta^{+\rho\nu}\big]
\big[(\beta^{+\mu\sigma})^\dagger\beta^+_{\sigma\nu}\big]+
(+\rightarrow-)\nonumber\\
&&+\frac{\tau_1}{16}
\big[(\beta^+_{\mu\nu})^\dagger\beta^{-\mu\nu}\big]\big[(\beta^-_{\rho\sigma})^\dagger
\beta^{+\rho\sigma}\big]\nonumber\\
&&+\frac{\tau_2}{16}\big[(\beta^+_{\mu\nu})^\dagger\vec{\tau}\beta^{-\mu\nu}\big]
\big[(\beta^-_{\rho\sigma})^\dagger\vec{\tau}\beta^{+\rho\sigma}\big]\nonumber\\
&&+\frac{\tau_3}{64}\big[(\beta^+_{\mu\nu})^\dagger\beta^{-\mu\nu}\big]
\big[(\beta^+_{\rho\sigma})^\dagger\beta^{-\rho\sigma}\big]+c.c.\nonumber\\
&&+\frac{\tau_4}{64}\big[(\beta^+_{\mu\nu})^\dagger\beta^{-\rho\sigma}\big]
\big[(\beta^{+\mu\nu})^\dagger\beta^-_{\rho\sigma}\big]+c.c.\nonumber\\
&&+\frac{\tau_5}{64}\big[(\beta^+_{\mu\nu})^\dagger\beta^{-\rho\sigma}\big]
\big[(\beta^+_{\rho\sigma})^\dagger\beta^{-\mu\nu}\big]+c.c. .
\end{eqnarray}
Here [~] indicates the contraction of $SU(2)_L$-indices and we have used the identity (for doublets $\chi$ with $\alpha$ etc. denoting different doublets)
\begin{equation}\label{No1a}
[\chi^\dagger_\alpha\vec{\tau}\chi_\beta][\chi^\dagger_\gamma\vec{\tau}\chi_\delta]=2
[\chi^\dagger_\alpha\chi_\delta][\chi^\dagger_\gamma\chi_\beta]-
[\chi^\dagger_\alpha\chi_\beta][\chi^\dagger_\gamma\chi_\delta]
\end{equation}
in order to bring ${\cal L}_{\beta,4}$ into a standard form. The couplings $\tau_\pm,\tau_1,\tau_2$ are real whereas $\tau_3,\tau_4,\tau_5$ are complex. The quartic interactions (\ref{QI}) are the most general ones consistent with the discrete symmetry $G_A:\beta^-\rightarrow-\beta^-,\beta^+\rightarrow\beta^+$. For $\tau_3=\tau_4=\tau_5=0$ our model exhibits a global $U(1)_A$ symmetry.

Finally, we add the usual gauge invariant kinetic terms for the gauge fields and fermions
\begin{eqnarray}\label{c2}
-{\cal L}_F&=&\frac{1}{4}F^{\mu\nu}F_{\mu\nu},\nonumber\\
-{\cal L}_{\psi,kin}&=&ie^\mu_m
\{\bar{\psi}_L\gamma^m D_\mu \psi_L+\bar{\psi}_R\gamma^mD_\mu\psi_R\}.
\end{eqnarray}
(We have suppressed indices and summations for the different species of fermions and gauge fields.) 
Our model involving fermions, gauge bosons and chiral tensor fields is specified by
\begin{equation}\label{c3}
{\cal L}={\cal L}^{ch}_{\beta,kin}+{\cal L}_{ch}+{\cal L}_{\beta,\psi}+{\cal L}_F+{\cal L}_{\psi,kin}.
\end{equation}
It does not contain a fundamental scalar field. 

We observe that our model contains only dimensionless couplings. There are three gauge couplings and thirteen real parameters in the chiral couplings $F_{U,D,L}$ that cannot be absorbed by chiral rotations (nine mass eigenvalues, three CKM-mixings and one CP-phase). Furthermore, we have ten real parameters for the quartic couplings $\tau_i$. If the influence of the quartic couplings $\tau_i$ is small, the number of relevant parameters is smaller than for the standard model with a ``fundamental'' Higgs scalar. This would lead to an enhanced predictive power. 

With respect to parity the tensor field transforms as
\begin{equation}\label{No3b}
\begin{array}{llcl}
P:&\beta^{+,0}_{kl}\rightarrow (\beta^{+,0}_{kl})^*&,&
\beta^{+,0}_{0k}\rightarrow -(\beta^{+,0}_{0k})^*~,\\
&\beta^{+,+}_{kl}\leftrightarrow -\beta^{-,+}_{kl}&,&
\beta^{+,+}_{0k}\leftrightarrow \beta^{-,+}_{0k}~,\\
&\beta^{-,0}_{kl}\rightarrow (\beta^{-,0}_{kl})^*&,&
\beta^{-,0}_{0k}\rightarrow -(\beta^{-,0}_{0k})^*~,
\end{array}
\end{equation}
where $\beta^{+,0}$ and $\beta^{+,+}$ denote the electrically neutral and charged components of $\beta^+$. (In addition, the spacelike coordinates are switched, 
$x^\mu=(t,\vec{x})\rightarrow x^{\prime\mu}=(t,-\vec{x})$.)
The chiral interactions are parity conserving in the quark sector only for $F_U=F_D=F_U^\dagger=F^\dagger_D$ whereas the lepton sector always violates parity since no right handed neutrino is present. Similarly, the transformation under charge conjugation
\begin{equation}\label{No3c}
C:~\beta^{\pm,0}_{\mu\nu}\rightarrow -\beta^{\pm,0}_{\mu\nu}~,~
\beta^{\pm,+}_{\mu\nu}\rightarrow \beta^{\mp,-}_{\mu\nu}
\end{equation}
implies $C$-invariance in the quark sector for $F_U=F_D=F^T_U=F^T_D$. The chiral interactions are invariant with respect to the combined CP symmetry 
\begin{equation}\label{No3a}
CP: \quad \beta^\pm_{kl}\rightarrow -(\beta^\pm_{kl})^*~,~(\beta^\pm_{0k})\rightarrow
(\beta^\pm_{0k})^*
\end{equation}
provided the chiral couplings $F_{U,D,L}$ are real. (It is sufficient that this holds in an appropriate flavor-basis.) The CP transformation of a gauge field with symmetric generator (i.e. the photon, $Z-$ and $W_1$-boson) reads $A_k(x)\rightarrow A_k(x'), A_0(x)\rightarrow -A_0(x')$, such that in terms of the parity reflected coordinates $x'$ the magnetic field strength $F_{kl}$ switches sign whereas the electric field strength $F_{0k}$ is invariant. (For gauge fields with antisymmetric generators like $W_2$ one has an additional minus sign.) The kinetic term (\ref{4}) as well as the gauge sector of the action (\ref{c2})  are CP-invariant in the chiral basis. 

Flavor violation arises in the chiral basis only through the chiral couplings (\ref{2}). Also CP-violation originates from the chiral couplings. (Additional CP-violation effects from the quartic couplings (\ref{QI}), i.e. the imaginary parts of $\tau_{3,4,5}$, will be neglected in this paper and can easily be added for a  more extended discussion.) By appropriate chiral rotations we may bring the matrices for the chiral couplings $\bar{F}_{U,D,L}$ to a real and diagonal form. In this new  basis of ``mass eigenstates'' all effects of generation mixing and CP-violation occur only in the non-diagonal CKM-matrix for the couplings of the weak gauge bosons to the fermions. This has important consequences for the strength of effective strangeness-violating neutral currents or observable CP-violation. Their suppression is very analogous to the standard model. In the limit of vanishing electroweak gauge couplings the generation mixing and CP violation disappears. 

\section{Classical Propagators}
The classical action $S$ is the starting point for the definition of the quantum field theory by a functional integral. The classical propagators for $\beta^\pm$ are found by inverting the second functional variation of $S$ with respect to $\beta^\pm$. They are crucial for the perturbative loop expansion and we will discuss them in detail in this section. The propagators show unusual features which are related to the classical and quantum instabilities of the free theory that are discussed in \cite{CWQ}. Nevertheless, we recall that the propagation properties of the physical excitations (particles) are related to the second functional derivative of the quantum effective action $\Gamma$ instead of the classical action $S$. The properties of $\Gamma$ differ substantially from $S$. In particular, we will argue in sect. VII that the physical particles are massive, in contrast to the massless fields described by the classical propagators. The material of the present section should therefore not be mistaken as a discussion of the physical spectrum of our model. 

For a detailed analysis of the propagator we represent $\beta^\pm_{\mu\nu}$ explicitely by three-vectors $B^\pm_k$, according to the three dimensional irreducible representations of the Lorentz group:
\begin{eqnarray}\label{No4}
\begin{array}{lcl}
\beta^+_{jk}=\epsilon_{jkl}B^+_l&,&\beta^+_{0k}=iB^+_k\quad,\\
\beta^-_{jk}=\epsilon_{jkl}B^-_l&,&\beta^-_{0k}=-iB^-_k.
\end{array}
\end{eqnarray}
In momentum space the kinetic term reads ($\Omega$: four-volume)
\begin{eqnarray}\label{No5}
-{\cal L}^{ch}_{\beta,kin}=\Omega^{-1}
\int&&\frac{d^4q}{(2\pi)^4}
\{B^{+*}_k(q)P_{kl}(q)B^+_l(q)\\
&&+B^{-*}_k(q)P^*_{kl}(q)B^-_l(q)\}\nonumber
\end{eqnarray}
with $(k,l,j=1\dots 3)$
\begin{equation}\label{No6}
P_{kl}=-(q^2_0+q_jq_j)\delta_{kl}+2q_kq_l-2i\epsilon_{klj}q_0q_j.
\end{equation}
The inverse propagator $P_{kl}$ obeys the relations $(q^2=q^\mu q_\mu$)
\begin{eqnarray}\label{No7}
P_{kl}P^*_{lj}&=&(q_kq_k-q^2_0)^2\delta_{mj}=q^4\delta_{mj}\quad,\nonumber\\
 P^*_{kl}(q)&=&P_{lk}(q)
\end{eqnarray}
and $\big(q_\mu=(q_0,\vec{q})~,~\tilde{q}_\mu=(-q_0,~\vec{q})\big)$
\begin{equation}\label{No8}
P_{kl}(-q)=P_{kl}(q)~,~P_{kl}(\tilde{q})=P_{lk}(q).
\end{equation}
In a $3\times 3$ matrix notation one has
\begin{equation}\label{No9}
P^\dagger=P~,~PP^*=q^4~,~P^{-1}=\frac{1}{q^4}P^*.
\end{equation}
Clearly, the propagator is invertible for $q^2\neq 0$. The only poles of the propagator occur for $q^2=0$, indicating massless excitations. The plane wave solutions of the free field equations
\begin{equation}\label{FG}
P_{kl}(q)B^+_l(q)=0\quad,\quad P^*_{kl}(q)B^-_l(q)=0
\end{equation}
obey $q^2B^\pm_k(q)=0$.

One can further decompose $B^\pm_k(q)$ into longitudinal modes $\alpha^\pm(q)$ and transversal modes $\gamma^\pm_k(q)$ according to
\begin{eqnarray}\label{No10}
B^\pm_k(q)&=&B^{\pm(l)}_k(q)+B^{\pm(t)}_k(q)~,~\nonumber\\
B^{\pm (l)} _k(q)&=&\hat{P}^{(l)}_{kj}B^\pm_j(q)=
\frac{q_kq_j}{\vec{q}^2}
B^\pm_j(q)=q_k\alpha^\pm(q)~,\nonumber\\
B^{\pm (t)}_k(q)&=&\hat{P}^{(t)}_{kj}B^\pm_j(q)=
\left(\delta_{kj}-\frac{q_kq_j}{\vec{q}^2}\right)B^\pm_j(q)\nonumber\\
&=&\epsilon_{kjm}
q_j\gamma^\pm_{m}(q).
\end{eqnarray}
Here the projectors $\hat{P}^{(l)},\hat{P}^{(t)}$ obey
\begin{eqnarray}\label{No11}
\hat{P}^{(l)}_{kj}&=&\frac{q_kq_j}{\vec{q}^2}\quad,\quad \hat{P}^{(l)}+\hat{P}^{(t)}=1\quad,\nonumber\\
\hat{P}^{(l)2}&=&\hat{P}^{(l)}\quad,\quad \hat{P}^{(t)2}=\hat{P}^{(t)}\quad,\nonumber\\
\hat{P}^{(l)}\hat{P}^{(t)}&=&\hat{P}^{(t)}\hat{P}^{(l)}=0
\end{eqnarray}
and should not be confused with the inverse propagator $P_{kl}$. The transversal vector $\vec{\gamma}$ is orthogonal to $\vec{q}$,
\begin{equation}\label{No12}
q_k\gamma^\pm_k(q)=0.
\end{equation}
In terms of these modes the Langrangian becomes $(P\hat{P}^{(l)}=\hat{P}^{(l)}P=q^2\hat{P}^{(l)})$
\begin{eqnarray}\label{No13}
-{\cal L}^{(ch)}_{\beta,kin}&=&\Omega^{-1}
\int \frac{d^4q}{(2\pi)^4}
(K^{(l)}_++K^{(t)}_++K^{(l)}_-+K^{(t)}_-)\nonumber\\
K^{(l)}_\pm&=&q^2\big(B^{\pm(l)}_k(q)\big)^*
B^{\pm(l)}_k(q)~,\nonumber\\
K^{(t)}_\pm&=&\big(B^{\pm(t)}_k(q)\big)^*\bar{P}^\pm_{kj}
B^{\pm(t)}_j(q)~,\nonumber\\
\bar{P}^\pm_{kj}&=&-(q^2_0+\vec{q}^2)\hat{P}^{(t)}_{kj}\mp 2i\epsilon_{kjm}q_0q_m.
\end{eqnarray}

It is instructive to choose a fixed direction for the three-momentum, 
$\vec{q}=(0,0,Q),Q>0,\vec{q}^2=Q^2$. We write
\begin{equation}\label{No14}
\vec{B}^\pm(q)=
\left(\begin{array}{c}
\frac{1}{\sqrt{2}}\big(b^\pm_1(q)+b^\pm_2(q)\big)\\
\mp\frac{i}{\sqrt{2}}\big(b_1^\pm(q)-b^\pm_2(q)\big)\\
b_3(q)\end{array}\right)
\end{equation}
and observe that $b_3$ coincides with the longitudinal mode. One finds $(q^2=-q^2_0+Q^2)$
\begin{eqnarray}\label{No15}
K^{(l)}_\pm +K^{(t)}_\pm&=&q^2b^{\pm *}_3(q)b^\pm_3(q)-
(q_0+Q)^2b^{\pm*}_1(q)b^\pm_1(q)\nonumber\\
&&-(q_0-Q)^2b^{\pm *}_2(q)b^\pm_2(q)
\end{eqnarray}
and we recover the tree real eigenvalues of $P_{kl}$, namely $q^2,-(q_0+|\vec{q}|)^2$, 
$-(q_0-|\vec{q}|)^2$ with corresponding eigenvectors $\sim b_3,b_1,b_2$. Correspondingly, the plane wave solutions of the field equations (\ref{FG}) are
\begin{equation}\label{No16}
(q_0+Q)b^\pm_1=0~,~
(q_0-Q)b^\pm_2=0~,~
q^2b^\pm_3=0.
\end{equation}
Whereas the longitudinal mode $b_3$ has standard propagation properties, the behavior of the transversal modes is unusual. Solutions with positive $q_0$ exist only for $b_2$, whereas the solutions with negative $q_0$ are all contained in $b_1$. Actually, the field equations for $b_1$ and $b_2$ admit further ``secular solutions'' that are not plane waves. They are discussed in \cite{CWQ}.

The plane wave solutions $b_i(q)$ are irreducible representations of the subgroup of the Lorentz group leaving the vector $q_\mu=(\pm Q,0,0,Q)$ invariant, the little group. Applying the Lorentz transformations on these plane wave states induces irreducible representations of the Lorentz group. The longitudinal mode $b_3$ is invariant under the little group of rotations in the $1-2$-plane and therefore corresponds to a complex scalar mode. With respect to a rotation $B_1=\cos\alpha B'_1+\sin\alpha B'_2~,~B_2=\cos\alpha B'_2-\sin\alpha B'_1$ the fields $b^\pm_{1,2}$ are helicity eigenstates with opposite helicity and tansform with a phase 
$b^{+\prime}_1=e^{i\alpha}b^+_1~,~b^{+\prime}_2=e^{-i\alpha}b^+_2~,~
b^{-\prime}_1=e^{-i\alpha}b^-_1~,~b^{-\prime}_2=e^{i\alpha}b^-_2$. The plane wave solutions with positive energy $(q^0>0~,~q_0<0)$ are $b^+_1~,~(b^+_2)^*~,~b^-_1~,~(b^-_2)^*$. We note that for $b^+_{1,2}$ they contain only one helicity state (both $b^+_1$ and $(b^+_2)^*$ transform with the same phase $e^{i\alpha}$) whereas $b^-_{1,2}$ contains the opposite helicity state. At this point the plane waves in $b^\pm$ account per isospin component for one complex massless vector (with two opposite helicities) and two complex massless scalars. The additional degrees of freedom in $b_{1,2}$ correspond to the secular solutions \cite{CWQ}. Altogether, the generic solution for the time evolution of each $b^\pm_k(\vec{q},t)$ (now the Fourier transform is taken only with respect to the spacelike coordinates) has two complex numbers as initial conditions. The total number of degrees of freedom in $b^\pm_k$ is the same as for four massive complex vector fields. We will describe later a mechanism for mass generation which indeed leads to four massive complex vector fields for the degrees of freedom contained in $B^\pm_k$ - two per isospin component. 

The unusual properties of the transversal modes are also reflected in the energy density. The energy momentum tensor $T^{\mu\nu}$ is defined from variation of $S_M$ with respect to the vielbein 
\begin{equation}\label{IVa}
T^{\mu\nu}=\frac{1}{e}e^{m\mu}\frac{\delta S_M}{\delta e^m_\nu}.
\end{equation}
One obtains for the energy density $\rho$
\begin{eqnarray}\label{4a}
\rho=-T^0_0=Z_+&&\{\partial_0B_{k}^{+*}\partial_0B_k^++2\partial_kB_{k}^{+*}\partial_lB_l^+
\nonumber\\
&&-\partial_lB_{k}^{+*}\partial_lB_k^+\}+(+\rightarrow -).
\end{eqnarray}
For a Lagrangian involving only first time-derivatives of the fields we may alternatively compute the Hamiltonian or $\rho$ in flat space $(e^m_\mu=\delta^m_\mu)$ as
\begin{eqnarray}\label{B4}
&&\rho=-{\cal L}+\left\{
\partial_0B^+_k\frac{\partial{\cal L}}{\partial(\partial_0B^+_k)}
+\partial_0B^{+*}_k
\frac{\partial{\cal L}}{\partial(\partial_0B^{+*}_k)}\right.\nonumber\\
&&\hspace{4.2cm}\left.+(+\rightarrow -)\right\}.
\end{eqnarray}
This yields the same result (\ref{4a})\footnote{We note that for deriving eq. (\ref{4a}) from eq. (\ref{IVa}) we have varied the vielbein while keeping the coordinate-scalar $\beta_{mn}$ fixed. Variation with a fixed tensor $\beta_{\mu\nu}$ yields a different result. Both versions, however, coincide for the solutions of the field equations.}.

For plane waves only the longitudinal component contributes to $\rho$
\begin{equation}\label{4e}
\rho=2Z_+\partial_kb^{+*}_3\partial_kb^+_3+(+\rightarrow -).
\end{equation}
Then the energy density is positive semidefinite. In contrast, the secular classical solutions for $b_{1,2}$ can also lead to negative $\rho$ \cite{CWQ}.  Classical or quantum instability of the free theory for chiral tensors is probably the main reason why these fields have not been investigated more intensively in the past. We note that the classical instability of the free theory appears here in a very particular form. The plane wave solutions have a stable longitudinal component and a transversal component that does not contribute to the energy density at all. In this sense the transversal modes are not standard ghosts (nor tachyons), but rather at the borderline between stability and instability. Due to this borderline situation secular growing solutions exist which cannot be described by plane waves. They are responsible for the classical instability of the free theory. However, due to interactions the quantum fluctuations may shift the stability behavior from the borderline into the stable domain. This is precisely what happens in the scenario of sect. VII. In presence of the fluctuation induced mass term the effective propagator for the massive spin one particles will no more show a classical instability. All plane waves have then positive energy density. The instability of the free theory can also be seen on the quantum level \cite{CWQ}. The cure is the same - the effective propagator precisely accounts for the quantum fluctuation effects in presence of interactions. We conclude that it is premature to discard chiral tensor fields because of instabilities of the free theory. In view of the borderline behavior a statement about the stability properties of the interacting theory can only be made once the fluctuation effects induced by the interactions are taken into account. 

We close this section by rewriting  the interactions discussed in sect. II in terms of the fields $B^\pm_k$. Since the fields $B^\pm_k$ are unconstrained and have well defined propagators for $q^2\neq 0$ they constitute a convenient basis for the loop calculation in the next section. In terms of $B^\pm_k$ the quartic interactions read
\begin{eqnarray}\label{No17}
-{\cal L}_{\beta,4}&=&
\frac{\tau_+}{4}\big[(B^+_k)^\dagger B^+_l\big]
\big[(B^+_k)^\dagger B^+_l\big]+(+\rightarrow -)\nonumber\\
&&+\tau_1\big[(B^+_k)^\dagger B^-_k\big]
\big[(B^-_l)^\dagger B^+_l\big]\nonumber\\
&&+\tau_2\big[(B^+_k)^\dagger\vec{\tau}B^-_k\big]
\big[(B^-_l)^\dagger\vec{\tau}B^+_l\big]\nonumber\\
&&+\frac{\tau_3}{4}\big[(B^+_k)^\dagger B^-_k\big]
\big[(B^+_l)^\dagger B^-_l\big]+c.c.\nonumber\\
&&+\frac{\tau_4}{4}\big[(B^+_k)^\dagger B^-_l\big]
\big[(B^+_k)^\dagger B^-_l\big]+c.c.\nonumber\\
&&+\frac{\tau_5}{4}\big[(B^+_k)^\dagger B^-_l\big]
\big[(B^+_l)^\dagger B^-_k\big]+c.c. .
\end{eqnarray}
Positivity of the energy density restricts the allowed values of the couplings $\tau_j$ since $-{\cal L}_{\beta,4}$ should be a positive semidefinite quartic form. 

Finally, we translate the chiral couplings (\ref{2}) to the basis of fields $B^\pm_k$ by using

\begin{eqnarray}\label{31AA}
\beta_+&=&-2B^+_k\sigma^k_+,\ \beta_-=-2B^-_k\sigma^k_-,\nonumber\\
\bar{\beta}_+&=&-2B^{+*}_k\sigma^k_-~,~\bar{\beta}_-=-2B^{-*}_k\sigma^k_+,
\end{eqnarray}
with $\sigma^k_\pm$ defined in terms of the Pauli matrices $\tau^k$ as

\begin{equation}\label{31AB}
\sigma^k_+=\left(\begin{array}{cc}
\tau^k&0\\0&0\end{array}\right)\ ,\ \sigma^k_-=\left(\begin{array}{cc}
0&0\\0&\tau^k\end{array}\right).
\end{equation}
With respect to the discrete transformations one has 
\begin{eqnarray}\label{19aa}
P&:&B^{++}_k\rightarrow-B^{-+}_k~,~B^{\pm 0}_k\rightarrow(B^{\pm 0}_k)^*~,\nonumber\\
C&:&B^{\pm+}_k\rightarrow(B^{\mp+}_k)^*~,~B^{\pm0}_k\rightarrow-B^{\pm0}_k~,\nonumber\\
CP&:&B^\pm_k\rightarrow-(B^\pm_k)^*.
\end{eqnarray}

\section{Chiral freedom}
The chiral couplings are asymptotically free - in contrast to the Yukawa couplings of the Higgs scalar. This will be a crucial point for the understanding of the dynamics and ground state of our model. In view of this we call our scenario ``chiral freedom''. The remarkable feature responsible for asymptotic freedom is the opposite sign of the fermion anomalous dimensions for the contribution from chiral tensors as compared to the one from the Higgs scalar. 

We introduce an infrared cutoff scale $k$ for the flow of the scale dependent effective action \cite{CWERGE} (or equivalently a renormalization scale $\mu$ defined by nonvanishing ``external momenta''). Neglecting effects involving the electroweak gauge couplings the one-loop vertex corrections for the chiral couplings vanish. The running of the renormalized chiral couplings is therefore given by the anomalous dimensions of the chiral tensor fields

\begin{eqnarray}\label{M1}
\eta_+&=&-\frac{\partial\ln Z_+}{\partial\ln k}=\frac{1}{2\pi^2}tr(F_U^\dagger F_U),\nonumber\\
\eta_-&=&-\frac{\partial\ln Z_-}{\partial\ln k}=\frac{1}{2\pi^2}[tr(F_D^\dagger F_D)+\frac{1}{3}tr(F_L^\dagger F_L)]
\end{eqnarray}
and similar (matrix valued) anomalous dimensions for the fermion fields

\begin{eqnarray}\label{M2}
\eta_Q&=&-\frac{3}{4\pi^2}(F^\dagger_U F_U+F^\dagger_D F_D)-\frac{g^2_s}{2\pi^2}\nonumber\\ \eta_U&=&-\frac{3}{2\pi^2} F_U F^\dagger_U-\frac{g^2_s}{2\pi^2}\nonumber\\
\eta_D&=&-\frac{3}{2\pi^2}F_D F_D^\dagger-\frac{g^2_s}{2\pi^2}.
\end{eqnarray}
Details of the computation of the anomalous dimensions can be found in the appendix A. From

\begin{eqnarray}\label{M3}
k\frac{\partial}{\partial k} F_U&=&\frac{1}{2}(\eta_++\eta_U)F_U+\frac{1}{2} F_U\eta_Q\nonumber\\
k\frac{\partial}{\partial k} F_D&=&\frac{1}{2}(\eta_-+\eta_D)F_D+\frac{1}{2} F_D\eta_Q
\end{eqnarray}
one derives
the running of the chiral couplings\footnote{Running couplings have been computed in a different model for abelian tensor fields in  \cite{ACH}. Translating to our normalization and comparing with our results we find structural agreement but an overall relative factor of two in the  $\beta$-functions.}
\begin{eqnarray}\label{4f}
k\frac{\partial}{\partial k}F_U&=&-\frac{9}{8\pi^2}F_UF^\dagger_UF_U-
\frac{3}{8\pi^2}F_UF^\dagger_DF_D\nonumber\\
&&+\frac{1}{4\pi^2}F_U ~tr(F^\dagger_UF_U)-\frac{1}{2\pi^2}g^2_sF_U\nonumber\\
k\frac{\partial}{\partial k}F_D&=&-\frac{9}{8\pi^2}F_DF^\dagger_DF_D-
\frac{3}{8\pi^2}F_DF^\dagger_UF_U\\
&&+\frac{1}{4\pi^2}F_D tr(F^\dagger_DF_D+\frac{1}{3}F^\dagger_LF_L)-
\frac{1}{2\pi^2}g^2_sF_D\nonumber\\
k\frac{\partial}{\partial k}F_L&=&-\frac{9}{8\pi^2}F_LF^\dagger_LF_L+
\frac{1}{4\pi^2}F_L ~tr(F^\dagger_DF_D+\frac{1}{3}F^\dagger_LF_L)\nonumber
\end{eqnarray}
Here $F_{U,D,L}$ are the renormalized chiral couplings
\begin{equation}\label{6}
F_U=Z^{-1/2}_u\bar{F}_UZ^{-1/2}_qZ^{-1/2}_+
\end{equation}
with $Z_u$ the wave function renormalization matrix of the right handed up-type quarks and $Z_q$ the one for the left-handed quarks. We note that the pointlike quartic couplings (\ref{QI}) do not affect the running of the chiral couplings in one-loop order.

Very similar to QCD, chiral freedom comes in pair with a characteristic infrared scale $\Lambda_{ch}$ where the chiral couplings grow large. Let us concentrate on the single chiral coupling of the top quark $f_t=(F_U)_{33}$ and neglect all other chiral couplings and $g_s$ for the moment. The solution of the evolution equation (\ref{4f}) 
\begin{equation}\label{8}
f^2_t(k)=\frac{4\pi^2}{7\ln(k/\Lambda^{(t)}_{ch})}
\end{equation}
exhibits by dimensional transmutation the ``chiral scale'' $\Lambda^{(t)}_{ch}$ where $f_t$ formally diverges. We propose that $\Lambda^{(t)}_{ch}$ sets also the characteristic mass scale for electroweak symmetry breaking. The solution of the gauge hierarchy problem is then very similar to the understanding of a small ``confinement scale'' $\Lambda_{QCD}$ in a setting with fundamental scales given by some grand unified scale or the Planck scale.

Indeed, we suggest that the strong coupling $f_t$ induces a top-antitop condensate $\langle\bar{t}_Lt_R\rangle\neq 0$. Here $\bar{t}_Lt_R$ plays the role of a composite Higgs field \cite{BL} - a nonvanishing expectation value leads to ``spontaneous breaking'' of the electroweak gauge symmetry, giving the $W$- and $Z$-bosons a mass while leaving the photon massless. The composite Higgs doublet fields $\varphi_t\sim\bar{q}_Lt_R,~\varphi_b\sim\bar{b}_Rq_L$ carry the same electroweak quantum numbers as the usual ``fundamental'' Higgs field. If $\varphi_t$ and $\varphi_b$ are normalized with a standard (gauge invariant) kinetic term the observed masses of the $W,Z$-bosons follow for a Fermi scale $\langle\varphi\rangle=(\langle\varphi_t\rangle^*\langle\varphi_t\rangle +\langle\varphi_b^*\rangle \langle\varphi_b\rangle)^{1/2}=174~GeV$. With $\langle\varphi\rangle=a\Lambda_{ch}$ (where $a$ remains to be computed) this can be achieved by a suitable value of $f_t$ at some appropriate short distance scale. The symmetries allow for effective Yukawa couplings of $\varphi_t$ to the up-type quarks and $\varphi_b$ to the down-type quarks and charged leptons. With respect to the axial 
$G_A$-symmetry $\varphi_t$ and $\varphi_b$ carry the same charges as $\beta_+$ and $\beta_-$, i.e. the $G_A$-parity of $\varphi_b$ is odd. For $\langle\varphi_t\rangle\neq 0,~\langle\varphi_b\rangle\neq 0$ all quarks and charged leptons can acquire a mass.

\section{Gap equation for the top quark mass}

For a first check if these ideas are reasonable we solve the coupled gap equations for the top-quark- and $\beta^+$-propagators. The lowest order Schwinger-Dyson equation \cite{SD} for a possible top quark mass $m_t$ reads 
\begin{equation}\label{9}
m_t=-4im^*_t\int\frac{d^4q}{(2\pi)^4}
\frac{R_{mn}(q)\delta^{mn}f^2_t(q)}{q^2+|m_t|^2}.
\end{equation}
It is depicted in Fig. 1. Here $R_{mn}(q)$ denotes the $\beta^+-\beta^+$component of the chiron-propagator-matrix. For $m_t=0$ one would have $R_{mn}(q)=0$ by virtue of conserved hypercharge. However, $U(I)_Y$ is spontaneously broken by $\langle\varphi_t\rangle\neq 0$ or $m_t\neq 0$. By inserting a top-quark loop we obtain the lowest order Schwinger-Dyson equation for the inverse propagator of the electrically neutral component of $\beta^+$ and infer
\begin{eqnarray}\label{10}
&&R_{mn}(q)\delta^{mn}=-\frac{72i\big(f^*_t(q)\big)^2 I(q)}{q^4},\\
&&I(q)=\int \frac{d^4q'}{(2\pi)^4}
\frac{m^2_t}{[(q'+\frac{q}{2})^2+|m_t|^2][(q'-\frac{q}{2})^2+|m_t|^2]}.\nonumber
\end{eqnarray}
We show the Schwinger-Dyson equation for the $\beta^+-\beta^+$ self-energy graphically in Fig. 2. 

\begin{figure}[htb]
\centering
\includegraphics[scale=0.45]{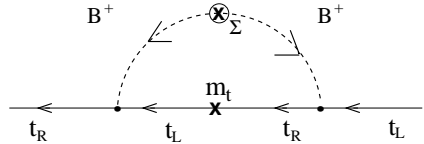}

\bigskip
\caption{Schwinger-Dyson equation for the top quark mass.\label{fig1a}}
\end{figure}

\begin{figure}[htb]
\centering
\includegraphics[scale=0.45]{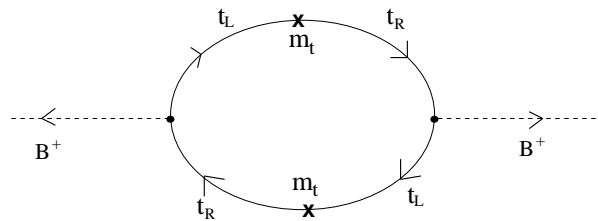}

\bigskip
\caption{Schwinger-Dyson equation for the $B^+B^+$ self-energy $\Sigma$.\label{fig2}}
\end{figure}

\begin{figure}[h!tb]
\centering
\includegraphics[scale=0.45]{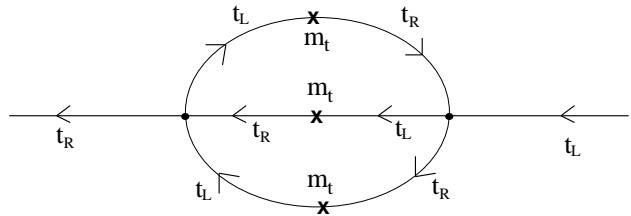}

\bigskip
\caption{Two-loop Schwinger-Dyson equation for the top quark mass.\label{fig3}}
\end{figure}

Alternatively, we could integrate out the fields $\beta^\pm$ in favor of a non-local four-fermion interaction. In this purely fermionic formulation (besides the gauge interactions) we have recovered the term corresponding to the combination of eqs. (\ref{9})(\ref{10}) in the Schwinger-Dyson equation at two loop order. Graphically, this two loop equation obtains by inserting Fig. \ref{fig2} for $\Sigma$ in Fig. \ref{fig1a} and replacing four fermion lines connected by a $B^+$-propagator by an effective vertex, as shown in Fig. \ref{fig3}. We note that due to the tensor structure of the effective four fermion interaction the mass generation for the top quark involves a two loop graph. This is an important difference as compared to effective interactions in the scalar channel \cite{BL}.

\noindent

\noindent

\noindent

For eqs. (\ref{9})(\ref{10}) we have made several simplifications. (i) We have replaced an effective momentum dependent mass term $m_t(p)$ by a constant even though $m_t(p)$ is supposed to vanish fast for large $p^2$. This momentum dependence would provide for an effective ultraviolet cutoff in the $q'$-integral for $I(q)$. We take this effect into account by implementing  a suitable UV-cutoff  $\Lambda_t$ in the integral $I(q)$ (\ref{10}) after continuation to euclidean signature. More precisely, we approximate 
\begin{equation}\label{11}
I(q)=\frac{im^2_t}{32\pi^2}\ln
\frac{\Lambda^4_t+(|m_t|^2+q^2/4)^2}{(|m_t|^2+q^2/4)^2}.
\end{equation}
With $x$ the analytic continuation of $q^2$ to euclidean space eq. (\ref{9}) then yields 
\begin{eqnarray}\label{45A}
m_t&=&\frac{9}{16\pi^4}|m_t|^2m_t\int^\infty_0\frac{dx}{x}\frac{|f_t(x)|^4}{x+|m_t|^2}\nonumber\\
&&\times \ln \frac{\Lambda^4_t+(|m_t|^2+x/4)^2}{(|m_t|^2+x/4)^2}.
\end{eqnarray}

(ii) We have used the classical $\bar{\beta}^+-\beta^+$ propagator for the inversion of the two point function, as needed for the computation of $R_{mn}$. This results in an IR-divergence due to the squared propagator $\sim q^{-4}$. Presumably, the full propagator will contain a mass term acting as an IR-regulator (sect. VII). Our first approach takes this into account by limiting the $q$-integral in eq. (\ref{9}) to $q^2>M^2_\beta>\Lambda^2_{ch}$. Combining eqs. (\ref{45A}) and (\ref{8}) results in the condition (after rescaling $x\rightarrow m^2_tx$ and with $m_t\hat{=}|m_t|)$
\begin{equation}\label{12}
\int^\infty_{M^2_\beta/m^2_t}\frac{dx}{x(x+1)}
\frac{\ln\left(\frac{\Lambda^4_t}{m^4_t(1+x/4)^2}+1\right)}
{\left(\ln\frac{m^2_t}{\Lambda^{(t)2}_{ch}}+\ln x\right)^2}
=\frac{49}{36}.
\end{equation}
For not too large $M^2_\beta/m^2_t$ this has indeed a nontrivial solution with $m_t>\Lambda^{(t)}_{ch}$. For the example $\Lambda_t=3m_t,M_\beta=m_t~(M_\beta=1.1 m_t)$ one finds $m_t=1.45~ \Lambda^{(t)}_{ch}~(m_t=1.27\Lambda^{(t)}_{ch})$. 

Our second approach introduces an explicit mass term in the chiron propagator, replacing the factor $1/q^4$ in eq. (\ref{10}) by $1/(q^2+m^2_+)^2$. Furthermore, we use an effective infrared cutoff $k^2=(q^4+m^4_t)^{1/2}$ in the running of the chiral coupling (\ref{8}).  Eq. (\ref{12}) is then replaced by
\begin{eqnarray}\label{47A}
&&\int^\infty_0 \frac{dxx}{(x+1)(x+m^2_+/m^2_t)^2}\nonumber\\
&&\frac{\ln\left(\frac{\Lambda^4_t}{m^4_t(1+x/4)^2}+1\right)}
{\left(\ln\frac{m^2_t}{\Lambda^{(t)2}_{ch}}+\frac{1}{2}\ln(1+x^2)\right)^2}
=\frac{49}{36}.
\end{eqnarray}
For $\Lambda_t/m_t=3,~m_+/m_t=(5;4;2)$ one finds $m_t=(1.0025\Lambda^{(t)}_{ch};
1.0055\Lambda^{(t)}_{ch};1.065\Lambda^{(t)}_{ch})$.

We conclude that the generation of a nonvanishing top-antitop condensate $\langle\varphi_t\rangle\neq 0$ seems plausible. From a different viewpoint, the gap equations have a nontrivial solution if the chiral coupling $f_t$ exceeds a critical value. This is always the case since $f_t$ grows to very large values unless an infrared cutoff $\sim m_t$ is generated by electroweak symmetry breaking. Our result corresponds to $f_t(m_t)=6.8(8.5)$ for the first approach and $f_t(m_t)=(83;56;16.5)$ for the second one. Not surprisingly, the uncertainty about $f_t(m_t)$ is rather large. 

A similar gap equation for the $W$-boson mass could determine $M_W/\Lambda^{(t)}_{ch}$ - in consequence the ratio $m_t/M_W$ would become predictable! Its computation amounts to the determination of the effective top quark Yukawa coupling $h_t=m_t/\langle\varphi_t\rangle$. The accuracy of the of the gap equations in a regime with strong interactions is rather limited, however. We consider the gap equations mainly as a tool to demonstrate the qualitative effects. The functional renormalized group could lead to a somewhat more reliable picture. 

We will assume that both $\varphi_t$ and $\varphi_b$ acquire nonvanishing expectation values such that all particles become massive except the photon - small neutrino masses can be generated by $L$-violating dimension five operators involving two composite scalar doublets. One possible mechanism for inducing $\langle\varphi_b\rangle\neq 0$ is the continued running of the chiral coupling $f_b$. Its increase is not stopped by $\langle\varphi_t\rangle\neq 0$ and may result in a second scale $\Lambda^{(b)}_{ch}$. Alternatively, effective interaction terms of the type $(\varphi^\dagger_t\varphi_b)(\varphi^\dagger_b\varphi_t)$ may trigger expectation values of a similar size  $\langle\varphi_b\rangle\approx\langle\varphi_t\rangle$.

\section{Effective low energy action and derivative expansion}
We learn from the preceeding section that due to the strong chiral couplings the vacuum properties are non-trivial. One may find the phenomenon of ``spontaneous breaking'' of the electroweak gauge symmetry by a composite order parameter which replaces the fundamental Higgs scalar in the standard model. In order to extract the particle-propagators and transition amplitudes we need to compute the effective action $\Gamma$ which incorporates all quantum fluctuations and generates the 1PI-irreducible vertices. Let us first restrict the discussion to the coupled system of chiral tensor fields and composite scalar bilinears $\varphi_t\sim\bar{q}_Lt_R,~\varphi_b\sim\bar{b}_Rq_L,~q_L=(t_L,b_L)$. Introducing sources for the chiral tensors and the scalar bilinears $\bar{q}_Lt_R,\bar{b}_Rt_L$ and performing a Legendre transform one finds the implicit relation
\begin{eqnarray}\label{a6}
&&\Gamma[\beta_{\mu\nu},\varphi_{t,b}]=i\ln \int D\hat{\beta}D\psi \exp\left\{iS_M[\beta+\hat{\beta},\psi]
\right.\nonumber\\
&&+i\int d^4x\left[\frac{\delta\Gamma}{\delta\varphi_t}\Big(c_t(\bar{q}_Li\tau_2 t_R)^T-\varphi_t \Big)
+\frac{\delta\Gamma}{\delta\varphi_b}(c_b\bar{b}_Rq_L-\varphi_b)\right.\nonumber\\
&&\left.\left.+\frac{\delta\Gamma}{\delta\beta^+_{\mu\nu}}\hat{\beta}^+_{\mu\nu}+
\frac{\delta\Gamma}{\delta\beta^-_{\mu\nu}}\hat{\beta}^-_{\mu\nu}+h.c.\right]\right\}.
\end{eqnarray}
Here the integration variable $\hat{\beta}^\pm_{\mu\nu}$ denotes the deviation of the tensor field from the ``background field'' $\beta^\pm_{\mu\nu}$ - the latter being the argument of $\Gamma$. The fermionic integration variables are collectively denoted by $\psi$. We also use the freedom in the normalization of $\varphi_{t,b}$ such that the scalar kinetic term in $\Gamma$ has standard normalization - this fixes the constants $c_{t,b}$. With the usual gauge fixing procedure this formula is easily extended to include gauge fields in the functional integral as well as in the arguments of $\Gamma$. Also fermionic Grassmann fields can be included in the arguments of $\Gamma$.

Eq. (\ref{a6}) may be a starting point for approximations, for example by truncating the effective action to a small number of terms. In this section we will only exploit the symmetry properties of $\Gamma$. The classical action (\ref{No1}) (\ref{c3}) is invariant under hermitean conjugation. (Hermitean conjugation includes a total reordering of all fermionic Grassmann variables.) Therefore the effective action (\ref{a6}) for the bosonic fields $\beta^\pm_{\mu\nu},\varphi_t,\varphi_b$ is real. (More generally, it is hermitean if ``background spinor fields'' are included). Furhermore, if the regularization preserves all symmetries (absence of anomalies) the effective action should have the same symmetries as the classical action.\footnote{The issue of gauge symmetries is somewhat involved but well understood.}

What is not guaranteed a priori is the locality of the effective action. For example, the effective action may become local at long distances in terms of suitable composite fields, while being nonlocal if expressed in the original fields of the microscopic action. Such a scenario is quite generic for a model with strong interactions, as exemplified by QCD. We will argue in the next section that $\Gamma[\beta,\varphi]$ indeed is nonlocal. We will demonstrate explicitely that a nonlocal term in the inverse propagator for the chiral tensor fields is generated by loops involving only local interactions. We will also show that this term appears as a simple local mass term in a different field basis where the degrees of freedom in $\beta_{\mu\nu}$ are directly associated to a massive vector field.

Nevertheless, we first investigate in this section the (temporary) assumption that the effective action $\Gamma[\beta,\varphi]$ is local and can be expanded in the number of derivatives. This will give us an idea about the possible new vertices in presence of electroweak symmetry breaking. In this section we will discuss a systematic expansion in the inverse mass dimension of the effective couplings. We will include all terms with dimension up to four, i.e. all effective couplings with positive mass dimension and dimensionless couplings. In this approximation $\Gamma[\beta,\varphi]$ includes the terms (\ref{c3}) discussed in sect. II, now with appropriate renormalized couplings replacing the bare or microscopic couplings. What is new are further terms which arise due to the presence of the additional scalar fields $\varphi_{t,b}$. 

One expects that the quantum fluctuations generate a kinetic term for a composite scalar fields $\varphi_{t,b}$. Gauge invariance requires the usual covariant derivatives
\begin{equation}\label{A7a}
-{\cal L}_{\varphi,kin}=D^\mu\varphi^\dagger_tD_\mu\varphi_t+D^\mu\varphi^\dagger_bD_\mu\varphi_b.
\end{equation}
For nonvanishing expectation values $\langle{\varphi}_t\rangle=\bar{\varphi}_t,\langle{\varphi}_b\rangle=\bar{\varphi}_b$ this induces mass terms for the $W$- and $Z$-bosons by the Higgs mechanism, e. g. 
\begin{equation}\label{A7b}
M^2_W=\frac{g^2}{2}(|\bar{\varphi}_t|^2+|\bar{\varphi}_b|^2).
\end{equation}

Indeed, the strong chiral couplings induce an effective potential for $\varphi_t$ and $\varphi_b$ from which both the expectation values and the masses of the composite scalars can be computed. Including terms quartic in $\varphi_{t,b}$ the most general form consistent with the $G_A$-symmetry reads
\begin{eqnarray}\label{A3}
U&=&\mu^2_t[\varphi^\dagger_t\varphi_t]+\mu^2_b[\varphi^\dagger_b\varphi_b]\nonumber\\
&&+\frac{1}{2}\lambda_t[\varphi^\dagger_t\varphi_t]^2+\frac{1}{2}\lambda_b
[\varphi^\dagger_b\varphi_b]^2\nonumber\\
&&+\lambda_m[\varphi^\dagger_t\varphi_t][\varphi^\dagger_b\varphi_b]
+\lambda_a[\varphi^\dagger_t\varphi_b][\varphi^\dagger_b\varphi_t]\nonumber\\
&&+\lambda_v[\varphi^\dagger_t\varphi_b][\varphi^\dagger_t\varphi_b]+\lambda^*_v[\varphi^\dagger_b\varphi_t][\varphi^\dagger_b\varphi_t]
\end{eqnarray}
with real $\mu^2_{t,b},\lambda_{t,b},\lambda_m$ and $\lambda_a$. The masses for the physical scalars obtain from the second derivatives of $U$ at the minimum. For the truncation (\ref{A3}) the spectrum of physical scalars comprises three neutral scalars and one charged scalar. The size of the scalar mass is characteristically $\sim m^2_s\sim \lambda\bar{\varphi}^2$. Since the scalars are composites induced by the strong chiral interactions one expects large $\lambda$ and characteristic scalar masses around $500$ GeV\footnote{Much higher scalar masses are less likely in view of the renormalization flow of strong scalar couplings (``triviality bound'').}.

For $\lambda_v=0$ the scalar potential exhibits a continuous $U(1)_A$ symmetry. On the other hand a nonvanishing $\lambda_v$ remains consistent with a $Z_8$-subgroup of $U(1)_A$ where 
$\psi_L $ $\rightarrow$~$\exp(-in\pi/4)$~$\psi_L,$ $\psi_R \rightarrow$ $\exp(in\pi/4)$ $\psi_R, $ $\varphi_t\rightarrow$ $\exp (in\pi/2)$ $\varphi_t,$ $\varphi_b\rightarrow$ $\exp(-in\pi/2)\varphi_b$, and with the $Z_2$-symmetry $G_A:$ $ d_R$$\rightarrow$ $ -d_R,e_R$$\rightarrow -e_R,$$\varphi_b\rightarrow -\varphi_b$. The generation of a nonzero $\lambda_v$ is expected if the quartic interactions of $\beta^\pm$ violate the continuous $U(1)_A$-symmetry while preserving the discrete $G_A$ symmetry. 

A continuous $U(1)_A$-symmetry has a QCD-anomaly and plays the role of a Peccei-Quinn symmetry \cite{PQ}. This would solve the strong CP-problem, but the corresponding axion \cite{WeWi} - the pseudo-goldstone boson from the spontaneous breaking of an anomalous global symmetry - is presumably excluded by experiment \cite{Kim}. Still, the scalars $\varphi_t$ and $\varphi_b$ are not fundamental. It seems worthwhile to investigate if an unexpected loophole opens up due to form factors. 

No ultra-light scalar is expected in the case of a large $\lambda_v$, as allowed by the $G_A$ symmetry. Nevertheless, if the 
quartic couplings of the chiral tensors (\ref{QI}) are small, this may lead to a value of $|\lambda_v|$ much smaller than the characteristic size of the other quartic scalar couplings. By virtue of the $U(1)_A$ symmetry $\lambda_v$ must be proportional to a linear combination of $\tau_3,\tau_4,\tau_5$. The mass $m_L$ of the lightest neutral scalar field - the ``would be Goldstone boson'' of the $U(1)_A$-symmetry - is then suppressed as compared to the typical size of the scalar masses $m_s$ by a factor $m^2_L/m^2_s\sim\lambda_v/\lambda_s$, with $\lambda_s$ a typical large quartic scalar coupling. For small $\lambda_v$ also the couplings of the lightest scalar to the quarks and leptons become close to the pure derivative couplings of a Goldstone boson. For a mass below the $e^+-e^-$-threshold $\sim 1 MeV$ the lightest scalar could not decay into charged particles anymore. Detailed phenomenology and bounds for the possibility of a light neutral scalar in our model wait for exploration. The differences as compared to the axion arise from the explicit $U(1)_A$-violation by $\lambda_v\neq 0$.

The quark- and charged lepton  masses are generated by effective Yukawa couplings $U,D,L$ of the composite scalar fields
\begin{eqnarray}\label{12a}
-{\cal L}_y=\bar{u}_RU\tilde{\varphi}_tq_L
&-&\bar{q}_LU^\dagger\tilde{\varphi}^\dagger_tu_R\nonumber\\
+\bar{d}_RD\varphi^\dagger_bq_L&-&\bar{q}_LD^\dagger\varphi_bd_R\nonumber\\
+\bar{e}_RL\varphi^\dagger_bl_L&-&\bar{l}_LL^\dagger\varphi_be_R
\end{eqnarray}
according to $M_U=U\bar{\varphi}_t, M_D=D\bar{\varphi}_b^*,M_L=L\bar{\varphi}_b^*$. (As usual, $\tilde{\varphi}_t=-i\tau_2\varphi^T_t$.) The solution of the gap equation in sect. V can be interpreted as a first calculation of the combination $U_{33}\bar{\varphi}_t=m_t$. 

Possible generation and mixing patterns in the chiral couplings $F_{U,D,L}$ will be reflected in the effective fermion masses. It is always possible to make $F_{U,D,L}$ diagonal and real by appropriate chiral transformations. This results in a Cabibbo-Kobayashi-Maskawa-type mixing matrix for the weak interactions and renders issues like particle decays, strangeness violating neutral currents or CP-violation very similar to the Higgs-mechanism in the standard model. Up to the small weak interaction effects the diagonal $F_{U,D,L}$ imply separately conserved quantum numbers for the different fermion species - in turn also the mass matrices $M_{U,D,L}$ must be diagonal in this basis. The transmission of generation structures from the chiral couplings $F$ to the mass matrices $M$ can be understood in terms of symmetries. As an example, consider diagonal $F_{U,D}$ with $f_t$ and $f_b$ as the only nonvanishing entries. Omitting again weak interactions this would result in an enhanced global chiral $SU(4)_L\times SU(4)_R$-symmetry acting on $u,d,s,c$. Such a flavor symmetry would not be affected by nonzero $\
\langle\bar{t}_Lt_R\rangle,\langle\bar{b}_Rb_L\rangle$ and forbids mass terms for the four ``light'' quark flavors. Switching on $f_c$ (and/or off diagonal $F_{tc},F_{ct}$) allows for the generation of a charm quark mass. Now the flavor symmetry is reduced to $SU(3)_L\times SU(3)_R$ and we recover the usual setting of chiral symmetries in QCD with three massless quarks, including the anomalous axial symmetry. Similar considerations hold for the different scales of $m_s$ and $m_{u,d}$ and for the hierarchies of the charged lepton masses.

For nonvanishing $\bar{\varphi_t},\bar{\varphi_b}$ mass-like terms for $\beta^\pm$ can be generated by effective interactions. The most general local interaction $\sim\beta^2\varphi^2$ which is consistent with the symmetry 
$G_A:\beta_-\rightarrow -\beta_-,~\varphi_b\rightarrow-\varphi_b$ reads
\begin{eqnarray}\label{13}
-{\cal L}_{M\beta}&=&
\frac{1}{8}tr\{\sigma_1[\varphi^\dagger_t\varphi_b][\bar{\beta}_-\beta_+]+\sigma_2[\varphi^\dagger_t\beta_+][\bar{\beta}_-\varphi_b]+\nonumber\\
&&+\sigma_+[\bar{\beta}_+\varphi_t][\bar{\beta}_+\varphi_t]
+\sigma_-[\bar{\beta}_-\varphi_b][\bar{\beta}_-\varphi_b)]\\
&&+\sigma_{v1}[\varphi^\dagger_b\varphi_t][\bar{\beta}_-\beta_+]
+\sigma_{v2}[\varphi^\dagger_b\beta_+][\bar{\beta}_-\varphi_t]\nonumber\\
&&+\sigma_{v+}[\bar{\beta}_+\varphi_b][\bar{\beta}_+\varphi_b]+\sigma_{v-}
[\bar{\beta}_-\varphi_t][\bar{\beta}_-\varphi_t]\}+c.c.\nonumber
\end{eqnarray}
The couplings $\sigma_{vk}$ violate the continuous $U(1)_A$-symmetry. Omitting the isospin structure one has
\begin{eqnarray}\label{40A}
\frac{1}{8}tr\bar{\beta}_-\beta_+&=&\frac{1}{4}\beta^{-*}_{\mu\nu}\beta^{+\mu\nu}
=B^{-*}_kB^+_k~,~\nonumber\\
\frac{1}{8}tr\beta_\pm\beta_\pm&=&\frac{1}{4}\beta^\pm_{\mu\nu}\beta^{\pm\mu\nu}
=B^\pm_kB^\pm_k~,
\end{eqnarray}
and $tr\beta_+\beta_-=4\beta^+_{\mu\nu}\beta^{-\mu\nu}=0,~tr\bar{\beta}_+\beta_+
=4\beta^{+*}_{\mu\nu}\beta^{+\mu\nu}=0$. 
We note that these masslike terms are off diagonal $(B^{-*}B^+)$ or ``Majorana-like'' $(B^+B^+)$. We show in appendix B that they are not sufficient to generate an acceptable mass pattern for stable particles. 

In addition, there are two possible couplings cubic in $\beta$ and linear in $\varphi$
\begin{eqnarray}\label{a7,1}
-{\cal L}_{3\beta}&=&\frac{\gamma_t}{8}\big[\varphi^\dagger_t\beta^{-\nu}_\mu\big]\Big\{
\big[(\beta^{+\rho}_\nu)^\dagger\beta^{-\mu}_\rho\big]
-\big[(\beta^{+\mu}_\rho)^\dagger\beta^{-\rho}_\nu\big]\Big\}\nonumber\\
&&+\frac{\gamma_b}{8}\big[\varphi^\dagger_b\beta^{+\nu}_\mu\big]\Big\{
\big[(\beta^{-\rho}_\nu)^\dagger\beta^{+\mu}_\rho\big]-\big[(\beta^{-\mu}_\rho)^\dagger
\beta^{+\rho}_\nu\big]\Big\}\nonumber\\
&&+c.c.
\end{eqnarray}
In terms of the fields $B_k$ they read
\begin{eqnarray}\label{a7,2}
-{\cal L}_{3\beta}&=&\gamma_t\epsilon_{klm}
\big[\varphi^\dagger_tB^-_k\big]
\big[(B^+_l)^\dagger B^-_m\big]\nonumber\\
&&+\gamma_b\epsilon_{klm}\big[\varphi^\dagger_bB^+_k\big]
\big[(B^-_l)^\dagger B^+_m\big]+c.c.
\end{eqnarray}
The couplings $\gamma_{t,b}$ are consistent with the discrete symmetry $G_A$ but violate the $U(1)_A$-symmetry. 
They may play an important role for the mechanism of mass generation discussed in the next section and app. C. 

Furthermore, there are ``cubic terms'' that mix the gauge fields and $\beta^\pm_{\mu\nu}$ for nonzero $\varphi_{t,b}$, i.e.
\begin{eqnarray}\label{A3A}
-{\cal L}_{F\beta}&=&\nu_{y+}[\varphi^\dagger_t\beta^+_{\mu\nu}\big]Y^{\mu\nu}+
\nu^*_{y+}\big[(\beta^+_{\mu\nu})^\dagger\varphi_t\big]Y^{\mu\nu}\nonumber\\
&&+\nu_{w+}[\varphi^\dagger_t\vec{\tau}\beta^+_{\mu\nu}]\vec{W}^{\mu\nu}+\nu^*_{w+}
\big[(\beta^+_{\mu\nu})^\dagger\vec{\tau}\varphi_t\big]\vec{W}^{\mu\nu}\nonumber\\
&&+\nu_{y-}[\varphi^\dagger_b\beta^-_{\mu\nu}]Y^{\mu\nu}+\nu^*_{y-}
\big[(\beta^-_{\mu\nu})^\dagger\varphi_b\big]Y^{\mu\nu}\nonumber\\
&&+\nu_{w-}[\varphi^\dagger_b\vec{\tau}\beta^-_{\mu\nu}]\vec{W}^{\mu\nu}+\nu^*_{w-}
\big[(\beta^-_{\mu\nu})^\dagger\vec{\tau}\varphi_b\big]\vec{W}^{\mu\nu}.
\end{eqnarray}
Here $Y_{\mu\nu}=\partial_\mu Y_\nu -\partial_\nu Y_\mu$ is the field strength for the hypercharge-gauge boson whereas $\vec{W}_{\mu\nu}$ denotes the gauge-covariant field strength for the $W$-bosons which includes terms quadratic in the gauge field $\vec{W}_\mu$. We observe that ${\cal L}_{F\beta}$ is invariant under CP-transformations if the  dimensionless couplings $\nu_i$ are real.\footnote{Invariants of the type $\int D_\mu\beta^{\mu\nu}D_\nu\varphi$ can be brought to the form (\ref{A3A}) by partial integration.} 

\begin{figure}[htb]
\centering
\includegraphics[scale=0.45]{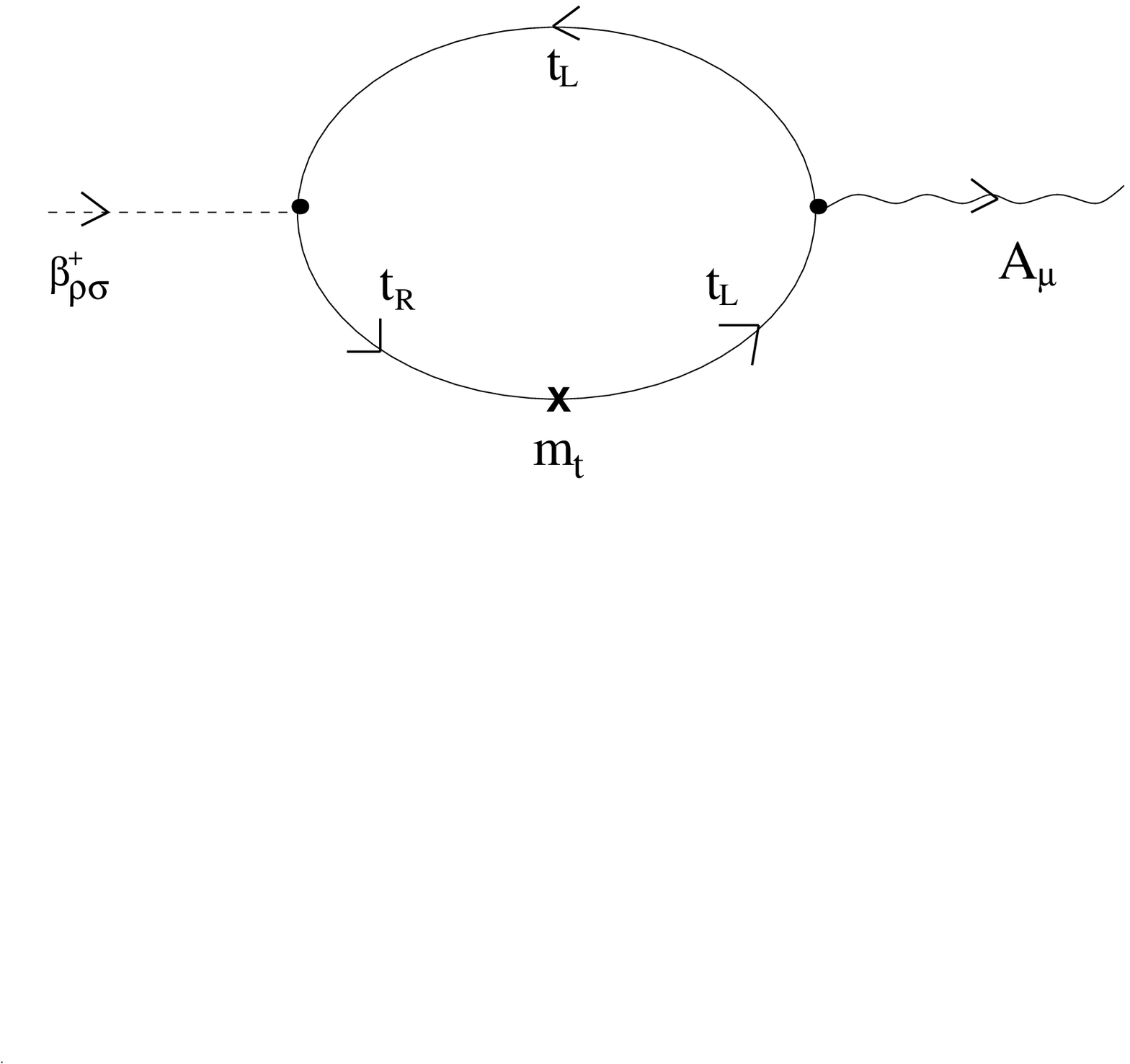}
\vspace{-4.0cm}
\caption{Chiron - photon mixing.\label{fig4}}
\end{figure}

The mixing between the gauge fields and the chirons in presence of nonzero $\bar{\varphi}_{t,b}$ will play an important role for the phenomenology of our model. Part of eq. (\ref{A3A}) is the mixing with the photon
\begin{equation}\label{N1}
-{\cal L}=-\frac{\epsilon^+_\gamma}{2\sqrt{2}}\left(\beta^{+0}_{\mu\nu}+(\beta^{+0}_{\mu\nu})^*\right)
F^{\mu\nu}
\end{equation}
where $\epsilon_\gamma\sim\bar{\varphi}_{t,b}$. Fig. \ref{fig4} shows a contribution to $\epsilon^+_\gamma$ from a top quark loop which is evaluated in app. A, yielding $\epsilon^+_\gamma\approx 0.2(f_t/5)~m_t$. A similar contribution for $\beta^-_{\mu\nu}$ involves a $b$-quark loop. Due to the smaller mass chiral coupling and electric charge of the $b$-quark one has 
\begin{equation}\label{N2}
\epsilon^-_\gamma\approx -\frac{1}{2}\frac{f_bm_b}{f_tm_t}\epsilon^+_\gamma.
\end{equation}

This concludes the list of dimension four terms,
\begin{eqnarray}\label{A8a}
&&\Gamma=-\int d^4x
\{{\cal L}^{ch}_{\beta,kin}+{\cal L}_{ch}+{\cal L}_{\beta,\psi}+{\cal L}_F+{\cal L}_{\psi,kin}\nonumber\\
&&+{\cal L}_{\phi,kin}+U
+{\cal L}_y+{\cal L}_{M,\beta}+{\cal L}_{3\beta}+{\cal L}_{F,\beta}\}.
\end{eqnarray}
We emphasize that, at least in principle, all the various coupling constants appearing in the effective action are computable in terms of the couplings of the action displayed in sect. II.

In principle, we may extend the discussion to terms with higher dimension. As an example of dimension six terms we discuss the modification of the kinetic term for the chiral tensor fields. Due to the spontaneous breaking of the hypercharge- and $G_A$-symmetries the $\beta$-field can also aquire kinetic terms with a structure different from eq. (\ref{4})
\begin{eqnarray}\label{14}
-{\cal L}^{ind}_{\beta,kin}&=&
\left\{ Z_s[\varphi^\dagger_t\varphi_b][(D_\rho\beta^-_{\mu\nu})^\dagger D^\rho\beta^{+\mu\nu}]\right.\nonumber\\
&&+Z_{t+}[\varphi^\dagger_t\partial_\rho\beta^{\mu\nu}_+]
[\varphi^\dagger_t\partial^\rho\beta_{+\mu\nu}]\nonumber\\
&&+Z_{b-}[\varphi^\dagger_b\partial_\rho\beta^{\mu\nu}_-]
[\varphi^\dagger_b\partial^\rho\beta_{-\mu\nu}]\nonumber\\
&&+Z_{t-}[\varphi^\dagger_t\partial_\rho\beta^{\mu\nu}_-]
[\varphi^\dagger_t\partial^\rho\beta_{-\mu\nu}]\nonumber\\
&&+Z_{b+}[\varphi^\dagger_b\partial_\rho\beta^{\mu\nu}_+]
[\varphi^\dagger_b\partial^\rho\beta_{+\mu\nu}]\nonumber\\
&&+Z_{m_1}[\varphi^\dagger_t\partial_\mu\beta^{\mu\nu}_+]
[\varphi^\dagger_b\partial_\rho\beta^\rho_{-\nu}]\nonumber\\
&&\left.+Z_{m2}[\varphi^\dagger_b\partial_\mu\beta^{\mu\nu}_+]
[\varphi^\dagger_t\partial_\rho\beta^\rho_{-\nu}]\right\}
+c.c.
\end{eqnarray}
All terms are consistent with the discrete $G_A$-symmetry while a continuous $U(1)_A$ is violated by nonvanishing $Z_{t-},Z_{b+}$. For real $Z_k$ the effective kinetic terms conserve CP.

\noindent

\section{Chiron propagator}

In the presence of strong interactions the momentum dependence of the propagator for the chiral tensor fields needs to be investigated carefully. We discuss here the inverse propagator of the $B^+$-field, neglecting for a moment the mixing between the various $B-$ fields due to the effective interactions (\ref{13}) or the mixing with the gauge fields from (\ref{A3A}). Lorentz-symmetry dictates the most general form of the full inverse propagator 
\begin{equation}\label{H1A}
\tilde{P}_{kl}(q)=Z_+(q^2)P_{kl}(q)=\tilde{P}_+(q^2)P_{kl}(q)/q^2.
\end{equation}
The possible poles of the propagator are given by the zeros of the function
\begin{equation}\label{H2a}
\tilde{P}_+(q^2)=Z_+(q^2)q^2.
\end{equation}
Stable particles correspond to a zero at negative $q^2=-m^2$ with $\partial\tilde{P}_+/\partial q^2> 0$. We show in appendix B that the energy density is then strictly positive and the B-fields describe massive spin one particles. An imaginary part of $P_+(q^2=-m^2)$ turns these states to resonances. Possible instabilities could arise from a tachyon corresponding to a zero at $q^2>0$ or to a ghost arising when $\partial\tilde{P}_+/\partial q^2<0$ at the zero $\tilde{P}_+$. Furthermore, secular instabilities \cite{CWQ} are possible if $\tilde{P}_+$ vanishes for $q^2=0$. The issue of stability and consistency of our model therefore dependends crucially on the form of $\tilde{P}_+(q^2)$. We will argue that $\tilde{P}_+(q^2)$ deviates substantially from the classical form $\tilde{P}_{+,cl}(q^2)=q^2$. This will cure the classical instabilities alluded to in sect. III and render our model consistent.

Below we will argue that the zero of the real part of $\tilde{P}_+(q^2)$ occurs for $q^2<0$ such that our model describes massive spin-one particles (resonances) without a problem of stability. A sufficient approximation for our purposes is
\begin{equation}\label{a12}
\tilde{P}(q^2)=q^2+m^2_+.
\end{equation}
Then the on-shell condition corresponding to the zero eigenvalues of $\tilde{P}_{kl}(q)$ is the standard relativistic condition for a massive particle
\begin{equation}\label{FF11}
q^2=-m^2_+~,~q^2_0=m^2_++\vec{q}^2.
\end{equation}
This follows directly from the squared inverse propagator
\begin{equation}\label{FF12}
\tilde{P}_{km}(q)\tilde{P}^*_{ml}(q)=(q^2+m^2_+)^2\delta_{kl}.
\end{equation}
Formally, we infer from eq. (\ref{H1A})
\begin{equation}\label{70C}
Z_+(q^2)=\frac{q^2+m^2_+}{q^2}
\end{equation}
and observe a nonlocal form of the effective action with $Z_+(q^2)$ diverging for $q^2\rightarrow 0$. 

The mass term appears in a nonlocal form only in the basis of fields $\beta_{\mu\nu}$ or $B_k$. There exists a different basis where this mass term is local\footnote{For a covariant formulation see appendix B.}, i.e. 
\begin{equation}\label{70A}
S^\pm_\mu=\frac{\partial_\nu}{\sqrt{\partial^2}}\beta^{\pm\nu}\ _\mu~,~\partial_\mu S^{\pm\mu}=0.
\end{equation}
In terms of $S_\mu$ the modified kinetic term (cf. eq. (\ref{4})) 
\begin{eqnarray}\label{70B}
\Gamma^{ch}_{\beta,kin}&=&-\int_q Z(q)q_\mu q_\nu\big(\beta^{\mu\rho}(q)\big)^\dagger\beta^\nu\ _\rho(q)\nonumber\\
&=&
\int_q(q^2+m^2)S^{\mu \dagger}(q) S_\mu(q)
\end{eqnarray}
turns out to be the standard action for a free massive vector field. This is not so surprising since the physical particles must correspond to irreducible representations of the inhomogeneous Lorentz group. For massive particles the little group is $SO(3)$, and with respect to $SO(3)$ antisymmetric tensors are equivalent to vectors. It seems a quite natural scenario that the particle spectrum for the strongly interacting chiral tensor fields consists of massive vector fields. This presumably even holds if the nonlocal factor $(\partial^2)^{-1/2}$ in the relation between $S^\mu$ and $\beta^{\mu\nu}$ is replaced by a more complicated function of momentum. We observe that the six complex fields $B^\pm_k$ precisely correspond to the twelve physical degrees for four massive spin one particles (two electrically charged and two neutral). 

We next present our arguments why the zero of $\tilde{P}_+(q^2)$ occurs for negative $q^2$. (We omit here the imaginary part which is due to the decay of the massive spin one bosons.) We begin with the range $q^2\geq0$ and argue that $\tilde{P}_+(q^2)$ remains strictly positive in this range, thus excluding a tachyonic pole in the propagator. Furthermore, also the  limit $\tilde{P}_+(q^2\rightarrow 0)=m^2_+$ yields a positive real number $m^2_+$. This implies that the  ``classical pole'' at $q^2=0$ is absent in the full propagator and secular instabilities can no longer occur. In terms of the wave function renormalization $Z_+$ this amounts to a nonlocal behavior
\begin{equation}\label{H3a}
\lim_{q^2\rightarrow 0}Z_+(q^2)=\frac{m^2_+}{q^2}.
\end{equation}
We will present two arguments for this nonlocal behavior. First we study in this section the ``running'' of $\tilde{P}_+(q^2)$ with $q^2$ and argue that $\tilde{P}(q^2=0)>0$ is plausible. Our second argument shows explicitely that a nonlocal term (\ref{H3a}) is generated in presence of electroweak symmetry breaking. It is due to a loop involving chiral tensors which becomes possible due to the effective cubic couplings (\ref{a7,1},\ref{a7,2}). More precisely, the presence of this loop excludes the classical behavior $\tilde{P}\sim q^2$ near $q^2=0$ and the explicit calculation gives $\tilde{P}(0)=m^2_+>0$. Since this loop contribution from the cubic couplings is presumably less important numerically we present this second argument in appendix C. 

Let us study the flow
\begin{equation}\label{H4a}
\frac{\partial\tilde{P}_+(q^2)}{\partial q^2}=Z_+(q^2)+q^2\frac{\partial Z_+(q^2)}{\partial q^2}
\end{equation}
from large positive $q^2$, where we normalize $Z_+(q^2=\Lambda^2)=1$, towards $q^2=0$. The second term in eq. (\ref{H4a}) is given by the anomalous dimension that we compute in app. A. We first neglect the influence of a nonvanishing top quark mass and find
\begin{equation}\label{H5a}
\frac{\partial\tilde{P}_+(q^2)}{\partial q^2}=\left(1-\frac{f^2_t(q^2)}{4\pi^2}\right)Z_+(q^2).
\end{equation}
For $Z_+(q^2)>0$ the inverse propagator $\tilde{P}_+$ has a minimum at $q^2_{min}$ where $f^2_t(q^2_{min})=4\pi^2$. It increases again for $q^2<q^2_{min}$. Since $\tilde{P}_+(q^2)>q^2$ (for $q^2<\Lambda^2)$ the value at the minimum is positive, $\tilde{P}_+(q^2_{min})>q^2_{min}$. More explicitely, we may insert eq. (\ref{8}), i.e. $f^2_t(q^2)=8\pi^2/[7\ln(q^2/\Lambda^2_{ch})]$ and solve eq. (\ref{H5a}) explicitely
\begin{equation}\label{H6a}
\tilde{P}_+(q^2)=q^2
\left(\frac{\ln(q^2/\Lambda^2_{ch})}{\ln(\Lambda^2/\Lambda^2_{ch})}\right)^{-2/7}.
\end{equation}
In this approximation $\tilde{P}_+(q^2)$ diverges as $q^2$ approaches $\Lambda^2_{ch}$.

It is, of course, not easy to figure out what happens precisely when $q^2$ gets close to the chiral scale $\Lambda^2_{ch}$ and below. A plausible hypothesis assumes that the  increase of $\tilde{P}_+(q^2)$ with decreasing $q^2$ is stopped and $\tilde{P}_+(q^2)$ reaches a constant $m^2_+$ for $q^2\rightarrow 0$. Indeed, for low enough $q^2$ the electroweak symmetry breaking slows down or stops the running of the couplings by providing an effective infrared cutoff through the top quark mass $m_t$. In a momentum range where $Z_+$ may diverge it is actually more appropriate to follow the flow of $\hat{f}^2_t=\bar{f}^2_t/Z^2_\psi=f^2_t Z_+$ instead of renormalizing the chiral coupling by dividing through $Z_+$. Only the momentum dependence of the wave function renormalization for the top quark, $Z_{t,R},Z_{t,L}$ contribute to the running of $\hat{f}^2_t(q^2)$. We discuss in app. A the dependence of $Z_t$ on $q^2$ and on the dynamically generated mass scales $m^2_+,m^2_t$ . 

We may distinguish between two basic scenarios for the mass generation for the chiral tensor fields. In the first one the electroweak symmetry breaking is a subleading effect such that $m_t$ can be neglected. The dominant infrared effect generated by the strong chiral couplings is then the mass generation for the chiral tensor. Correspondingly, the masses $m^2_\pm$ are the largest non-perturbative mass scales induced by dimensional transmutation. In turn, the mixing effects are small corrections since they involve spontaneous electroweak symmetry breaking by nonvanishing $\varphi_{t,b}$ in eqs. (\ref{13}) (\ref{A3A}). At the scale where the chiral tensors get massive the electroweak symmetry remains unbroken and the  quarks and leptons are still massless. 

Let us consider next the effective theory for the massless fields with low momenta $|q^2|\ll m^2_\pm$. This is still a strongly interacting theory due to the four fermion interactions induced by the exchange of the chiral tensors. If the corresponding four fermion coupling is above a critical value it grows as the momentum scale is lowered. In turn, the scale where the effective four fermion coupling ``diverges'' can typically by associated with the scale of electroweak symmetry breaking, very similar to \cite{BL},\cite{GJW}. This scenario could induce a ``little hierarchy'' $m_t\ll m_+$. However, the growth of the effective four fermion interaction is typically rather fast and we do not expect a ratio $m_t/m_+$ smaller than $10^{-2}$. (A ratio $m_t/m_+\approx 1/10$ may be a natural value for this scenario, i.e. $m_+\approx 2TeV$.) 

In the second scenario the electroweak symmetry breaking is a crucial ingredient for the mass generation for the chiral tensor fields. In this case one expects $m_\pm$ to be only mildly larger than $m_t$. Our phenomenological discussion below shows that chiral tensor masses around $1$ TeV (or larger) may be favored, thus lending some advantage to the first scenario. Of course, the transition between both scenarios is smooth. 

The first scenario seems quite plausible, even though we cannot present a convincing theoretical proof at this stage of our investigation. The increase of the inverse wave function renormalization for the top quark, $Z^{-1}_t$, is the key ingredient for the increase of the renormalized chiral coupling $f_t$. In app. A we have approximated the momentum dependence of $Z^{-1}_t(p^2)$ by
\begin{equation}\label{A13a} 
Z^{-1}_t(p^2)=1+\frac{\mu^2_t}{p^2}.
\end{equation}
Solving the one loop Schwinger-Dyson equation for this form yields indeed a contribution
\begin{equation}\label{A13b} 
\Delta P_{kl}(q)=\frac{3\hat{f}^2_t\mu^2_t}{2\pi^2}
\frac{P_{kl}(q)}{q^2}
\end{equation}
and therefore
\begin{equation}\label{A13c} 
m^2_+=\frac{3\hat{f}^2_t\mu^2_t}{2\pi^2}.
\end{equation}
Of course, this is only a plausibility argument since the solution of the Schwinger-Dyson equation depends sensitively on the shape of $Z^{-1}_t(p^2)$ in the non-perturbative region. 

If a mass generation of this type does not operate we have to turn the second scenario. Electroweak symmetry breaking generates new effective couplings. In particular, the inverse propagator $P_{kl}(q)$ receives now a new loop contribution with intermediate chiral tensor fields. We show in app. C that this yields indeed a contribution $m^2_\pm\sim m^2_t$, where the proportionality coefficient involves the so far undetermined effective couplings $\gamma$ in eq. (\ref{a7,2}). In any case, the positivity of $\tilde{P}$ for $q^2>0$ with $\tilde{P}(q^2=0)=m^2>0$ implies by continuity that $\tilde{P}(q^2)$ (or its real part) should have a zero at $q^2<0$. This follows from the limiting behavior $\tilde{P}=q^2$ for $q^2\rightarrow-\infty$.

We conclude that for a suitable form of the effective action the massive spin one particles arising from the chiral tensor field can be consistently described without tachyons, ghosts or other problems of stability. This suggests that our model with chiral tensor fields is an acceptable physical model, without problems of stability or unitarity. The key ingredient for this cure of the diseases of the free chiral tensor propagator are the modifications of the propagator through the strong effective interactions. In particular, this concerns the generation of a mass term.

\section{Gauge boson - \\chiral tensor - mixing}
\label{massive}

In this section we explore a simple effective action for the massive chirons. It is based on the mass term in eq. (\ref{70B}). We include the mixing between the various spin one states and present a first short discussion of the phenomenological consequences of this effect.

\begin{itemize}
\item [a)] {\bf  Propagator for charged chiral tensors}
\end{itemize}

We concentrate first on the $B^\pm$ components with nonzero electric charge that we denote by 
\begin{equation}\label{70D} 
B^{+,+}_k=C_{1k}~,~B^{-,+}_k=C_{2k}.
\end{equation}
We use for these fields the nonlocal wave function renormalization suggested by the preceeding section
\begin{equation}\label{70E} 
Z_\pm(q^2)=Z_\pm\left(1+\frac{m_\pm}{q^2}\right).
\end{equation}
Taking also into account the off-diagonal masslike terms from eq. (\ref{13}) one finds for the terms quadratic in $C$ in momentum space

\begin{eqnarray}\label{L1}
\Gamma_2&=&\int_q\{Z_+ C_{1k}^* P_{k\ell}(q)C_{1k}+Z_- C^*_{2k}
P^*_{k\ell}(q)C_{2\ell}\nonumber\\
&&+Z_+ m^2_+ C^*_{1k}\frac{P_{k\ell}(q)}{q^2}C_{1\ell}+Z_- m^2_-
C^*_{2k}\frac{P^*_{k\ell}(q)}{q^2}C_{2\ell}\nonumber\\
&&+\tilde m^2(C^*_{2k} C_{1k}+C^*_{1k}C_{2k})\}.
\end{eqnarray}
Here $\tilde{m}^2$ is given by 
\begin{equation}\label{61a}
\tilde m^2=\sigma_1\bar{\varphi}^*_t\bar{\varphi}_b+\sigma_{\nu 1}\bar{\varphi}^*_b\bar{\varphi}_t.
\end{equation}
(We have not included the dimension six term (\ref{14}), i.e. $\tilde
Z=0$ in app. B.) 

In the absence of mixing with the gauge fields the inverse
propagator 

\begin{equation}\label{L2} 
\Gamma_C^{(2)}=\left(\begin{array}{ccc}
Z_+ P+Z_+m^2_+ P/q^2&,&\tilde m^2\\
\tilde m^2&,&Z_-P^*+Z_-m^2_-P^*/q^2\end{array}\right)
\end{equation}
has vanishing eigenvalues for
\begin{equation}\label{L3}
Z_+Z_-(q^2+m^2_+)(q^2+m^2_-)-\tilde m^4=0.
\end{equation}
The corresponding renormalized mass terms
\begin{eqnarray}\label{L4}
m^2_{R1,2}&=&\bar m^2\pm\sqrt{\hat m^4+\Delta^2},\nonumber\\
\bar m^2&=&\frac{1}{2}(m^2_++m^2_-),\quad
\Delta=\frac{1}{2}(m^2_+-m^2_-),\ \nonumber\\
\hat m^2&=&\frac{\tilde m^2}{\sqrt{Z_+Z_-}}
\end{eqnarray}
are both positive provided that
\begin{equation}\label{L5}
\hat m^4<m^2_+m^2_-.
\end{equation}
When no distinction between the different renormalized mass eigenvalues (e.g. $m_{R1},m_{R2}$) is made we will denote the chiron mass by $M_c$. 

The energy density in the rest frame $(\vec q=0,\ P/q^2=1)$
\begin{eqnarray}\label{L6}
\rho=&&\int_x\{Z_+(\dot C_{1k}^*\dot
C_{1k}+m^2_+C^*_{1k}C_{1k})\nonumber\\
&&+Z_-(\dot C^*_{2k}\dot
C_{2k}+m^2_-C^*_{2k}C_{2k})\nonumber\\
&&+\tilde m^2(C_{2k}^* C_{1k}+C^*_{1k}C_{2k})\}
\end{eqnarray}
is positive if the condition (\ref{L5}) holds. The effective action
(\ref{L1}) therefore describes the massive antisymmetric tensor fields in
a fully consistent way as massive spin one particles.

\begin{itemize}
\item [b)] {\bf  Mixing with $W$-boson}
\end{itemize}

We next include the cubic coupling to the gauge bosons
(\ref{A3A}). For the charged fields the spontaneous symmetry breaking
$(\bar{\varphi}_{t,b}\ne 0)$ generates a mixing with the $W^\pm$-gauge
bosons. By partial integration the relevant pieces in $-{\cal
  L}_{F\beta}$ become in quadratic order in $\beta_{\mu\nu}$ and
$W_{d\mu}^\pm=\frac{1}{\sqrt{2}}(W_{1\mu}\mp iW_{2\mu})$
\begin{eqnarray}\label{L7}
-{\cal L}_{F,\beta}&=&-2\sqrt{2}\left(\nu_{w+}\bar{\varphi}^*_t\partial^\mu\beta^{+,+}_{\mu\nu}W^{-\nu}_d
\right.\nonumber\\
&&\left.+\nu_{w-}\bar{\varphi}^*_b\partial^\mu\beta^{-,+}_{\mu\nu}W^{-\nu}_d\right)+c.c.
\end{eqnarray}
(We have introduced a subscript $d$ for ``diagonal'' in order to facilitate the
distinction from the physical fields $W^{\pm\nu}$.) The mixing is most
easily discussed in terms of the field $S^\pm_\mu$ (\ref{70A})
\begin{eqnarray}\label{L8}
-{\cal
 L}_{F\beta}&=&-2\sqrt{2}W_d^{-\mu}\sqrt{\partial^2}\left(\frac{\nu_{w+}\bar{\varphi}^*_t}{\sqrt{Z_+}}S^{+,+}_{d,\mu}+\frac{\nu_{w-}\bar{\varphi}^*_b}{\sqrt{Z_-}}S^{-,+}_{d,\mu}\right)\nonumber\\&&+c.c.
\end{eqnarray}

Combining with eq. (\ref{L1}) and the usual kinetic and mass terms for
the gauge fields this yields for the inverse propagator matrix for the
vector fields $(S^{+,+}_{d,\mu},S^{-,+}_{d,\mu}, W^+_{d,\mu}$)
\begin{equation}\label{L9}
\Gamma^{(2)}_C=\left(\begin{array}{ccc}
q^2+m^2_+,&\hat m^2,&\varepsilon^*_+\sqrt{-q^2}\\
\hat m^2,&q^2+m^2_-,&\varepsilon^*_-\sqrt{-q^2}\\
\varepsilon_+\sqrt{-q^2}, &\varepsilon_-\sqrt{-q^2},&q^2+\bar
M^2_W\end{array}
\right)
\end{equation}
with
\begin{equation}\label{L10}
\varepsilon_+=-2\sqrt{2}\frac{\nu_{w+}\bar{\varphi}^*_t}{\sqrt{Z_+}},\
\varepsilon_-=-2\sqrt{2}\frac{\nu_{w-}\bar{\varphi}_b^*}{\sqrt{Z_-}}.
\end{equation}
(For convenience we use here the gauge $\partial_\mu W^\mu_d=0$ such
that all three vectors are divergence free.) The squared renormalized mass eigenvalues $\lambda$ obey

\begin{eqnarray}\label{L11}
&&(m^2_+m^2_--\hat m^4)(\bar M^2_W-\lambda)-\lambda(m^2_++m^2_-)(\bar  M^2_W-\lambda)\nonumber\\
&&+\lambda^2(\bar M^2_W-\lambda)\nonumber\\
&&-\lambda[m^2_-\varepsilon^*_+\varepsilon_++m^2_+\varepsilon^*_-
\varepsilon_--\hat
  m^2(\varepsilon^*_+\varepsilon_-+\varepsilon_+\varepsilon^*_-)]\nonumber\\
&&+\lambda^2(\varepsilon^*_+\varepsilon_++\varepsilon^*_-\varepsilon_-)=0.
\end{eqnarray}

The mixing between the chirons and the gauge bosons is small if $|\varepsilon\sqrt{-q^2}|$ is  small as
compared to the mass differences $|m^2_{R1}-m^2_{R2}|,|m^2_{R1}-\bar
M^2_W|,|m^2_{R2}-\bar M^2_W|$ (with $m^2_{R1,2}$ given by
eq. (\ref{L4})). We may use for $\epsilon$ an estimate similar to appendix A, i.e. $\epsilon\approx 0.2(f_t/10)m_t$. For $\sqrt{-q^2}=M_W$ and $M_c>m_t$ the mixing is small (unless $f_t$ is extremely large). For small mixing one may then expand in powers of $\varepsilon$ and
note that the corrections to the mass eigenvalues appear only in second order
in $\varepsilon$. Let us consider a situation where
$m^2_{R1},m^2_{R2}\gg\bar M^2_W$. The shift in the smallest eigenvalue, $\Delta=\lambda-\bar M^2_W$,
obeys in linear order
\begin{eqnarray}\label{L12}
&&\Delta[m^2_+m^2_--\hat m^4-(m^2_++m^2_-)\bar M^2_W+\bar M^4_W]=\nonumber\\
&&-\bar
  M^2_W[m^2_-\varepsilon^*_+\varepsilon_++m^2_+\varepsilon^*_-\varepsilon_--\hat m^2(\varepsilon^*_+\varepsilon_-+\varepsilon_+\varepsilon^*_-)]\nonumber\\
&&+\bar
  M^4_W[\varepsilon^*_+\varepsilon_++\varepsilon^*_-\varepsilon_-].
\end{eqnarray}
As a consequence, the physical $W$-boson mass gets a correction
\begin{equation}\label{L13}
M^2_W=\bar M^2_W+\Delta,\ \Delta\sim\varepsilon^2\bar M^2_W/M^2_c.
\end{equation}

Besides the shift in the mass eigenvalues the mixing also modifies the
coupling of the physical $W$-boson $W_\mu$ (eigenstate of
$\Gamma^{(2)}_C$ (\ref{L9})) to the fermions. Indeed, the ``diagonal'' fields
$S^{+,+}_{d,\mu},S^{-,+}_{d,\mu}$ contain a small admixture of $W^+_\mu$. In
linear order in $\varepsilon$ one can replace
\begin{eqnarray}\label{L14}
S^{+,+}_{d,\mu}&\rightarrow& S^{+,+}_\mu-\alpha_+\sqrt{\partial^2}
W^+_\mu,\nonumber\\
S^{-,+}_{d,\mu}&\rightarrow& S^{-,+}_\mu-\alpha_-\sqrt{\partial^2}W^+_\mu,
\end{eqnarray}
where $\alpha_\pm$ are of the order $\varepsilon/M^2_c$ according  to
the diagonalization of the matrix (\ref{L9}). In consequence, the
chiral coupling of $S^\pm_{d,\mu}$ to the fermions induces a new effective
coupling of the $W$-boson. For the example of the leptons one obtains
from eq. (\ref{2})

\begin{eqnarray}\label{L15}
-{\cal L}_{ch}&=&-\frac{1}{2}\bar\ell_L\bar
 F^\dagger_L\sigma^{\mu\nu}\beta^-_{\mu\nu} e_R+h.c.\nonumber\\
&&\rightarrow -\{\bar\ell_L F^\dagger_L \sigma^{\mu\nu}
 e_R\}\frac{\partial_\mu}{\sqrt{\partial^2}} S^-_{d,\nu}+h.c.\nonumber\\
&&\rightarrow\alpha_-\{\bar\nu_L F^\dagger_L\sigma^{\mu\nu}
 e_R\}\partial_\mu W_\nu^++h.c.\nonumber\\
&&\rightarrow\frac{\alpha_-}{2}\{\bar\nu_L F^\dagger_L\sigma^{\mu\nu}
 e_R\} W^+_{\mu\nu}+h.c.
\end{eqnarray}
The new interaction  is of the Pauli type. As compared to the standard
weak  four-fermion  V-A interaction for low momenta the new effective
interaction is suppressed by a relative factor  $\alpha_-q_\mu$ with
$q_\mu$ the momentum of the exchanged $W$-boson.  This holds in
analogy for the quarks. One may use the observational constraints on
the general form of the weak four-fermion interaction 
  in order to derive bounds on $\alpha$ and therefore on $\varepsilon$.

\begin{itemize}
\item [c)] {\bf Mixing with neutral gauge bosons}
\end{itemize}

The discussion of masses and mixings in the neutral sector is more involved since there are six independent neutral vector states: the photon and $Z$-boson (or $Y_\mu$ and $W^3_\mu$) and four states $S^{+,0}_\mu,\ S^{-,0}_\mu,\ (S^{+,0}_\mu)^*,\ (S^{-,0}_\mu)^*$. Let us first neglect the mixing with the gauge bosons. The nonlocal momentum dependence of $Z_{\pm}(q^2)$ (\ref{70C}) induces again ``diagonal terms'' $\sim\ Z_\pm(q^2+m^2_\pm)$, similar to the charged vector bosons in eq. (\ref{L9}). 
The off-diagonal term $\sim\ \tilde m^2$ is now replaced by
\begin{equation}\label{L16}
\tilde m^2_0=(\sigma_1+\sigma_2)\bar{\varphi}_t^*\bar{\varphi}_b+(\sigma_{v 1}+\sigma_{v 2})\bar{\varphi}_b^*\bar{\varphi}_t.
\end{equation}
Further off-diagonal terms $\sim \sigma_+\bar{\varphi}^2_t+\sigma_{v_+}\bar{\varphi}^2_b$ and 
$\sigma_-\bar{\varphi}^2_b+\sigma_{v-}\bar{\varphi}^2_t$ mix the real and imaginary parts of the complex fields 
$S^{+,0}$ and $S^{-,0}$. One finds a second positivity constraint similar to eq. (\ref{L5}). It is obviously obeyed if the off-diagonal entries $\sim \sigma_i$ are small as compared to $m^2_\pm$. 

The cubic terms (\ref{A3A}) induce now a mixing with the $Z$-boson and modify the physical $Z$-boson mass $m^2_Z$. This can be worked out in a straightforward manner in analogy to the charged vector mesons. Of particular interest is the mixing with the photon. Since the electromagnetic $U(1)$-gauge symmetry remains conserved by the expectation values $\bar{\varphi}_{t,b}$ we are guaranteed that the physical photon remains massless. However, the photon may acquire new effective couplings. Let us demonstrate this effect by reducing the $6\times 6$ matrix $\Gamma^{(2)}_c$ to a $2\times 2$ matrix where one vector state $A^\mu_d$ corresponds to the photon after the Higgs-mechanism, but without the mixing effects with $\beta_{\mu\nu}$, whereas the other state is a suitable linear combination of neutral chiral tensor fields. In presence of the mixing $\sim \beta$ the inverse propagator is of the form\footnote{Due to the presence of four more states the coefficients $M^2_c$ and $\beta$ depend actually mildly on $q^2$. For small $|q^2|$ the resulting corrections are $\sim q^2/M^2cR$.}
\begin{equation}\label{L17}
\Gamma^{(2)}_c=\left(\begin{array}{lll}
q^2+M^2_c&,&\epsilon_\gamma\sqrt{-q^2}\\
\epsilon_\gamma\sqrt{-q^2}&,&q^2\end{array}\right)\cdot 
\end{equation}
As it should be, the determinant 
\begin{equation}\label{L18}
\det=q^2(q^2+M^2_c+\epsilon^2_\gamma)
\end{equation}
vanishes for a photon on mass shell $(q^2=0)$. Also the mixing $\sim\epsilon_\gamma\sqrt{-q^2}/M^2_c$ vanishes on-shell, guaranteeing the universal electric charges of the physical states. 

\begin{itemize}
\item [d)] {\bf Anomalous magnetic moment of muon and electron}
\end{itemize} 

However, for a virtual photon exchange with $q^2\neq0$ we have to take into account the new Pauli-type coupling similar to eq. (\ref{L15})
\begin{equation}\label{L19}
-{\cal L}_{ch}\rightarrow \alpha_\gamma\{\bar{e}_RF_L\sigma^{\mu\nu}e_L
+\bar{d}_RF_D\sigma^{\mu\nu}d_L\}
F_{\mu\nu}+h.c.
\end{equation}
with 
\begin{equation}\label{A16A}
\alpha_\gamma=-\frac{\epsilon^-_\gamma}{2\sqrt{2}M^2_c}.
\end{equation}
For simplicity we associate in eq. (\ref{A16A}) the neutral tensor field with the real part of $\beta^-$, or more precisely with $(\beta^-+\beta^{-*})/\sqrt{2}$, since this component couples to the leptons. The mixing coefficient $\epsilon_\gamma$ in eqs. (\ref{L17}), (\ref{A16A}) is then related to a piece in effective action
\begin{equation}\label{A16B}
\Gamma_{m}=-\frac{\epsilon^-_\gamma}{2\sqrt{2}}\int_x\left[\beta^{-0}_{\mu\nu}+(\beta^{-0}_{\mu\nu})^*\right]F^{\mu\nu}.
\end{equation}
We estimate $\epsilon^-_\gamma$ in appendix A 
\begin{equation}\label{A16C}
\epsilon^-_\gamma=-0.1 c_\beta f_b m_b.
\end{equation}
Here $c_\beta$ is a coefficient of the order unity which takes into account that a more accurate treatment should involve appropriate linear combinations of the four neutral tensor fields and the $Z$-boson.

We observe that the relative strength of the tensor coupling of the photon to $e,\mu, \tau, d,s,b$ is given by the respective size of the chiral couplings (same $\alpha_\gamma$), whereas the tensor coupling to $u, c,t$ involves also a different mixing term $\tilde{\alpha}_\gamma\neq\alpha_\gamma$.  Since the Pauli-type interaction (\ref{L19}) is linear in the mixing parameter $\epsilon_\gamma$ it is one of the dominant effects that may decide on the phenomenological viability of our model. 

We work in a basis for the fermion fields where the chiral matrices $F_{U,D,L}$ are all diagonal. As in the standard model, the only flavor violation arises then from the CKM-matrix for the coupling of the charged $W$-bosons to the quarks. Small flavor-changing neutral currents can only be induced by loop effects. In particular, the lepton numbers $L_{e,\mu,\tau}$ are conserved separately for each lepton 
generation and the Pauli-interaction (\ref{L19}) does not contribute to rare decays like $\mu\rightarrow e\gamma$. It gives, however, a contribution to the anomalous magnetic moment $g-2$ for the muon and electron. The precision tests for the anomalous magnetic moments can directly be related to $\alpha_\gamma f_{\mu,e}$. We recall that the generation structure in the lepton sector suggests for the chiral couplings of the electron and muon $f_e\ll f_\mu$.  Given that the anomalous magnetic moment is also proportional to the lepton mass $m_e,m_\mu$ and in view of the present relative experimental precision the muon magnetic moment is by far the more sensitive probe. 

More explicitely, for real diagonal $F_L$ and $M_L$ the  new Pauli-type interaction of the muon reads $(\alpha_\gamma=\alpha_{\gamma,R}+i\alpha_{\gamma,I})$
\begin{equation}\label{M1a}
-{\cal L}_{ch}=f_\mu\bar{\mu}\sigma^{\mu\nu}(\alpha_{\gamma,R}\gamma^5+i\alpha_{\gamma,I})\mu F_{\mu\nu}.
\end{equation}
The real part of $\alpha_\gamma$ contributes to the anomalous magnetic moment of the muon\footnote{In our conventions the mass term for the muon reads $-{\cal L}=m_\mu\bar{\mu}\gamma^5\mu$ with real $m_\mu$.}
\begin{eqnarray}\label{M2a}
\Delta(g-2)&=&-\frac{8m_\mu}{e}f_\mu\alpha_{\gamma,R}\nonumber\\
&=&3\cdot 10^{-3}f_\mu\frac{\epsilon^-_\gamma}{m_t}\left(\frac{m_t}{M_c}\right)^2.
\end{eqnarray}
Using the estimate (\ref{A16A}) this yields
\begin{equation}\label{A16D}
\Delta(g-2)=-4\cdot 10^{-7}c_\beta\sigma f^2_b\left(\frac{m_t}{M_c}\right)^2~,~\sigma=\frac{f_\mu m_b}{f_b m_\mu}.
\end{equation}
We expect $\sigma$ to be of order one, with the sign of $\sigma$ given by the unknown relative sign between $f_\mu$ and $f_b$. (Our basis of fermion fields is such that all masses are real and positive.)

In order to get a feeling on the bounds on the chiron mass $M_c$ let us compare eq. (\ref{M2a}) with the difference between  the central value of the recent $g-2$-experiment \cite{G2} and the theoretical estimates for the standard model, $\Delta(g-2)=5\cdot 10^{-9}$. From eq. (\ref{A16D}) we infer that the bound depends strongly on the unknown chiral coupling of the $b$-quark. For
 $f_b\lesssim 1/20$ the chiron-contribution to $g-2$ seems too small to be presently observable even for $M_c$ in the vicinity of $m_t$.
 For $M_c =300 GeV, f_b\approx 1/6$ the chiral contribution could account for the experimental value if the sign of
 $c_\beta\sigma$ is negative. For even larger $f_b$ chirons with mass around $1TeV$ could still influence the anomalous magnetic moment at an observable level. 

The imaginary part $\alpha_{\gamma,I}$ violates CP and shows up in the electric dipole moment of the muon (or electron). It is much smaller than $\alpha_{\gamma,R}$, being generated only by loops involving the CKM matrix in the $W$-boson couplings. We have not yet estimated its size. 

\begin{itemize}
\item [e)] {\bf Tensor coupling of the photon and $Z$-boson and $e^+-e^-$-scattering}
\end{itemize} 

The tensor coupling of the photon (\ref{L19}) could also affect the $e^+e^-$-scattering at LEP. The relevant effective interaction obtains by replacing in eq. (\ref{L19})
\begin{equation}\label{b14a}
F_{\mu\nu}\rightarrow \frac{e}{2(-\partial^2)}\big(\partial_\mu\{\bar{e}_j\gamma_\nu e_j\}-\partial_\nu
\{\bar{e}_j\gamma_\mu e_j\}\big)
\end{equation}
(with $e\approx 0.3$ in the prefactor denoting the electric charge, $\alpha_{em}=e^2/4\pi$, whereas $e_j$ denotes the charged leptons $(e,\mu,\tau)$.) Thus the effective interaction contributing to $e^+e^-\rightarrow b\bar{b}$ reads in leading order 
\begin{eqnarray}\label{EEA}
-{\cal L}&=&-\frac{\alpha_\gamma}{2}ef_b
\{\bar{b}_R\sigma^{\mu\nu}b_L\}
\frac{1}{\partial^2} \big(\partial_\mu\{\bar{e}\gamma_\nu e\}-\partial_\nu \{\bar{e}\gamma_\mu e\}\big)\nonumber\\
&&+h.c.
\end{eqnarray}
In the center of mass system this results in an effective (nonlocal) four fermion interaction
\begin{equation}\label{EEB}
-{\cal L}=-i\frac{\alpha_\gamma e f_b}{\sqrt{s}}
\{\bar{e}_L\tau_ke_L-\bar{e}_R\tau_k e_R\}
\{\bar{b}_R\tau_k b_L\}+h.c.
\end{equation}
with $s=4E^2=-(p_1+p_2)^2$ the center of mass energy. Interestingly, this effective interaction can be related to the  $g-2$ - corrections for the muon: for the relative strength of both corrections $\alpha_\gamma$ drops out such that it depends only on the ratio of chiral couplings $f_b/f_\mu$. 

Similar interactions arise from the exchange of a $Z$-boson, with $s^{-1/2}$ replaced by $s^{1/2}/(s-m^2_Z)$. (We include the decay width of the $Z$-boson in a complex value of $m^2_Z$.) Also the photon coupling to $e^+e^-$ is replaced by the appropriate $Z$-coupling and $\alpha_\gamma$ by $\alpha_Z$. 

The dominant contribution is expected for the scattering $e^+e^-\rightarrow \bar{b}b$, since the other channels are suppressed by small chiral couplings. In principle, the new tensor coupling of the $Z$-boson could lead to a deviation of the forward-backward asymmetry, the left-right asymmetry, or the partial cross section and decay width into $b$-quarks from the standard model value. 
 
The particular structure of the tensor-vector-interaction (\ref{EEB}) involves an odd number of left handed (or right handed) fields. This distinguishes it from the normal vector-vector-interactions mediated by the photon and $Z$-boson and suppresses the interference between normal and tensor-vector contributions by a factor $\sim m_b/\sqrt{s}$. The characteristic strength of the anomalous tensor coupling $g_T$ of the $Z$-boson is given by $2\alpha_Z f_b \sqrt{s}$. On the $Z$-resonance, and with $\alpha_Z\approx f_b m_b/(30M^2_c)$, this amounts to $g_T\approx 8\cdot10^{-4}f^2_b(m_t/M_c)^2$, to be compared with $g$ for the vector/axial vector coupling. In view of the suppression of the interference term and the constraint from the anomalous magnetic moment of the muon,  $f^2_b(m_t/M_c)^2\lesssim 1/80$,  it seems unlikely that this effect is visible in the LEP-experiments, the best chances being for large $f_b/f_\mu$.

Besides the interference term there are also direct terms where the effective interaction (\ref{EEB}) enters quadratically. On the $Z$-resonance the suppression is now $\sim g^2_T$ and therefore even stronger, albeit the ``interference suppression'' $\sim m_b/m_Z$ is now absent. 

 Obviously, all these estimates are only rough order of magnitude estimates and a detailed computation is needed for reliable bounds on the effective interactions beyond the standard model.

So far, the effects of the new spin one particles seem to be consistent with observation. If their mass is not too high the  chirons may lead to interesting observable effects which distinguish our model from the standard model. In particular, the mixing with the gauge bosons affects the effective interactions of the standard model particles even in an energy range far below the mass of the chiral tensors. Further phenomenological consequences will be sketched in the next section.

\section{Phenomenology of chiron exchange}
In this section we compute the effects of the exchange of the antisymmetric tensor fields $\beta^\pm_{\mu\nu}$ in a simple and systematic way. For this purpose we solve the field equations for $\beta_{\mu\nu}$ in the presence of the fermion, gauge and scalar fields and reinsert the (field-dependent) solutions into the effective action. The new effective interactions for the fields of the electroweak standard model (now without the antisymmetric tensor fields $\beta_{\mu\nu}$) reflect all $\beta$-exchange diagrams. For processes without external chiral tensor fields this procedure is, in principle, exact. The approximations arise through the truncation of the effective action that we take here up to quadratic order in $\beta_{\mu\nu}$ or $S_\mu$. (For this section it will be most convenient to work in the basis of fields $S_\mu$. Note that for the solution of the field equations derived from the effective action arbitrary changes of basis are allowed.) We also should emphasize that contributions from chiron loops to the standard model vertices are not contained in this procedure. They are incorporated into $\Gamma[\beta_{\mu\nu}=0]$. We will not consider these explicit contributions from chiron loops in this paper.

\begin{itemize}
\item [a)] {\bf Exchange of charged chiral tensors}
\end{itemize} 

We concentrate first on the effects of the (virtual) exchange of the charged fields $S^{\pm,+}_\mu$. In lowest order we need the terms linear and quadratic in $S$ (we omit from now on the index for the electric charge)
\begin{eqnarray}\label{Q1}
-{\cal L}&=&(J^{+\mu})^\dagger S^+_\mu+(J^{-\mu})^\dagger S^-_\mu+h.c.\nonumber\\
&&+(\partial^\mu S^{+\nu})^*\partial_\mu S^+_\nu+(\partial^\mu S^{-\nu})^*\partial_\mu  S^-_\nu\nonumber\\
&&+m^2_+(S^\mu_+)^*S_{+\mu}+m^2_-(S^\mu_-)^* S_{-\mu}\nonumber\\
&&+\hat{m}^2\big((S^\mu_+)^*S_{-\mu}+(S^\mu_-)^*S_{+\mu}\big).
\end{eqnarray}
The currents are divergence free, $\partial_\mu J^{\pm\mu}=0$, and are given by
\begin{eqnarray}\label{Q2}
(J^{+\mu})^\dagger&=&\epsilon_+\sqrt{\partial^2}W^{\mu*}+
\frac{\partial_\nu}{\sqrt{\partial^2}}\bar{u}_R 
F_U\sigma^{\nu\mu}d_L\nonumber\\
(J^{-\mu})^\dagger&=&\epsilon_-\sqrt{\partial^2}W^{\mu*}\nonumber\\
&&+\frac{\partial_\nu}{\sqrt{\partial^2}}
(\bar{u}_LF^\dagger_D\sigma^{\nu\mu}d_R+
\bar{\nu}_LF^\dagger_L\sigma^{\nu\mu}e_R).
\end{eqnarray}
The solution of the field equations $(\alpha,\beta=+,-)$ reads
\begin{equation}\label{Q3}
(S^{\alpha,\mu})^\dagger=-(J^{\beta\mu})^\dagger G^{\beta\alpha}
\end{equation}
with (cf. eq. (\ref{L4}))
\begin{eqnarray}\label{Q4}
(G^{-1})&=&\left(\begin{array}{lll}
-\partial^2+m^2_+&,&\hat{m}^2\\
\hat{m}&,&-\partial^2+m^2_-\end{array}\right)
\end{eqnarray}
and
\begin{eqnarray}\label{Q4a}
G&=&(-\partial^2+m^2_{R1})^{-1}
(-\partial^2+m^2_{R2})^{-1}\nonumber\\
&&\left(\begin{array}{lll}
-\partial^2+m^2_-&,&-\hat{m}^2\\
-\hat{m}^2&,&-\partial^2+m^2_+\end{array}\right).
\end{eqnarray}
Therefore the resulting effective Lagrangian becomes
\begin{equation}\label{Q5}
-{\cal L}=-(J^{\beta\mu})^\dagger G^{\beta\alpha}J^\alpha_\mu.
\end{equation}
The interactions (\ref{Q5}) reflect the new physics beyond the standard model that is induced by the chiral tensor fields. Experimental tests of our model proceed by the test of these new interactions.

\begin{itemize}
\item [b)] {\bf Corrections to the $W$-propagator}
\end{itemize} 

Let us discuss the different pieces of the quadratic form (\ref{Q5}) separately. For the contribution quadratic in $W$ one obtains $(W_{\mu\nu}=\partial_\mu W^+_\nu-\partial_\nu W^+_\mu)$
\begin{eqnarray}\label{Q6}
-{\cal L}_{WW}&=&\frac{1}{2}W^{\mu\nu^*}p_W(-\partial^2)W_{\mu\nu},\nonumber\\
p_W&=&|\epsilon_+|^2G^{++}+|\epsilon_-|^2G^{--}+2Re[\epsilon_+\epsilon^*_-G^{+-}].
\end{eqnarray}
This modifies the inverse propagator of the $W$-boson according to
\begin{equation}\label{Q7}
q^2+\bar{M}^2_W\rightarrow q^2\big(1+p_W(q^2)\big)+\bar{M}^2_W.
\end{equation}
In particular, the pole of the propagator at $q^2=-\lambda$ occurs for
\begin{eqnarray}\label{Q8}
&&\bar{M}^2_W=\lambda\Bigg(1+{[(m^2_{R1}-\lambda)(m^2_{R2}-\lambda)]^{-1}}\\
&&{[|\epsilon_+|^2(m^2_--\lambda)+(|\epsilon_-|^2(m^2_+-\lambda)-(\epsilon_+\epsilon^*_-+\epsilon_-\epsilon^*_+)\hat{m}^2]}
\Bigg)\nonumber
\end{eqnarray}
which coincides with eq. (\ref{L11}). Besides the location of the pole the correction $p_W(q^2)$ also modifies the momentum dependence of the effective weak interactions by replacing 
\begin{equation}\label{Q9}
\frac{g^2}{M^2_W+q^2}\rightarrow \frac{g^2}{\bar{M}^2_W+q^2\big(1+p_W(q^2)\big)}
=\frac{g^2_{eff}(q^2)}{M^2_W+q^2}.
\end{equation}
This modification can be cast into the form of a momentum dependent weak mixing angle $\sin^2 \theta_W(q)=e^2/g_{eff}(q)$, i.e.
\begin{eqnarray}\label{Q10}
R^{-1}_W(q)&=&\frac{\sin^2\theta_W(\mu)}{\sin^2\theta_W(q)}=
\frac{g^2_{eff}(q)}{g^2_{eff}(\mu)}\nonumber\\
&=&\frac{(M^2_W+q^2)\Big[\bar{M}^2_W+\mu^2\big(1+p_W(\mu^2)\big)\Big]}
{\Big[\bar{M}^2_W+q^2\big(1+p_W(q^2)\big)\Big](M^2_W+\mu^2)}.
\end{eqnarray}
The factor $R_W(q)$ multiplies the momentum dependence of $\sin^2\theta_W$ in the standard model which is induced by radiative corrections.

For low momenta we may further approximate
\begin{eqnarray}\label{Q13}
G^{++}&=&\frac{m^2_-}{m^2_+m^2_--\hat{m}^4}~,~G^{--}=\frac{m^2_+}{m^2_+m^2_--\hat{m}^4}~,\nonumber\\
G^{+-}&=&G^{-+}=-\frac{\hat{m}^2}{m^2_+m^2_--\hat{m}^4}
\end{eqnarray}
such that $p_W$ becomes a momentum independent constant. For $M^2_W\ll m^2_{R1}, m^2_{R2}$ it is related to $\Delta$ by 
\begin{equation}\label{a15}
\Delta = -p_W\bar{M}^2_W = M^2_W -\bar{M}^2_W.
\end{equation}
In lowest order in $p_W$ one infers $\bar{M}^2_W=M^2_W(1+p_W)$ and we conclude that in this approximation $R_W$ has no correction linear in $p_W, R^{-1}_W(q^2)=1+0(p^2_W)$. More generally, in the limit of constant $p_W$ the shift in the wave function renormalization for the $W$-boson (\ref{Q7}) can entirely be absorbed by a multiplicative change in the weak coupling constant $g^2_{eff}=g^2/(1+p_W)$. Since the value of $g$ is a free parameter the effects of the spin-one-boson mixing become unobservable. They only influence the relation between the measured value of $\sin^2\theta_W$ and the value of $g$ that may be predicted in a unified theory of all gauge interactions (GUT). 

\begin{itemize}
\item [c)] {\bf Effective tensor interactions for quarks and leptons}
\end{itemize} 

Next, the effective four fermion interaction reflects the direct exchange of the antisymmetric tensor field
\begin{eqnarray}\label{Q11}
-{\cal L}_{4Fch}=&-&\{\bar{u}_RF_U\sigma^{\nu\mu}d_L\}G^{++}
(-\-\partial^2)\nonumber\\
&&\tilde{P}_{\nu\rho}\{\bar{d}_LF^\dagger_U\sigma^\rho\ _\mu u_R\}\nonumber\\
&-&\{\bar{u}_LF^\dagger_D\sigma^{\nu\mu}d_R+\bar{\nu}_LF^\dagger_L\sigma^{\nu\mu}e_R\}G^{--}(-\partial^2)\nonumber\\
&&\tilde{P}_{\nu\rho}\{\bar{d}_RF_D\sigma^\rho \ _\mu u_L+\bar{e}_RF_L\sigma^\rho\ _\mu\nu_L\}\nonumber\\
&-&\{\bar{u}_RF_U\sigma^{\nu\mu}d_L\}G^{+-}{(-\partial^2)}\nonumber\\
&&\tilde{P}_{\nu\rho}
\{\bar{d}_RF_D\sigma^\rho\ _\mu u_L+\bar{e}_RF_L\sigma^\rho \ _\mu\nu_L\}\nonumber\\
&-&\{\bar{u}_LF^\dagger_D\sigma^{\nu\mu}d_R+\bar{\nu}_LF^\dagger_L\sigma^{\nu\mu}e_R\}G^{-+}(-\partial^2)\nonumber\\
&&\tilde{P}_{\nu\rho}\{\bar{d}_LF^\dagger_U\sigma^\rho\ _\mu u_R\}.
\end{eqnarray}
Here we use the projector
\begin{equation}\label{Q12}
\tilde{P}_{\nu\rho}=\frac{\partial_\nu\partial_\rho}{\partial^2}~,~\tilde{P}_{\nu\rho}\tilde{P}^\rho\ _\mu=\tilde{P}_{\nu\mu}.
\end{equation}
The interaction (\ref{Q11}) reflects the direct production of the resonances associated with the chiral vector fields. Obviously, this direct production at LHC would be the most direct signal for our model. The poles in $G(q^2)$ occur at the physical masses of the new spin one bosons and are regulated by their finite decay width (not computed here). We observe that the coupling of the chiral tensor fields to electrons or quarks of the first generation is suppressed by the small chiral couplings. If they are of a similar size as the Yukawa couplings of the first generation they may be several orders of magnitude smaller than the weak gauge coupling. This suppression, combined with the presumably large width of the chiron resonance may make a detection difficult.  

Chiral tensor exchange also affects the low-momentum physics. Effective interactions of a similar type as eq. (\ref{Q11}) have been discussed\footnote{We note, however, that our description of massive antisymmetric tensor fields differs in crucial aspects from \cite{CHD}.} by Chizhov \cite{CHD} in order to explain ``anomalies'' \cite{PIBETA} in the decay $\pi^+\rightarrow e^+\nu\gamma$. However, comparing the overall size of the effective four fermion interaction ${\cal L}_{4Fch}$ with the usual weak interaction, we find a strong suppression $\sim(f^2_1/g^2)(M^2_W/M^2_{c})$ with $f_1$ and $M_{c}$ typical values the first generation chiral coupling and the mass of the chiral tensor fields. It seems not easy to limit this suppression to about $1\%$ as needed for the suggestion of \cite{CHD}.    We also note that in the limit where the chiral couplings $F_{U,D}$ are diagonal in the basis of quark mass eigenstates the interaction (\ref{Q11}) does not contribute to the kaon decay.

\begin{itemize}
\item [d)] {\bf Pauli interactions of the $W$-boson}
\end{itemize} 

Finally, the mixed pieces are responsible for the new Pauli-type interaction of the $W$-boson
\begin{eqnarray}\label{Q14}
-{\cal L}_p&=&\frac{1}{2}\bar{u}_RF_U
\sigma^{\mu\nu}d_L(\epsilon^*_+G^{++}+\epsilon^*_-G^{+-})W_{\mu\nu}\nonumber\\
&&+\frac{1}{2}(\bar{u}_LF^\dagger_D\sigma^{\mu\nu}d_R+\bar{\nu}_LF^\dagger_L\sigma^{\mu\nu}e_R)\nonumber\\
&&(\epsilon^*_-G^{--}+\epsilon^*_+G^{-+})W_{\mu\nu}+h.c..
\end{eqnarray}
Comparison with eq. (\ref{L15}) shows $\alpha_-=\epsilon^*_-G^{--}\ +\epsilon^*_+G^{-+}$ and we also define $\alpha_+=\epsilon^*_+G^{++}+\epsilon^*_-G^{+-}$. For the first generation the new spin and momentum structures are severely suppressed.   As compared to the usual coupling of the $W$-bosons the suppression factor is given by $(f_1\nu/g^2)(M_Wq/M^2_{c})$ with $f_1$ a chiral coupling to the first generation and $q$ a typical momentum.

For the low momentum transfer relevant for meson decays we can use the approximation (\ref{Q13}) and compute the effective four fermion interaction arising from the $W$-exchange in presence of the Pauli term (\ref{Q14}). The term linear in $\epsilon$ yields
\begin{eqnarray}\label{Q15}
-{\cal L}_{4FW}&=&\frac{g}{2\sqrt{2}M^2_W}\Big\{\alpha_-\big[\bar{u}_LF^\dagger_D
\sigma^{\mu\nu}d_R+\bar{\nu}_LF^\dagger_L\sigma^{\mu\nu}e_R\big]\nonumber\\
&&+\alpha_+\bar{u}_RF_U\sigma^{\mu\nu}d_L\Big\}\nonumber\\
&&\partial_\nu\{\bar{d}'_L\gamma_\mu u_L+\bar{e}_L\gamma_\mu\nu_L\}+h.c..
\end{eqnarray}
Here $d'$ contains the CKM-matrix for the standard coupling of the $W$-boson. As in the standard model, the weak decays only proceed via the off-diagonal couplings of the $W$-bosons. Neglecting the third generation we  may use 
$\bar{d}'u=\cos\theta_c\bar{d}u+\sin\theta_c\bar{s}u+\cos\theta_c\bar{s}c-\sin\theta_c\bar{d}c$.  In particular, we note the tensor contribution to the semileptonic decay $K^+\rightarrow\pi^0e^+\nu$ from 
$\big(f_e=(F_L)_{11}\big)$
\begin{eqnarray}\label{Q16}
-{\cal L}_{4FW}&=&-\kappa\sin\theta_c\partial_\nu(\bar{\nu}_L\sigma^{\mu\nu}e_R)(\bar{s}_L\gamma_\mu u_L)+\dots,\nonumber\\
\kappa&=&\frac{g\alpha_-f^*_e}{2\sqrt{2}M^2_W}
\sim\frac{f_e\nu}{g^2}\frac{M_W q}{M^2_c}G_F.
\end{eqnarray}
On the leptonic side of the standard matrix element is supplemented by an additional tensor piece $\sim\kappa$
\begin{equation}\label{Q17}
\sim -\left(\frac{g^2}{8M^2_W}\bar{\nu}_L\gamma^\mu e_L+\kappa\partial_\nu(\bar{\nu}_L\sigma^{\mu\nu}e_R)\right).
\end{equation}
In principle, this can be tested by the momentum distribution in the charged kaon decay. A similar piece, with 
$sin\theta_c s_L$ replaced by $\cos\theta_cd_L$, influences the leptonic pion decay. Concerning the anomalies  in the radiative decay $\pi^+\rightarrow e^+\nu\gamma$ \cite{PIBETA} we observe that for small momentum exchange the overall size of ${\cal L}_{4FW}$ is suppressed by a factor $q/M_W$ as compared to ${\cal L}_{4Fch}$. On the other hand, it involves only one power of the small chiral coupling $f_1$ instead of two. In any case, the typical strength of the new interaction $\sim\kappa$ seems to be much weaker than the usual weak interaction, making a detection in meson decays difficult. 

\begin{itemize}
\item [e)] {\bf Exchange of neutral chiral tensors}
\end{itemize} 

The tree exchange of the neutral chiral vector mesons $S^{\pm,0}_\mu$ can be discussed in complete analogy to the charged exchange. The source terms are similar to eqs. (\ref{Q1}), (\ref{Q2}) with an additional contribution from the hypercharge gauge field. The quadratic terms are more involved due to the existence of terms $\sim (S^{+,0}_\mu)^2$ etc. from the terms $\sim\sigma_+,\sigma_{v +}$ in eq. (\ref{13}). The problem is easily formulated in terms of a four-component vector
$S^\gamma_\mu=\big(S^{+,0}_\mu,S^{-,0}_\mu,(S^{+,0}_\mu)^*,(S^{-,0}_\mu)^*\big), \gamma=1\dots 4,$ such that eq. (\ref{Q1}) is replaced by
\begin{equation}\label{Q18}
-{\cal L}=j^{\gamma\mu}S^\gamma_\mu+h.c.+\frac{1}{2}S^{\gamma\mu}(G^{-1})^{\gamma\delta}S^\delta_\mu.
\end{equation}
The resulting effective Lagrangian in analogy to eq. (\ref{Q5}) reads
\begin{equation}\label{Q19}
-{\cal L}=-\frac{1}{2}j^{\gamma\mu}G^{\gamma\delta}j^\delta_\mu.
\end{equation}
One encounters possible modifications of the photon - and $Z^0$-propagators - they are severely restricted by the precision experiments at LEP. The Pauli term for the photon is constrained by the $g-2$ measurement, as discussed above. The effective tensor interactions induced by the direct chiron exchange are suppressed by the small chiral couplings. For example, the contribution to the amplitude $e^+e^-\rightarrow b\bar{b}$ is $\sim f_ef_b/M^2_c$. 

\begin{itemize}
\item [f)] {\bf Electroweak precision tests}
\end{itemize} 

The electron-positron scattering at LEP has tested the effective propagator of the $Z$-boson and the mixing between the isotriplet $(W_3)$ and isosinglet $(Y)$ gauge boson to high accuracy. Without discussing the details it is useful to make some general considerations on the predictions for the electroweak precision experiments for our model. Due to the factor $\sqrt{\partial^2}$ in the current involving the gauge bosons the mixing of the spin one bosons only contributes to the derivative terms for the effective propagators of the gauge bosons. In particular, the lowest contribution to the full inverse gauge boson propagator is of the order $q^2$. 

The $T$-parameter \cite{Peskin} describes a momentum independent mixing term $\sim W^\mu_3 Y_\mu$ in the effective action. It receives no ''tree'' correction within the approximation used here. In consequence, a possible shift of $T$ away from the standard model value is dominated by loops involving the chirons or the richer scalar sector of our model. If the chiron mass $M_c$  is high, the dominant addition to the standard model in the effective theory below $M_c$ may actually be the richer scalar sector. In particular, a loop involving a second ``low mass'' scalar field $\varphi_b$ in addition to $\varphi_t$ would slightly increase the $T$-parameter \cite{Barbieri}. This could lead to a good fit with the experimental value for scalar masses around $500$ GeV which are expected in our model. Further new loop corrections involve the chiral tensors. 

Besides $T$, the primary candidates for modifications concern the parameters $S$ and $U$, involving the isospin violation through kinetic terms mixing $W^\mu_3$ and $Y^\mu (S)$ or a difference in the kinetic terms for $W^\pm$ and $W_3(U)$. Contributions to $V,X,Y,W,Z$ \cite{BPRS} arise only through the momentum dependence of $p_W$ (\ref{Q6}) and similar quantities in the neutral sector. They are therefore further suppressed $\sim  \epsilon^2 M^2_Z/M^4_c$.  

For a discussion of isospin it is useful to include the charged components in a formula similar to eq. (\ref{Q19}) with indices running now from 1 to 8, reflecting the eight real components of the two doublets $S^\pm$. Since the chiral tensors belong to isospin doublets the propagator $G^{\gamma\delta}$ has isospin-singlet and isospin-triplet contributions, $G=G^{(1)}+\vec{G}^{(3)}\vec{\tau}$. In turn, the currents belong to isospin doublets, where the part involving the gauge bosons reads (\ref{A3A}) 
\begin{equation}\label{c18a}
j_\mu=\sqrt{-q^2}(\nu^*_yY_\mu\varphi+\nu^*_w\vec{W}_\mu\vec{\tau}\varphi).
\end{equation}
Insertion into eq. (\ref{Q19}) yields the structure 
\begin{eqnarray}\label{c18b}
-{\cal L}&=&-\frac{1}{2}G^{(1)}j^\dagger_\mu j^\mu-\frac{1}{2}j^\dagger_\mu\vec{G}^{(3)}\vec{\tau}j^\mu\nonumber\\
&=&\frac{q^2}{2}|\varphi|^2
\Big\{G^{(1)}\big[|\nu_y|^2Y^\mu Y_\mu +|\nu_w|^2\vec{W}^\mu\vec{W}_\mu\nonumber\\
&&-(\nu_y\nu^*_w+\nu^*_y\nu_w)Y^\mu W_{3\mu}\big]\nonumber\\
&&+G^{(3)}\big[-|\nu_y|^2Y^\mu Y_\mu +(\nu_y\nu^*_w+\nu^*_y\nu_w)Y^\mu W_{3\mu}\nonumber\\
&&+|\nu_w|^2(\vec{W}^\mu\vec{W}_\mu-2 W^\mu_3 W_{3\mu})\big]\Big\}.
\end{eqnarray}
(Here we use 
$\vec{G}^{(3)}\vec{\tau}=G^{(3)}\tau_3,\varphi^\dagger\tau_a\varphi=\varphi^\dagger\tau_3\varphi\delta_{a3}=-|\varphi|^2\delta_{a_3}$.) We conclude that the isospin singlet part of the propagator $G^{(1)}$ only contributes to $S$, not to $U$. Since the isospin triplet part $G^{(3)}$ is suppressed as compared to $G^{(1)}$ by a factor $|\varphi|^2/M^2_c$, the corrections to $S$ dominate. In the notation of \cite{BPRS} one finds
\begin{equation}\label{c18c}
\Delta\hat{S}=\frac{1}{119}\Delta S=-\frac{g'}{g}|\varphi|^2(\nu_y\nu^*_w+\nu^*_y\nu_w)G^{(1)}.
\end{equation}
With $|\varphi|=174$ GeV, $G^{(1)}\approx M^{-2}_c$, and using an estimate similar to app. A, $|\nu\varphi|\approx 0.2 m_t$, this yields an order of magnitude $\Delta\hat{S}\approx-0.05(m_t/M_c)^2$. For $M_c\approx 1$ TeV one finds $\Delta\hat{S}\approx 1.4\cdot 10^{-3}$, well within the experimental bounds. In view of the uncertainties, the bounds on $M_c$ from the electroweak precision tests are of a similar magnitude as the bound from the anomalous magnetic moment of the muon discussed in the preceeding section. We will assume as a reasonable approximate bound $M_c\gtrsim 300$ GeV. 

As discussed in the preceeding section, only small corrections to the effective four fermion interactions arise from the tensor couplings of the gauge bosons which are induced by their mixing with the chiral tensors. Also the effective tensor interactions from the direct exchange of the massive spin one particles are small. For example, their effect on the scattering $e^+e^-\rightarrow b\bar{b}$ is of the order $f_ef_b/M^2_c$, to be compared with $g^2/(s-m^2_Z)$ for the $Z$-boson exchange.

\begin{itemize}
\item [g)] {\bf Mixing with $\rho$-meson and radiative pion decay}
\end{itemize} 

We finally mention that a detailed phenomenological discussion may also have to include a tiny 
effective mixing between the chiral bosons $S^\mu$ and the vector mesons that arise as bound states of the usual (QCD)  
strong interactions, in particular the $\rho$-meson. Effective cubic terms replace in eq. (\ref{A3A}) $\vec{W}_{\mu\nu}$ by the field strength for the $\rho$-mesons. This adds a term 
$\epsilon_\rho\sqrt{\partial^2}\rho^{\mu*}$ to the currents (\ref{Q2}) and similar for the neutral mesons. An effective interaction of the type (\ref{Q16}) replaces $\kappa$ by a momentum dependent function reflecting the resonance in the $\rho$-propagator. Also $\bar{d}_L^\prime\gamma_\mu u_L$ is replaced in eq. (\ref{Q15}) by a vector-current $\bar{d}\gamma_\mu u$. (Correspondingly, a tiny mixing with the $K^*$-meson results in $\bar{s}\gamma_\mu u$.) The result is an effective action of the type
\begin{equation}\label{110}
-{\cal L}^{(\rho)}_{4F2}=-\kappa^{(\rho)}\partial_\nu(\bar{\nu}_L\sigma^{\mu\nu}e_R)(\bar{d}\gamma_\mu u)+c.c.
\end{equation}
It will be interesting to investigate if the size and structure of the interaction (\ref{110}) can account for the anomaly in the $\pi^+\rightarrow e^+\nu\gamma$ decay. A characteristic size of $\kappa^{(\rho)}$ is 
\begin{equation}\label{b13a}
\kappa^{(\rho)}\sim\frac{\nu_\rho}{g}\frac{M_W}{M^2_{c}}\frac{f_e g_\rho}{M^2_\rho}
\end{equation}
with $M_\rho$ and $g_\rho$ the rho-meson mass and its effective (strong) coupling to $\bar{u}\gamma^\mu d$. As compared to the standard weak four-fermion interaction (with $q\sim M_\pi$) the relative suppression is
\begin{equation}\label{b13b}
\frac{\kappa^{(\rho)}q}{G_F}\sim \frac{\nu_\rho f_e g_\rho}{g^3}\frac{M^3_W M_\pi}{M^2_{c}M^2_\rho}.
\end{equation}
Even for small $\nu_\rho$ and $f_e$ and large $M_{ch}$ the interaction (\ref{b13b}) may substantially affect the radiative pion decay. If this type of mixing dominates the modifications of the (semi-)leptonic meson decays it would not show up in the precision tests at LEP. Since the $\rho$ coupling to quarks conserves parity the radiative corrections cannot induce the observationally strongly forbidden matrix elements involving leptons and a pseudoscalar in the $\bar{d}u$ channel.

\section{Conclusions}
\label{Conclusions}
Let us close this note by a few remarks on the predictivity of our model and possible future tests. So far we have not found a major influence of the quartic couplings $\tau_j$ (\ref{QI}). Let us suppose that they are small and result only in minor corrections. Besides the three gauge couplings our model is then characterized by the chiral couplings $F_{U,D,L}$. These couplings can be mapped to the quark and lepton mass eigenvalues and the CKM-matrix. Just as for the Yukawa couplings in the standard model the chiral couplings determine the effective quark and lepton mass matrices $M_{U,D,L}$, although the connection is less direct. We note that the chiral coupling $f_t$ is used to set the Fermi scale.

In contrast to the standard model we have now two parameters less, i.e. the quadratic and quartic terms in the effective potential for the Higgs field. In consequence, several dimensionless ratios become predictable: the top quark and the Higgs scalar masses in units of the Fermi scale do not involve free parameters and are therefore computable, at least in principle. In other words, we may use the generation of a scale by dimensional transmutation in order to fix $m_t$ by a suitable choice of the microscopic chiral coupling $f_t$. For a given weak gauge coupling the ratio $M_W/m_t$ becomes then predictable. Its computation amounts to a calculation of the wave function renormalization which relates the $\bar{t}_Lt_R$ operator to a composite scalar field $\varphi_t$ with standard normalization of the kinetic term\footnote{Note that for $\varphi_t=Z^{-1/2}_{\varphi,t} \bar{t}_Lt_R$ the dimension of $Z_{\varphi,t}$ is mass$^4$. More generally, the scalar doublet with nonzero expectation value is a linear combination of various quark and lepton bilinears.} and the determination of $\langle\bar{t}_Lt_R\rangle/m_t$. This fixes $\langle\varphi_t\rangle$ in terms of $m_t-$ then $m_W\sim g\langle\varphi_t\rangle$ follows from the gauge invariance of the covariant scalar kinetic term.

An important future task will therefore be the estimate of the effective renormalized Yukawa coupling of the top quark $h_t$. A possible approach could introduce a composite field in the functional renormalization group equation \cite{GiWe} and follow the flow of $h_t$. Its low energy value yields the wanted relation $m_t=h_t\langle\varphi_t\rangle$. Typically, $h_t$ is large in the vicinity of the electroweak scale since it is induced by the strong interactions due to $f_t$. One expects a partial fixed point behavior, making the prediction presumably rather robust with respect to many details. In this sense our model resembles the top-condensation scenarios \cite{BL}. However, for the models of \cite{BL} predictivity is related to the upper bound of an infrared interval of allowed couplings \cite{CWIR}, whereas for the present model it simply follows from parameter counting. We also emphasize that due to the additional degrees of freedom the infrared interval is located at different values of $h_t$ as compared to the standard model, therefore leading to a value of $m_t$ different from \cite{BL}. 

Having fixed the fermion masses and the gauge couplings all other quantities involve no further free parameters. The masses of the composite scalars and the massive chiral tensors are, in principle, computable. Since we have to deal with strong interactions near the Fermi scale the number and location of the possible spin zero and spin one resonances is not necessarily given by a perturbative counting. For example, it is not clear a priori if we have only one scalar resonance consisting dominantly of $(\bar{t}_Lt_R, \bar{b}_L t_R)$ with a small admixture of  other quark bilinears, or also a second one with $t_R$ replaced by $b_R$, or even more. (The different parity with respect to $G_A$ suggests the presence of at least two scalar doublets $\varphi_t$ and $\varphi_b$.) In any case, the effective quartic couplings between the composite scalars are unlikely to be small and one expects rather high masses for scalar excitations, say around $500$ GeV. The chiral tensors presumably appear in the spectrum as heavy spin one resonances and the mass terms $m^2$ can be computed if the effective strongly interacting chiral tensor model can be solved. Our phenomenological investigation suggests chiral tensor masses well above the top quark mass. In view of the uncertainties we estimate a bound $M_c\gtrsim 300$ GeV.

In principle, all the effective couplings discussed in sect. VI and therefore the so far unknown constants in the phenomenological predictions of sect. IX are computable. Their computation will require a major effort in understanding the non-perturbative physics of the strong chiral couplings. Lattice simulations will be difficult since both chiral fermions and chiral tensors play a dominant role. We note in this respect that the analytic continuation of the chiral tensors to euclidean space encounters difficulties similar to the chiral fermions: the irreducible antisymmetric tensor representations of $SO(4)$ are pseudoreal and not complex conjugate to each other as for $SO(1,3)$.

The effort of understanding the chiral tensors seems worthwhile, however: our model is, in principle, highly predictive and clearly can be falsified. The quantum field theory is consistent provided the mass generation for the chiral tensors operates in the way discussed in sect. VII. Our model could solve the long standing gauge hierarchy problem. It explains the origin of mass entirely in terms of dimensional transmutation from dimensionless parameters. Interestingly, the deviations from the standard model predictions are expected to be most visible for processes involving heavy particles in view of their large chiral couplings. The forward-backward asymmetry in $e^+e^-\rightarrow b\bar{b}$ and the anomalous magnetic moment of the muon  are actually found at present at values somewhat different from the standard model prediction, even though the statistical significance is so far not very high. It is well conceivable that the LHC could give direct hints for the presence of chiral tensors.

\bigskip
\noindent
{\bf Acknowledgment}
The author would like to thank J. J\"ackel for discussions and finding a numerical error.

\section*{Appendix A: Loop formulae}

In this appendix we display some useful one-loop formulae. We explicitely introduce appropriate mass terms as infrared cutoffs. Variation with respect to these mass terms accounts for the infrared-running of the corresponding couplings.

\medskip
\noindent
{\bf 1)\quad Propagator for chiral tensor fields}

The contributions to the effective action which involve one fermion loop obtain from
\begin{equation}\label{A22A}
\Gamma^{(F)}_{1l}=i~Tr\ln P_F
\end{equation}
with $P_F$ the inverse fermion propagator in the presence of background fields. We will use here as background fields the chirons and photons such that in momentum space
\begin{eqnarray}\label{A22B}
&&P_F(q,q')_{ab}=\big\{(-\qslash+m_a\gamma^5)\delta(q-q')\nonumber\\
&&+eQ_a\gamma^\mu A_\mu(q-q')\big\}\delta_{ab}+B_{ab}(q,q')
\end{eqnarray}
with
\begin{eqnarray}\label{A22C}
B_{ab}(q,q')&=&\hat{f}_a\big\{(\tilde{\beta}_+)_{ab}-
(\bar{\tilde{\beta}}_+)_{ab}+(\bar{\beta}_-)_{ab}-(\beta_-)_{ab}\big\},\nonumber\\
(\beta_-)_{ab}&=&\frac{1}{2}\sigma^{\mu\nu}_-
(\beta^{-+}_{\mu\nu}\delta_{aU}+\beta^{-0}_{\mu\nu}\delta_{aD})\delta_{bD},\nonumber\\
(\bar{\beta}_-)_{ab}&=&\frac{1}{2}\sigma^{\mu\nu}_+\big((\beta^{-+}_{\mu\nu})^*\delta_{bU}+(\beta^{-0}_{\mu\nu})^*
\delta_{bD}\big)\delta_{aD},\nonumber\\
(\tilde{\beta}_+)_{ab}&=&\frac{1}{2}\sigma^{\mu\nu}_+
\big(-\beta^{++}_{\mu\nu}\delta_{bD}+\beta^{+0}_{\mu\nu}\delta_{bU}\big)\delta_{aU},\nonumber\\
(\bar{\tilde{\beta}}_+)_{ab}&=&\frac{1}{2}\sigma^{\mu\nu}_-\big(-(\beta^{++}_{\mu\nu})^*\delta_{aD}+(\beta^{+0}_{\mu\nu})^*\delta_{aU}\big)\delta_{bU}.
\end{eqnarray}
The momentum arguments are $\beta(q-q')$ and $\beta^*(q'-q)$ and we recall the relation (\ref{31AA}), i.e.
\begin{eqnarray}\label{A22D}
\frac{1}{2}\sigma^{\mu\nu}_\pm\beta^\pm_{\mu\nu}&=&-2\sigma^k_\pm B^\pm_k,\nonumber\\
\frac{1}{2}\sigma^{\mu\nu}_\mp (\beta^\pm_{\mu\nu})^*&=&-2\sigma^k_\mp(B^\pm_k)^*.
\end{eqnarray}
For the quarks we treat every generation separately with flavor indices $(a,b)=(U,D)$. The trace reads $Tr=N_c\int_q\sum\limits_a tr$, with $tr$ the trace over spinor indices and $N_c=3$. (The treatment of leptons is similar, without the color factor $N_c$.) The one loop contributions to propagators and vertices can be computed from appropriate derivatives of $\Gamma^{(F)}_{1l}$ with respect to the background fields. In particular, the contribution to the chiron propagator needs the term quadratic in $B$
\begin{equation}\label{A22E}
\Gamma=-\frac{i}{2}Tr P^{-1}_FB P^{-1}_F B +\dots
\end{equation}

\begin{figure}[htb]
\centering
\includegraphics[scale=0.45]{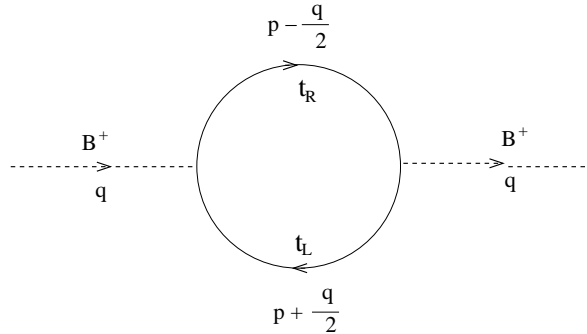}
\caption{Fermion loop contribution to the chiral tensor propagator.\label{figxa}}
\end{figure}

More precisely, the one loop correction to the inverse propagator in the $B^{+*}B^+$ channel adds to $P_{kl}(q)$ in eq. (\ref{No5}) a piece $\Delta P_{kl}(q)$. The contribution from the top quark loop (with $m_t$ as IR-cutoff, $m_t, \hat{f}_t$ real) is shown in Fig. \ref{figxa} and yields
\begin{eqnarray}\label{A1}
\Delta&& P_{kl}(q)=12i\hat{f}^2_t\int\frac{d^4 p}{(2\pi)^4}\\
&&tr\big\{\sigma^k_-(\pslash-\frac{\qslash}{2}-m_t\gamma^5)^{-1}\sigma^l_+
(\pslash+\frac{\qslash}{2}-m_t\gamma^5)^{-1}\big\}\nonumber\\
&&=24i\hat{f}_t\ ^2\int\frac{d^4p}{(2\pi)^4}\nonumber\\
&&\big[(p^2+\frac{q^2}{4}+m^2_t)^2-(pq)(pq)\big]^{-1}
\big(P_{kl}(p)-\frac{1}{4}P_{kl}(q)\big).\nonumber
\end{eqnarray}
Here we have used the trace formula 
\begin{equation}\label{B22}
tr\sigma^k_-\qslash \sigma_+^l\qslash=2 P_{kl}(q).
\end{equation}

\noindent
The chiral coupling appearing in this loop is given by $\hat{f}^2_t=\bar{f}^2_t/Z^2_\psi=f^2_t Z_+$. 

With the use of the identities (for arbitrary functions $f(p^2)$)
\begin{eqnarray}\label{A2}
&&\int\frac{d^4p}{(2\pi)^4}f(p^2)P_{kl}(p)=0~,\\
&&\int\frac{d^4p}{(2\pi)^4}f(p^2)(qp)^2P_{kl}(p)=\frac{1}{12}P_{kl}(q)
\int\frac{d^4p}{(2\pi)^4}p^4f(p^2)\nonumber
\end{eqnarray}
one finds for an expansion to quadratic oder in $q~($ with $x$  the analytic continuation of $p^2$ to euclidean signature)
\begin{eqnarray}\label{A.3}
\Delta P_{kl}(q)=\frac{\hat{f}^2_t}{8\pi^2} P_{kl}(q)\int^\infty_0 dx x\nonumber\\
\left[\frac{3}{(x+m^2_t)^2}-\frac{x^2}{(x+m^2_t)^4}\right].
\end{eqnarray}
Associating the IR-cutoff $k$ with $m_t$ this contributes to the anomalous dimension of $\beta_+$
\begin{equation}\label{A4}
\eta_+=-\frac{\partial ln Z_+}{\partial ln k}=\frac{f^2_t}{2\pi^2}.
\end{equation}
The generalization to three generations is straightforward and leads to eq. (\ref{M1}).

For $m_t=0$ the integral (\ref{A.3}) becomes infrared divergent. A Taylor expansion in $q$ is no longer justified. For a better understanding of the momentum dependence of the propagator correction we will next evaluate the integral (\ref{A1}) for arbitrary $q$, using 
\begin{eqnarray}\label{b17a}
&&\Delta P_{kl}(q)=-24i\hat{f}^2_t\int^\infty_0 dz_1\int^\infty_0 dz_2\int\frac{d^4 p'}{(2\pi)^4}\nonumber\\
&&\left\{\frac{1}{2}P_{kl}\left(p'-\frac{\gamma}{2}q\right)+\frac{1}{2} P_{kl}\left(p'+\frac{\gamma}{2}q\right)
-\frac{1}{4}P_{kl}(q)\right\}\nonumber\\
&&\exp\left\{-i(z_1+z_2)(p^{\prime 2}+m^2)-i\frac{z_1z_2}{z_1+z_2}q^2\right\},\nonumber\\
&&\gamma=(z_1-z_2)/(z_1+z_2).
\end{eqnarray}
With
\begin{eqnarray}\label{b17b}
\frac{1}{2}P_{kl}\left(p'+\frac{\gamma}{2}q\right)+\frac{1}{2}P_{kl}\left(p'-\frac{\gamma}{2}q\right)
-\frac{1}{4}P_{kl}(q)\nonumber\\
=P_{kl}(p')-\frac{z_1z_2}{(z_1+z_2)^2}P_{kl}(q)
\end{eqnarray}
we can perform the $p'$-integration
\begin{eqnarray}\label{b17c}
&&\Delta P_{kl}(q)=\frac{3\hat{f}^2_t}{2\pi^2}P_{kl}(q)A,\nonumber\\
&&A=\int^\infty_0dz_1\int^\infty_0dz_2\frac{z_1z_2}{(z_1+z_2)^4}\nonumber\\
&&\qquad\exp\left\{-i(z_1+z_2)m^2-i\frac{z_1z_2}{z_1+z_2}q^2\right\}\\
&&=\int^1_0dz\int^\infty_0\frac{d\lambda}{\lambda}z(1-z)\exp
\{-i\lambda(m^2+z(1-z)q^2\}.\nonumber
\end{eqnarray}
For the last equation we insert
\begin{equation}\label{b17d}
1=\int^\infty_0 \frac{d\lambda}{\lambda}\delta\left(1-\frac{z_1+z_2}{\lambda}\right),
\end{equation}
perform a rescaling $z_i\rightarrow \lambda z_i$ and carry out the $z_2$-integration. 

For the derivative of $A$ with respect to $m^2$ the $\lambda$-integration becomes Gaussian such that
\begin{eqnarray}\label{b17e}
\frac{\partial A}{\partial m^2}&=&-\frac{1}{q^2}\int^1_0 dz\frac{z(1-z)}{z(1-z)+\tilde{m}^2}=-\frac{B(\tilde{m}^2)}{m^2},
\nonumber\\
\tilde{m}^2&=&\frac{m^2}{q^2},\nonumber\\
B(\tilde{m}^2)&=&\tilde{m}^2-4\tilde{m}^4\int^1_0dy(4\tilde{m}^2+1-y^2)^{-1}\nonumber\\
&=&\tilde{m}^2\left(1-\frac{2\tilde{m}^2}{\sqrt{4\tilde{m}^2+1}}\ln
\frac{\sqrt{4\tilde{m}^2+1}+1}{\sqrt{4\tilde{m}^2+1}-1}\right)\nonumber\\
&&\textup{for}\quad\tilde{m}^2\geq 0.
\end{eqnarray}
For $q^2\ll m^2(\tilde{m}^2\rightarrow \infty)$ and $q^2\gg m^2 (\tilde{m}^2\rightarrow 0)$ we find the limits
\begin{equation}\label{b17f}
\lim_{\tilde{m}^2\rightarrow\infty}B(\tilde{m}^2)=\frac{1}{6}~,~\lim_{\tilde{m}^2\rightarrow 0}B(\tilde{m}^2)
=\tilde{m}^2.
\end{equation}

Let us define a momentum dependent wave function renormalization 
\begin{equation}\label{s19}
P_{kl}(q)+\Delta P_{kl}(q)=Z_+(q)P_{kl}(q)
\end{equation}
and a corresponding anomalous dimension
\begin{equation}\label{b17g}
\eta^{(m^2)}_+(q)=-2m^2\frac{\partial}{\partial m^2}\ln Z_+(q)
=\frac{3}{\pi^2}f^2_t B(\tilde{m}^2).
\end{equation}
We recover eq. (\ref{A4}) for $q^2\ll m^2$. On the other hand, in the opposite limit for $q^2\gg m^2$ we may 
set $m^2=0$ and evaluate
\begin{equation}\label{AA1}
\frac{\partial A}{\partial q^2}=-\frac{1}{6q^2},
\end{equation}
or 
\begin{equation}\label{AA2}
\eta^{(q^2)}_+(q)=-2q^2\frac{\partial}{\partial q^2}\ln Z_+(q)=\frac{f^2_t}{2\pi^2}.
\end{equation}
This yields the same value as $\eta^{(m^2)}_+$ in eq. (\ref{b17g}). 

We should stress, however, that the perturbative calculation neglects the momentum-dependence of the chiral coupling $\hat{f}^2_t$. For the solution of a Schwinger-Dyson equation the detailed momentum dependence of the vertex functions and propagators can be crucial. We demonstrate this by the approximate form
\begin{equation}\label{AA3}
\hat{f}^2_t(q,p)=\bar{f}^2_t\left(1+\frac{\mu^2_{t,R}}{\left(p-\frac{q}{2}\right)^2+m^2_t}
+\frac{\mu^2_{t,L}}{\left(p+\frac{q}{2}\right)^2+m^{2}_t}\right),
\end{equation}
which corresponds to the momentum dependence of the wave function for the right-and left-handed top quark. In an appropriate momentum range this will be a reasonable fit. For the purpose of demonstration of the qualitative effect we set $\mu^2_{t,R}=\mu^2_{t,L}=\mu^2_t$. Then the $z$-integral $A$ (\ref{b17c}) receives a further contribution which involves an additional factor $i(z_1+z_2)$, i.e. 
\begin{eqnarray}\label{AA4}
\Delta A&=&i\mu^2_t\int^1_0 dz\int^\infty_0 d\lambda z(1-z)\nonumber\\
&&\exp \big\{-i\lambda(m^2+z(1-z)q^2)\big\}\nonumber\\
&=&\mu^2_t B(\tilde{m}^2)/m^2.
\end{eqnarray}
For $m^2\ll|q^2|$ this yields
\begin{eqnarray}\label{AA5}
\Delta P_{kl}=\frac{3\bar{f}^2_t\mu^2_t}{2\pi^2}\frac{P_{kl}(q)}{q^2}.
\end{eqnarray}
The nonlocal momentum dependence $P_{kl}(q)/q^2$ arises from the dominance of low momentum modes in the $p$-integral and will be the same if $\mu^2_{t,L}\neq \mu^2_{t,R}$. 

Loops involving the pointlike quadratic coupling (\ref{QI}) do not contribute to the momentum dependence of the $\beta$-propagator in one loop order. There is therefore no contribution to the anomalous dimensions $\eta_\pm$. Since the tensor fields carry no strong charges there is also no contribution from the strong gauge coupling. Effects involving the weak and hypercharge gauge couplings have not yet been computed and are neglected, just as for the vertex correction.

\medskip
\noindent
{\bf 2)\quad Fermion anomalous dimension}

\begin{figure}[h!tb!]
\centering
\includegraphics[scale=0.45]{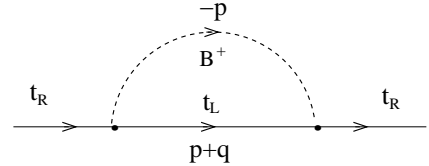}
\caption{Propagator correction for right handed top quark.\label{figxb}}
\end{figure}

The loop with emission and reabsorption of a virtual tensor fluctuation yields a correction to the kinetic term of the right handed top-quark as shown in fig. \ref{figxb}
\begin{equation}\label{A5}
\Delta {\cal L}_{kin}=-8i|f_t|^2\int\frac{d^4p}{(2\pi)^4}\bar{t}_R\sigma^k_+(\pslash+\qslash)^{-1}
\sigma^l_-P^{-1}_{lk}(-p)t_R.
\end{equation}

\noindent

\noindent
(A factor 2 is included from the two isospin components of the virtual tensor field and we neglect all other chiral couplings except the one for the top quark.) We extract the term linear in $q$
\begin{eqnarray}\label{A6}
&&\Delta{\cal L}_{kin}=-8|f_t|^2\bar{t}_RJ_t t_R~,\nonumber\\
&&J_t=-i\int\frac{d^4p}{(2\pi)^4}\frac{P_{kl}(p)}{p^6}\sigma^k_+\frac{\pslash+\qslash}{1+\frac{2(pq)}{p^2}+\frac{q^2}{p^2}}\sigma^l_-\nonumber\\
&&\rightarrow-i\int\frac{d^4p}{(2\pi)^4}\frac{P_{kl}(p)}{p^6}\sigma^k_+\gamma^\mu\sigma^l_-
\{q_\mu-\frac{2p_\mu(pq)}{p^2}\}
\end{eqnarray}
and use the identity
\begin{eqnarray}\label{A7}
&&P_{kl}(p)\sigma^k_+\gamma^\mu\sigma^l_-=-p^2\gamma^\mu_-+4p^\mu p_\nu\gamma^\nu_-\nonumber\\
&&P^*_{kl}(p)\sigma^k_-\gamma^\mu\sigma^l_+=-p^2\gamma^\mu_++4p^\mu p_\nu\gamma^\nu_+
\end{eqnarray}
with
\begin{equation}\label{A8}
\gamma^\mu_\pm=\gamma^\mu(1\pm\gamma^5).
\end{equation}
In view of eq. (\ref{A2}) only the second term in the curled bracket in eq. (\ref{A6}) contributes
\begin{eqnarray}\label{A9}
J_t&=&\frac{3i}{2}\int\frac{d^4p}{(2\pi)^4}\frac{1}{p^4}\qslash\left(\frac{1-\gamma^5}{2}\right)
\nonumber\\
&=&-\frac{3}{16\pi^2}\ln\left(\frac{\Lambda}{k}\right)\qslash\left(\frac{1-\gamma^5}{2}\right),
\end{eqnarray}
where we employ (euclidean) UV and IR cutoffs $k^2<p^2<\Lambda^2$. This yields the perturbative wave function renormalization
\begin{equation}\label{A10}
Z_t=1-\frac{3|f_t|^2}{2\pi^2}\ln\frac{\Lambda}{k}
\end{equation}
and the one loop anomalous dimension
\begin{equation}\label{A11}
\eta_t=-\frac{\partial\ln Z_t}{\partial\ln k}=-\frac{3}{(2\pi)^2}|f_t|^2.
\end{equation}

The relative minus sign as compared to a loop with a virtual scalar doublet field can be traced back to the relative minus sign in the first term on the r.h.s. of eq. (\ref{A7}). Similar computations for the left handed quark and for the case of three generations yields eq. (\ref{M2}), where we have also added the standard contribution to the quark anomalous dimension from the virtual gluon-quark loop. 

We may investigate the dependence of the top-wave function renormalization $Z_{t,R}$ on the various infrared cutoffs in more detail. We evaluate $\big(\qslash_+=\qslash(1+\gamma^5)/2\big)$ 
\begin{eqnarray}\label{x1a}
\Delta Z_{t,R}&=&4\bar{f}^2_t Z^{-1}_q tr\left\{\frac{\qslash_+}{q^2}J_t(q)\right\},\nonumber\\
J_t(q)&=&-i\int\frac{d^4p}{(2\pi)^4}\frac{P_{kl}(p)}{p^2}\frac{1}{\tilde{P}(p^2)}\sigma^k_+
\frac{\pslash+\qslash}{(p+q)^2+m^2_t}\sigma^l_-
\end{eqnarray}
in presence of a top quark mass $m_t$ and for the full tensor propagator characterized by eqs. (\ref{H1},\ref{H2}). For the derivative with respect to some parameter $x$ (e.g. $x=m^2_t$ or $q^2$) this yields
\begin{eqnarray}\label{x2a}
\frac{\partial \ln Z_{t,R}(q)}{\partial x}&=&-8i\hat{f}^2_t K_t\\
K_t&=&\int\frac{d^4 p}{(4\pi)^4}\frac{\partial}{\partial x}
\frac{4(pq)(pq)-p^2q^2+3p^2(pq)}{q^2p^2\tilde{P}(p^2)\big[(p+q)^2+m^2_t\big]}.\nonumber
\end{eqnarray}
For $|q^2|\ll m^2_t$ we may again expand
\begin{equation}\label{x3a}
K_t=-\frac{3}{2}\int\frac{d^4p}{(2\pi)^4}\frac{\partial}{\partial x}\frac{p^2}{\tilde{P}(p^2)(p^2+m^2_t)^2}.
\end{equation}
with $\tilde{P}(p^2)=p^2+M^2_c$ it is obvious that both $m^2_t$ and $M^2_c$ act as effective infrared cutoffs. Thus $k^2$ in eq. (\ref{A10}) may be identified with the largest IR-cutoff. 

\medskip
\noindent
{\bf 3)\quad Chiron-photon mixing}

In order to evaluate the diagram Fig. \ref{fig4} we compute the top quark loop in presence of a nonzero top quark mass $m_t\sim \varphi_t$ and external photon and tensor fields. We concentrate on the real part of $\beta^{+0}$. 
Expanding in linear order in $\beta^+_{\rho\sigma}$ and $A_\mu$ yields a contribution to the effective action
\begin{equation}\label{A201}
\Gamma_m=\int_p\big(\beta^{+0}_{\rho\sigma}(p)+(\beta^{+0}_{\rho\sigma})^*(-p)\big)h^{\rho\sigma\mu}A_\mu(-p)
\end{equation}
with $(N_c=3)$
\begin{eqnarray}\label{A202}
h^{\rho\sigma\mu}&=&-2iN_c\int\frac{d^4q}{(2\pi)^4}tr
\Big\{\frac{\qslash-\frac{\pslash}{2}-m_t\gamma^5}{\left(q-\frac{p}{2}\right)^2+m^2_t}\nonumber\\
&&\frac{f_t}{4}\sigma^{\rho\sigma}\gamma^5\frac{\qslash+\frac{\pslash}{2}-m_t\gamma^5}{\left(q+\frac{p}{2}\right)^2+m^2_t}
\frac{2e}{3}\gamma^\mu\Big\}\nonumber\\
&=&-\frac{i}{2}ef_tm_t p_\nu tr \{\sigma^{\rho\sigma}[\gamma^\mu,\gamma^\nu]\}I_1(p),
\end{eqnarray}
and
\begin{eqnarray}\label{A202a}
I_1(p)&=&\int\frac{d^4q}{(2\pi)^4}
\left[\left( q+\frac{p}{2}\right)^2+m^2_t\right]^{-1}\nonumber\\
&&\hspace{1.0cm}\left[\left(q-\frac{p}{2}\right)^2+m^2_t\right]^{-1}.
\end{eqnarray}
In eq. (\ref{A202}) we have included an additional factor two which arises from a similar term involving the imaginary part of $\beta^{+0}_{\rho\sigma}$ with $\sigma^{\rho\sigma}$ replacing $\sigma^{\rho\sigma}\gamma^5$. Using eqs. 
(\ref{1}), (\ref{C25}) it can be brought to the form (\ref{A202}).

The ultraviolet divergence of the momentum integral $I_1$ will be cut off by the momentum dependence of $f_t$ and $m_t$ that are not included explicitely. Similar to sect. V we use an effective UV-cutoff $\Lambda_t$ such that 
\begin{equation}\label{A203}
I_1(p)=-\frac{i}{16\pi^2}\int^1_0 dz\ln \frac{z(1-z)p^2+m^2_t}{\Lambda^2_t}.
\end{equation}
For $|p^2|\ll m^2_t$ we may approximate
\begin{equation}\label{A204}
I_1=\frac{i}{8\pi^2}\ln\frac{\Lambda_t}{m_t}.
\end{equation}
Evaluating the traces
\begin{eqnarray}\label{A205}
&&-2i tr \{\sigma^{\rho\sigma}\sigma^{\mu\nu}\}
=-8i(\eta^{\rho\mu}\eta^{\sigma\nu}-\eta^{\rho\nu}\eta^{\sigma\mu})\nonumber\\
&&-2itr\{\sigma^{\rho\sigma}\sigma^{\mu\nu}\gamma^5\}=8\epsilon^{\rho\sigma\mu\nu},
\end{eqnarray}
we obtain
\begin{eqnarray}\label{A206}
\Gamma_m&=&\int_p\big[\beta^{+0}_{\rho\sigma}(p)+(\beta^{+0}_{\rho\sigma})^*(-p)\big]
h^{\rho\sigma\mu}A_\mu(-p)\\
&=&4ie f_t m_t I_1 \int_p\big[\beta^{+0}_{\mu\nu}(p)+(\beta^{+0}_{\mu\nu})^*(-p)\big]
F^{\mu\nu}(-p).\nonumber
\end{eqnarray}
This yields in eq. (\ref{N1}) 
\begin{equation}\label{A207}
\epsilon^{(+)}_\gamma=\frac{\sqrt{2}ef_tm_t}{\pi^2}\ln\frac{\Lambda_t}{m_t}\approx 0.2\left(\frac{f_t}{5}\right)~m_t.
\end{equation}

Evaluating a similar expression for the real part of $\beta^{-0}$ results in 
\begin{equation}\label{C25}
\Gamma_{m}=\frac{ef_bm_b}{4\pi^2}\ln \frac{\Lambda_b}{m_b}\int_p
\big[\beta^{-0}_{\mu\nu}(p)+(\beta^{-0}_{\mu\nu})^*(-p)\big] F^{\mu\nu}(-p).
\end{equation}
The loop involves now the $b$-quark and we have to replace in eq. (\ref{A16A}) 
$f_t\rightarrow f_b, m_t\rightarrow m_b, 2e/3\rightarrow -e/3$. Comparison with eq. (\ref{A16B}) yields
\begin{equation}\label{C25A}
\epsilon^-_\gamma=-\frac{ef_bm_b}{\sqrt{2}\pi^2}\ln\frac{\Lambda_b}{m_b}
\end{equation}
We note that $\epsilon^-_\gamma$ is negative if $f_b$ and $m_b$ have the same sign.

\medskip
\noindent
{\bf 4)\quad Propagator corrections from effective cubic couplings}

The diagram contributing to the propagator of the antisymmetric tensor through a loop with two B-propagators involves the integral
\begin{equation}\label{H1}
J_{ll'}(q)=\int\frac{d^4p}{(2\pi)^4}\left[\left(p-\frac{q}{2}\right)^2
\left(p+\frac{q}{2}\right)^2\right]^{-2}B_{ll'}(p,q)
\end{equation}
with
\begin{equation}\label{H2}
B_{ll'}(p,q)=P_{kk'}\left(p+\frac{q}{2}\right)P_{mm'}\left(p-\frac{q}{2}\right)\epsilon_{kml}\epsilon_{k'm'l'}.
\end{equation}
By virtue of the Lorentz symmetry and dimensional analysis we expect
\begin{equation}\label{H5}
J_{ll'}(q)=\alpha\frac{P_{ll'}(q)}{q^2}.
\end{equation} 
We use the identity
\begin{eqnarray}\label{H3}
&&P_{kk'}(q)P_{mm'}(p)\epsilon_{kml}\epsilon_{k'm'l'}=
(pq)\big[P_{ll'}(q+p)-P_{ll'}(q-p)\big]\nonumber\\
&&\hspace{4.5cm}-p^2P_{ll'}(q)-q^2P_{ll'}(p),\nonumber\\
&&P_{kk'}(q)P_{mm'}(q)\epsilon_{kml}\epsilon_{k'm'l'}=2q^2P_{ll'}(q)
\end{eqnarray}
in order to infer
\begin{eqnarray}\label{H4}
&&B_{ll'}(p,q)=\left(p^2-\frac{q^2}{4}\right)\big[4P_{ll'}(p)-P_{ll'}(q)\big]\nonumber\\
&&-\left(p-\frac{q}{2}\right)^2P_{ll'}\left(p+\frac{q}{2}\right)-\left(p+\frac{q}{2}\right)^2
P_{ll'}\left(p-\frac{q}{2}\right).
\end{eqnarray}
The integral $J_{ll'}$ (\ref{H1}) is ultraviolet finite and infrared finite for $q\neq 0$. For $q=0$ it vanishes by virtue of eq. (\ref{A2}). Nevertheless, the limit $q_\mu\rightarrow 0$ depends on the direction of $q_\mu$. 

Using
\begin{eqnarray}\label{A15}
\frac{1}{p^2}&=&i\int^\infty_0dz~e^{-izp^2}~,~
\frac{1}{p^4}=-\int^\infty_0dzz e^{-izp^2}~,\nonumber\\
\gamma&=&\frac{z_1-z_2}{z_1+z_2}
\end{eqnarray}
we write
\begin{eqnarray}\label{A16}
&&J_{ll'}=\frac{1}{2}\int^\infty_0dz_1\int^\infty_0 dz_2 z_1 z_2 \nonumber\\
&&\int \frac{d^4p'}{(2\pi)^4}\exp \left\{-ip^{\prime 2}(z_1+z_2)-iq^2\frac{z_1z_2}{z_1+z_2}\right\} \nonumber\\
&&\Big[B_{ll'}(p'+\frac{\gamma}{2}q,q)+B_{ll'}(p'-\frac{\gamma}{2}q,q)\Big].
\end{eqnarray}
Here we have employed the symmetry $z_1\leftrightarrow z_2$ and we use the identities
\begin{eqnarray}\label{A17}
&&\frac{1}{2}\Big[P_{ll'}(p+q)+P_{ll'}(p-q)\Big]=P_{ll'}(p)+P_{ll'}(q)~,\nonumber\\
&&\frac{1}{4}\Big[P_{ll'}(p+q)-P_{ll'}(p-q)\Big]
=A_{ll'}(p,q)\nonumber\\
&&\qquad\quad=-\delta_{ll'}(p_0q_0+\vec{p}\vec{q})\nonumber\\
&&\qquad\qquad+p_lq_{l'}+q_lp_{l'}-i\epsilon_{ll'j}(p_0q_j+q_0p_j)
\end{eqnarray}
and
\begin{eqnarray}\label{A17a}
&&\frac{1}{2}\Big(B_{ll'}(p+\frac{\gamma}{2}q,q)+B_{ll'}(p-\frac{\gamma}{2}q,q)\Big)=2p^2P_{ll'}(p)\nonumber\\
&&+\frac{1}{2}(\gamma^2-3)\Big[q^2P_{ll'}(p)+p^2P_{ll'}(q)\Big]+\frac{1}{8}(\gamma^2-1)^2q^2P_{ll'}(q)\nonumber\\
&&+2(1+\gamma^2)(pq)A_{ll'}(p,q).
\end{eqnarray}
We next evaluate the Gaussian integrals
\begin{eqnarray}\label{A18}
&&\int\frac{d^4p}{(2\pi)^4}e^{-izp^2}
\{1;p^2;P_{ll'}(p);p^2P_{ll'}(p);(pq)A_{ll'}(p,q)\}\nonumber\\
&&=-\frac{i}{16\pi^2 z^2}\left\{1;-\frac{2i}{z};0;0;-\frac{i}{2z}P_{ll'}(q)\right\}
\end{eqnarray}
and obtain
\begin{eqnarray}\label{A19}
J_{ll'}&=&\frac{1}{2\pi^2}\int dz_1 dz_2 \exp
\left\{-iq^2\frac{z_1z_2}{z_1+z_2}\right\}\nonumber\\
&&\frac{z^2_1z^2_2}{(z_1+z_2)^5}
\left(1-\frac{i}{4}\frac{z_1z_2}{z_1+z_2}q^2\right) P_{ll'}(q),\nonumber\\
&=&\frac{1}{2\pi^2}P_{ll'}(q)\left(1+\frac{1}{4}q^2\frac{\partial}{\partial q^2}\right)
\frac{A}{q^2}=\frac{3A}{8\pi^2}\frac{P_{ll'}(q)}{q^2},\nonumber\\
A&=&\int dz_1 dz_2\exp\left\{-i\frac{z_1z_2}{z_1+z_2}\right\}
\frac{z^2_1z^2_2}{(z_1+z_2)^5}.
\end{eqnarray}
For the computation of $A$ we first replace the $z_2$-integration by an integration over 
$y=z_1z_2/(z_1+z_2)$, resulting in 
\begin{equation}\label{A20}
A=\int^\infty_0dz z^{-4}\left\{e^{-iz}(6+4iz-z^2)-6+2iz\right\}
\end{equation}
where the integrand approaches $1/12$ for $z\rightarrow 0$. Then the $z$-integration yields
\begin{equation}\label{A21}
A=\lim_{z\rightarrow 0}\Big[(2z^{-3}+iz^{-2})e^{-iz}-2z^{-3}+iz^{-2}\Big]=-\frac{i}{6}
\end{equation}
or
\begin{equation}\label{A22}
J_{ll'}(q)=-\frac{i}{16\pi^2}\frac{P_{ll'}(q)}{q^2}.
\end{equation}

\section*{Appendix B: massive antisymmetric fields}
The consistent description of massive antisymmetric tensor fields needs some care. We will require the stability of the solutions of the linear field equations (absence of tachyons) and the positivity of the energy density. In sect. VII we have discussed a consistent scenario where the mass term appears in a non-local form in the basis of fields $\beta_{\mu\nu}$ or $B_k$. In this appendix we first demonstrate that no consistent theory exists that is based on local interactions for $\beta_{\mu\nu}$ with operators up to dimension four. We will then sketch the possible ingredients which can make a model for antisymmetric tensor fields consistent. At this place we repeat that our model of chiral tensors has no additional gauge symmetry which could remove some field components from the physical spectrum. 

The problem with purely local interactions may be demonstrated for the components with nonzero electric charge that we denote
by $B^{+,+}_k=C_{1k}~,~B^{-,+}_k=C_{2k}.$ 
The most general effective action consistent with the Lorentz symmetry and involving up to two derivatives reads in quadratic order in $C$
\begin{eqnarray}\label{61}
\Gamma_2&=&\int d^4x\Big\{
Z_+\big[-\partial_0C^*_{1k}\partial_0 C_{1k}\nonumber\\
&&-\partial_lC^*_{1k}\partial_l C_{1k}+2\partial_kC^*_{1k}\partial_lC_{1l}
-2i\epsilon_{jkl}\partial_0C^*_{1k}\partial_jC_{1l}\big]\nonumber\\
&&+Z_-[-\partial_0C^*_{2k}\partial_0C_{2k}-\partial_lC^*_{2k}\partial_lC_{2k}
+2\partial_kC^*_{2k}\partial_lC_{2l}\nonumber\\
&&+2i\epsilon_{jkl}\partial_0C^*_{2k}\partial_jC_{2l}]\nonumber\\
&&+\tilde{Z}\partial^\mu C^*_{2k}\partial_\mu C_{1k}+\tilde{Z}^*\partial^\mu
C^*_{1k}\partial_\mu C_{2k})\nonumber\\
&&+\tilde m^2C^*_{2k}C_{1k}+\tilde m^{*2} C^*_{1k}C_{2k}\Big\}.
\end{eqnarray}
Here $\tilde{Z}$ and $\tilde{m}^2$ are real if CP is conserved (CP:
$C_{\alpha k}\rightarrow-C^*_{\alpha k})$.
We may determine $\tilde m^2$ and $Z_\pm,\tilde Z$ in terms of the effective action  (\ref{4}), (\ref{13}), (\ref{14})  of section VI

\begin{equation}\label{A6a}
\Gamma_{\beta,2}=-
\int d^4x
({\cal L}^{ch}_{\beta,kin}+{\cal L}^{ind}_{\beta,kin}
+{\cal L}_{M\beta}).
\end{equation}
It is evaluated for nonzero expectation values $\bar{\varphi}_t,\bar{\varphi}_b$.

We neglect first the mixing with the gauge fields (\ref{A3A}). Then the inverse propagator (second functional derivative of $\Gamma$) is derived from eq. 
(\ref{61}) and 
can be written in momentum space in terms of $P_{kl}(q)$ (\ref{No6}) 
\begin{equation}\label{62}
\Gamma^{(2)}_c=\left(\begin{array}{lll}
Z_+P&,&\tilde{Z}q^2+\tilde m^2\\
\tilde{Z}q^2+\tilde m^2&,&Z_-P^*\end{array}\right).
\end{equation}
The solutions of the field equations correspond to vanishing eigenvalues of $\Gamma^{(2)}_c$ and obey
\begin{equation}\label{63}
Z_+Z_-q^4-(\tilde{Z}q^2+\tilde m^2)^2=0.
\end{equation}
Eq. (\ref{63}) has two solutions for $q^2$ which describe massive particles with renormalized mass $m_R$ and a standard dispersion relation $q^2_0=m^2_R+\vec{q}^2$
\begin{equation}\label{64}
m^2_{R1}=\frac{\tilde m^2}{\tilde{Z}+\sqrt{Z_+Z_-}}~,~m^2_{R2}=
\frac{\tilde m^2}{\tilde{Z}-\sqrt{Z_+Z_-}}.
\end{equation}
This holds provided $\tilde m^2$ and $\tilde{Z}$ have the same sign and $\sqrt{Z_+Z_-}<|\tilde{Z}|$. In contrast, for $\sqrt{Z_+Z_-}>|\tilde{Z}|$ one of the excitations is a tachyon, indicating an instability of the ``vacuum''. 

Already at this point we note that the absence of tachyons requires  sizeable dimension six operators (\ref{14}) (within a framework based on local effective interactions). Still, the absence of tachyons is not enough - consistency also requires the absence of ghosts. This issue is most easily studied by investigating the energy density for plane waves. For the quadratic effective action (\ref{61}) the latter is given by
\begin{eqnarray}\label{65}
\rho&=&Z_+(\partial_0C^*_{1k}\partial_0C_{1k}
-\partial_lC^*_{1k}\partial_lC_{1k}+2\partial_kC^*_{1k}\partial_lC_{1l})\nonumber\\
&&+Z_-(\partial_0C^*_{2k}\partial_0C_{2k}
-\partial_lC^*_{2k}\partial_lC_{2k}+2\partial_kC^*_{2k}\partial_lC_{2l})\nonumber\\
&&+\tilde{Z}(\partial_0C^*_{2k}\partial_0C_{1k}
+\partial_lC^*_{2k}\partial_lC_{1k}\nonumber\\
&&\hspace{0.4cm}+\partial_0C^*_{1k}\partial_0C_{2k}+\partial_lC^*_{1k}
\partial_lC_{2k})\nonumber\\
&&+\tilde m^2(C^*_{2k}C_{1k}+C^*_{1k}C_{2k}).
\end{eqnarray}
A potential consistency problem is related to solutions with a negative energy density, as can be seen in the rest frame of the massive particle where $\partial_lC_k=0$. Whereas the terms $\sim Z_\pm$ are positive, the off diagonal terms $\sim \tilde{Z},\tilde m^2$ could indeed lead to negative $\rho$. 

Let us consider a plane wave which obeys the field equation 
\begin{equation}\label{66}
(\tilde{Z}q^2+\tilde m^2)C_2=-Z_+PC_1
\end{equation}
and therefore, using the dispersion relation $q^2=-m^2_R$, 
\begin{eqnarray}\label{67}
C_2&=&\frac{Z_+}{\tilde{Z}m^2_R-\tilde m^2}PC_1=
-\frac{Z_+}{\tilde m^2}\frac{1+\kappa}{\kappa}PC_1~,~\nonumber\\
\kappa&=&\pm\frac{\sqrt{Z_+Z_-}}{\tilde{Z}}.
\end{eqnarray}
The two values of $\kappa$ correspond to the two values of $m^2_R$ according to eq. (\ref{64}) 
\begin{equation}\label{52A}
1+\kappa=\frac{\tilde m^2}{\tilde{Z}m^2_R}.
\end{equation}
Insertion into eq. (\ref{65}) yields
\begin{equation}\label{68}
\rho=2Z_+q^2_0C^*_{1k}C_{1l}
\left\{2\delta_{kl}-\frac{1}{\kappa m^2_R}(P_{kl}+P^*_{kl})\right\}.
\end{equation}
For a massive particle we may choose the rest frame where $\vec{q}=0~,~P_{kl}=-q^2_0\delta_{kl}~,~q^2_0=m^2_R$. This yields
\begin{eqnarray}\label{69}
\rho&=&4Z_+m^2_R(1+\frac{1}{\kappa})C^*_{1k}C_{1k}\nonumber\\
&=&\pm 4\sqrt{\frac{Z_+}{Z_-}}\tilde m^2C^*_{1k}C_{1k}.
\end{eqnarray}
We observe that the energy density is negative for the mode with $\kappa<0$.

We conclude that for $Z_+Z_-\neq\tilde{Z}^2$ one finds waves with negative energy density.  Furthermore, for $Z_+Z_->\tilde{Z}^2$ a tachyonic instability occurs. At the boundary $Z_+Z_-=\tilde{Z}^2$ only one mode propagates, with $m^2_{R1}=\tilde m^2/(2\tilde{Z})$, and it has positive energy density. At first sight, it may seem that a consistent description of the low energy theory exists only for this particular point. One should recall, however, that this result is based on the lowest order in the derivative expansion which may not be valid in the presence of strong couplings. For example, for $|\kappa|$ close to one the renormalized mass $|m^2_{R2}|$ becomes very large and the problems occur in a momentum range where an effective action based on a derivative expansion is certainly no longer valid. For a more complete effective action we may replace the constants $Z_\pm,\tilde{Z}$ by functions of $q^2$. It is possible that a pole in the propagator is present for $q^2=-m^2_{R1}$
within the range of validity of the derivative expansion. Then the discussion above can be applied by using instead of $Z_\alpha(q^2)$
 the constants $Z_\alpha(q^2=-m^2_R$). In contrast, the second solution with renormalized mass term $m^2_{R2}$ may not exist anymore for a more general momentum dependence of the inverse propagator matrix (\ref{62}). The lesson is that a pole or an instability outside the region of validity of the derivative expansion should not be taken at face value and therefore not be considered as a serious problem. On the other hand, we have shown that no satisfactory solutions seem to be compatible with a lowest order derivative expansion for the field $\beta_{\mu\nu}$ (except for $Z_+Z_-=\tilde{Z}^2$).

In the presence of strong interactions there is actually no obvious argument that a derivative expansion should hold for $\beta_{\mu\nu}$. It is conceivable that composite operators become relevant and a derivative expansion rather holds for the composite fields. As a further alternative a derivative expansion may become valid in a different field basis, as in our scenario presented in sect. VII. In this case a consistent effective field theory can be obtained rather easily as we will demonstrate  by a nonlocal momentum dependence of the wave function renormalization
\begin{equation}\label{54A}
Z_\pm(q^2)=Z_\pm+\frac{Z_\pm m^2_\pm}{q^2}.
\end{equation}
The propagator corresponding to eq. (\ref{62}) reads now
\begin{eqnarray}\label{189}
&&(\Gamma^{(2)}_c)^{-1}=Det^{-1}
\left(\begin{array}{lll}
Z_-(q^2) P^*&,&-(\tilde{Z}q^2+\tilde m^2)\\
-(\tilde{Z}q^2+\tilde m^2)&,&Z_+(q^2)P\end{array}\right)\nonumber\\
&&Det=Z_+(q^2)Z_-(q^2)q^4-(\tilde{Z}q^2+\tilde m^2)^2.
\end{eqnarray}
In the limit where the off-diagonal terms $\sim(\tilde{Z}q^2+\tilde{m}^2)$ can be neglected one has
\begin{equation}\label{54C}
Det=Z_+Z_-(q^2+m^2_+)(q^2+m^2_-).
\end{equation}
The propagating waves again correspond to the two poles of the propagator. They describe now two particles with renormalized masses $m_\pm$. 

The ansatz (\ref{54A}) leads to a nonlocality in the effective action (\ref{61}) only for the fields $C_k$ or, equivalently, $\beta^\pm_{\mu\nu}$. This nonlocality is absent if we use projected fields $(D^2=D^\mu D_\mu)$
\begin{equation}\label{54D}
S^\pm_m=\sqrt{Z_\pm}\frac{D_\nu}{\sqrt{D^2}}\beta^\pm_{nm}e^{n\nu}.
\end{equation}
Indeed, in terms of the four-vectors $S^\pm_\mu=e^m_\mu S^\pm_m$ the effective action  
(\ref{4}) describes massive vector fields in a standard formulation
\begin{equation}\label{54E}
\Gamma_2=\int_x\big\{(D^\mu S^{+\nu})^\dagger D_\mu S^+_\nu+m^2_+(S^{+\nu})^\dagger
S^+_\nu+(+\rightarrow -)\big\}.
\end{equation}
(Here we use eq. (\ref{4}) with the ansatz (\ref{54A}) such that $Z_\pm(-D^2)$ acts on the second field $\beta^\pm_{nq}$.) The nonlocal term $\sim m^2_\pm$ in eq. (\ref{54A}) appears now as a simple local mass term for the fields $S^\pm_\mu$. 

We may infer the energy density by variation with respect to the metric at fixed Lorentz-vectors $S^\pm_m$. For vanishing gauge fields 
one finds the standard form for massive vector fields
 (in flat space)
\begin{eqnarray}\label{54F}
&&\rho=
\partial_0(S^{+\mu})^\dagger\partial_0 S^+_\mu+\partial_i(S^{+\mu})^\dagger
\partial_iS^+_\mu+m^2_+(S^{+\mu})^\dagger S^+_\mu\nonumber\\
&&\qquad\qquad +(+\rightarrow -).
\end{eqnarray}
We also observe that for vanishing gauge fields the vector $S^\pm_\mu$ is divergence free, $\partial^\mu S^\pm_\mu =0$, using the antisymmetry of $\beta^\pm_{mn}$ in eq. (\ref{54D}). In the rest frame of the particle one therefore has $\partial_0 S^\pm_0=0$ and the solution of the field equations implies $S^\pm_0=0$. In consequence, only the spacelike components $S^\pm_k$ contribute to eq. (\ref{54F}) and the energy density is manifestly positive. The sign of the energy density does not change under Lorentz-transformations and one concludes that the positivity of $\rho$ holds as well in a Lorentz-boosted inertial system with nonzero $\vec{q}$.

Our example demonstrates that for momentum dependent $Z_\pm(q^2)$ a satisfactory effective action with positive energy density and without tachyons is possible. By continuity, this is maintained if the parameters $\tilde{Z}$ and $\tilde{m}^2$ in eq. (\ref{189}) are not too large. The effective action (\ref{54E}) is local in terms of $S_\mu$, but $S_\mu$ obeys a nonlocal constraint
\begin{equation}\label{x1}
D_\mu S^{\pm\mu}=\frac{\sqrt{Z_\pm}}{2\sqrt{D^2}}F^\pm_{\mu\nu}\beta^{\pm\nu\mu}.
\end{equation}
Here $F^\pm_{\mu\nu}$ involves the field strength of the gauge fields, $Y_{\mu\nu}$ and $\vec{W}_{\mu\nu}$, according to
\begin{equation}\label{x2}
[D_\mu, D_\nu]\beta^\pm_{\rho\sigma}=F^\pm_{\mu\nu}\beta^\pm_{\rho\sigma}.
\end{equation}
For the inversion of eq. (\ref{54D}) one may use the identity
\begin{eqnarray}\label{x3}
\beta^\pm_{\mu\nu}&=&\frac{1}{2\sqrt{Z_\pm}}\frac{1}{\sqrt{D^2}}
\{D_\mu S^\pm_\nu-D_\nu S^\pm_\mu\nonumber\\
&&\pm \frac{i}{2}\epsilon_{\mu\nu}\ ^{\rho\sigma}
(D_\rho S^\pm_\sigma -D_\sigma S^\pm_\rho)\}\nonumber\\
&&-\frac{1}{2D^2}P^{\pm\alpha\beta}_{\mu\nu}F^{\pm\rho}_\alpha P^{\pm\sigma\tau}_{\rho\beta}\beta^\pm_{\sigma\tau}
\end{eqnarray}
which involves the projectors
\begin{equation}\label{x4}
P^{\pm\alpha\beta}_{\mu\nu}=\delta^\alpha_\mu\delta^\beta_\nu-\delta^\beta_\mu\delta^\alpha_\nu\pm i\epsilon_{\mu\nu}\ ^{\alpha\beta}.
\end{equation}
Thus, the price for a local description in terms of $S_\mu$ is a nonlocal Yukawa coupling to the fermions, which follows from the insertion of eq. (\ref{x3}) into eq. (\ref{2}).

\section*{Appendix C: Chiral tensor propagator in presence of cubic couplings}
\label{chiraltensor}
\begin{figure}[htb]
\centering
\includegraphics[scale=0.45]{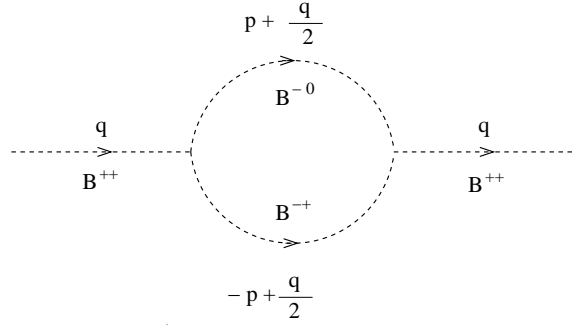}
\caption{Propagator correction through chiral tensor loop in presence of cubic couplings.\label{figx}}
\end{figure}

In this appendix we argue that in presence of spontaneous electroweak symmetry breaking a contribution to the ``nonlocal'' mass term (\ref{H3a})  $\sim m^2_\pm$ is generated. One source of this effect are the effective cubic couplings $\sim\beta^3$ which are generated from ${\cal L}_{3\beta}$ (\ref{a7,2}) in presence of spontaneous electroweak symmetry breaking $\bar{\varphi}_t\neq 0,\bar{\varphi}_b\neq 0$. Such couplings have an important influence on the structure of the $\beta$-propagator. Indeed, a one loop contribution with an intermediate pair of chiral tensor fields becomes now possible (cf. fig. \ref{figx}) and induces an effective mass term for the chiral tensor fields. This shows that the assumption of a local effective action in sect. VI is not self-consistent. Indeed, our calculation is permformed in absence of a mass term, i.e. for $\tilde{P}(q^2=0)=0$. It results in a contribution to $\tilde{P}(q^2=0)=\Delta m^2_\pm>0$, implying inconsistency of the hypothesis $\tilde{P}(q^2=0)=0$. An improved Schwinger-Dyson version of this computation could use nonzero $m^2_\pm$ in the chiral tensor propagators in the loop. 

The inverse propagator for the field $B^{++}$ acquires a correction

\noindent

\begin{eqnarray}\label{FF1}
&&-\Delta {\cal L}^+_{kin}=i\Omega^{-1}\nonumber\\
&&\int_q\Big\{(|\gamma_t\varphi^*_t|^2+|\gamma_b\varphi^*_b|^2)
\big(B^{++}_k(q)\big)^* J_{kl}(q) B^{++}_l(q)\nonumber\\
&&+|\gamma_b\varphi^*_b|^2\big(B^{+0}_k(q)\big)^* J_{kl}(q)B^{+0}_l(q)\Big\}.
\end{eqnarray}
Here the momentum integral for the loop is given by
\begin{eqnarray}\label{FF2}
&&J_{kl}(q)=\int\frac{d^4p}{(2\pi)^4}\epsilon_{mnk}\epsilon_{m'n'l}\nonumber\\
&&P^{*-1}_{mm'}\left(p+\frac{q}{2}\right)P^{*-1}_{nn'}\left(-p+\frac{q}{2}\right).
\end{eqnarray}
Similarly, the correction to the $B^{-}$ propagator reads \footnote{The poles of propagator for $B^+,~G_+=-iP^*_{kl}(q)(q^2-i\epsilon)^{-2}$ and the one for $B^-,~G_-=-iP_{kl}(q)(q^2-i\epsilon)^2$ are located at the same position. This implies $J_{lk}=-J^*_{kl}$.}
\begin{eqnarray}\label{FF3}
-\Delta{\cal L}^-_{kin}&=&i\Omega^{-1}\int_q
\Big\{\big(|\gamma_t\varphi^*_t|^2+|\gamma_b\varphi^*_b|^2\big)\nonumber\\
&&\big(B^{-+}_k(q)\big)^* J_{lk}(q)B^{-+}_l(q)\nonumber\\
&&+|\gamma_t\varphi^*_t|^2\big(B^{-0}_k(q)\big)^*
J_{lk}(q)B^{-0}_l(q)\Big\}
\end{eqnarray}
and a mixing between the neutral fields $B^{+0}$ and $(B^{-0})^*$ is induced via an effective term
\begin{eqnarray}\label{FF4}
&&-\Delta{\cal L}^{m}_{kin}=-i\Omega^{-1}\int_q\gamma_b\gamma_t\varphi^*_b\varphi^*_t~B^{-0}_k(q)J_{kl}(q)B^{+0}_l(q)\nonumber\\&&+c.c.
\end{eqnarray}

Actually, only one linear combination of $B^{+0}$ and $(B^{-0})^*$ gets a correction from this particular loop involving the cubic couplings. Indeed, inserting the neutral expectation values $\langle\varphi^0_t\rangle=\bar{\varphi}_t,\langle\varphi^0_b\rangle=\bar{\varphi}_b$ the cubic vertex (\ref{a7,2}) can be written in the form
\begin{eqnarray}\label{FF5}
-{\cal L}_{3\beta}&=&\epsilon_{klm}\big[\gamma^*_t\bar{\varphi}_t(B^{-0}_k)^*-\gamma_b\bar{\varphi}^*_b
B^{+0}_k\big]B^{++}_l(B^{-+}_m)^*\nonumber\\&&+c.c.
\end{eqnarray}
Defining the orthogonal linear combinations
\begin{eqnarray}\label{FF6}
&&B^0_k=\big(|\gamma^*_t\bar{\varphi}_t|^2+|\gamma^*_b\bar{\varphi}_b|^2\big)^{-1/2}
\big[\gamma^*_t\bar{\varphi}_t(B^{-0}_k)^*-\gamma_b\bar{\varphi}^*_bB^{+0}_k\big]\nonumber\\
&&R^0_k=\big(|\gamma^*_t\bar{\varphi}_t|^2+|\gamma^*_b\bar{\varphi}_b|^2\big)^{-1/2}
\big[\gamma^*_b\bar{\varphi}_b(B^{-0}_k)^*+\gamma_t\bar{\varphi}^*_tB^{+0}_k\big],\nonumber\\
\end{eqnarray}
we find that the inverse propagator for the fields $B^{++},(B^{-+})^*$ and $B^0$ receives the same correction, replacing $P_{kl}(q)$ by
\begin{equation}\label{FF7}
\tilde{P}_{kl}(q)=P_{kl}(q)+i\big(|\gamma^*_t\bar{\varphi}_t|^2+|\gamma^*_b\bar{\varphi}_b|^2\big)J_{kl}(q).
\end{equation}
The propagator for the field $R^0_k$ remains unaffected.

We have evaluated $J_{kl}(q)$ in appendix A and find (\ref{A22})
\begin{equation}\label{FF8}
iJ_{kl}(q)=\frac{1}{16\pi^2}\frac{P_{kl}(q)}{q^2}.
\end{equation}
This induces an effective nonlocal term in the inverse propagator which acts like a mass term with
\begin{equation}\label{FF9}
m^2=\frac{1}{16\pi^2}\big(|\gamma^*_t\bar{\varphi}_t|^2+|\gamma^*_b\bar{\varphi}_b|^2\big).
\end{equation}
Indeed, the inverse propagator takes the form
\begin{equation}\label{FF10}
\tilde{P}_{kl}(q)=\frac{P_{kl}(q)}{q^2}(q^2+m^2).
\end{equation}
We expect other contribution to $m^2_\pm$, also producing a mass term for $R^0_k$. In fact, the contribution from the loop in Fig. \ref{figx} may well be subleading, but it demonstrates a direct contribution to the nonlocal mass term.

\end{document}